\documentclass[aps,pre,twocolumn,groupedaddress,floatfix,longbibliography,superscriptaddress,10pt]{revtex4-2}

\usepackage{amsmath,amssymb,amsfonts,mathrsfs,bm,bbold,braket,slashed,cancel}

\usepackage{graphicx}
\usepackage{epsfig}
\usepackage{subfigure}
\usepackage{multirow}
\usepackage{feynmf}
\usepackage{subcaption}
\usepackage{soul}
\usepackage{cancel}
\usepackage{times}

\usepackage[dvipsnames,svgnames]{xcolor}

\usepackage{hyperref}
\hypersetup{colorlinks=true,linkcolor=Blue,citecolor=Blue,urlcolor=Blue}

\usepackage{accents}
\usepackage{ulem}
\usepackage{simpler-wick}
\usepackage{orcidlink}

\DeclareUnicodeCharacter{039B}{\ensuremath{\Lambda}}

\newcommand{\bea}{\begin{eqnarray}}
\newcommand{\eea}{\end{eqnarray}}
\newcommand{\be}{\begin{equation}}
\newcommand{\ee}{\end{equation}}

\newcommand{\nn}{\nonumber}
\newcommand{\ii}{\mathrm{i}}

\newcommand{\sumint}{\mathop{\vcenter{\hbox{\ooalign{\hfil$\displaystyle\sum$\hfil\cr\hfil$\displaystyle\int$\hfil\cr}}}}}

\begin{document}

\title{Effective Fermion-Photon Vertex in a Rotating Relativistic Medium: Tree-Level Structure}

\author{Jorge David Casta\~no-Yepes~\orcidlink{0000-0002-8654-1304}}
\email{jorge.yepes@correounivalle.edu.co}
\affiliation{ Departamento de Física, Universidad del Valle, Ciudad Universitaria Meléndez,
Santiago de Cali 760032, Colombia}
\author{Alejandro Ayala~\orcidlink{0000-0003-3929-9209}}
\email{ayala@nucleares.unam.mx}
\affiliation{Instituto de Ciencias Nucleares, Universidad Nacional Autónoma de México, Apartado Postal 70-543, CdMx 04510, México.}
\author{J. E. Martínez-Herrera~\orcidlink{0009-0006-8054-0219}}
\affiliation{Programa de Física, Universidad del Quindío, Armenia 630004, Colombia}
\author{Enrique Mu\~noz~\orcidlink{0000-0003-4457-0817}}
\email{ejmunozt@uc.cl}
\affiliation{Facultad de F\'isica, Pontificia Universidad Cat\'olica de Chile, Vicu\~{n}a Mackenna 4860, Santiago, Chile}
\affiliation{Center for Nanotechnology and Advanced Materials CIEN-UC, Avenida Vicuña Mackenna 4860, Santiago, Chile}

\begin{abstract}
We study the tree-level fermion-photon effective vertex for on-shell particles in a relativistic environment undergoing rigid rotation at zero temperature. The rotational background breaks translational invariance in the transverse plane and introduces a preferred direction, leading to a distinct decomposition of the vertex into longitudinal and transverse structures with respect to the rotation axis. Using fermion and photon modes constructed in the rotating frame, we derive the corresponding selection rules and show that the energy and the longitudinal component of the momentum are conserved, whereas no independent transverse momentum-conservation condition arises. Instead, the transverse dependence is governed by radial integrals involving three Bessel functions, whose support constrains the magnitudes of the transverse momentum projections. The vertex can be organized in a circular basis for the transverse components and a Cartesian basis along the rotation axis, and expressed in terms of six Dirac structures. Their coefficients define 24 form-factor components associated with the photon-mode components, physical photon states, and spin-dependent structures induced by the rotating background. We also verify current conservation and gauge invariance of the resulting tree-level matrix element under the boundary conditions considered. These results provide the tree-level structure of the fermion-photon interaction in rotating relativistic matter and establish the basis for the study of radiative corrections in vortical backgrounds.
\end{abstract}

\maketitle 

\section{Introduction}
\label{secI}

The large orbital angular momentum generated in noncentral relativistic heavy-ion collisions provides a natural setting in which quantum fields evolve in a vortical background. The resulting vorticity is not merely a macroscopic property of the fluid, but introduces a preferred spatial direction and modifies the kinematics and symmetry structure relevant for microscopic processes. In particular, measurements of global hyperon polarization have provided evidence for large vorticity in the matter produced at RHIC~\cite{STAR:2017ckg}, with angular velocities reaching values of order $9\times10^{21}\,\text{s}^{-1}$. Subsequent measurements of global and local polarization at RHIC, the LHC, and lower collision energies have further established the relevance of vorticity for the spin dynamics of the produced matter~\cite{STAR:2018gyt,STAR:2021beb,STAR:2023nvo,ALICE:2019onw,HADES:2022enx,STAR:2020xbm,STAR:2019erd,STAR:2023eck,Sarkar:2022uyt,ALICE:2026eje,STAR:2022fan}.

The coupling between vorticity and microscopic degrees of freedom has  been investigated in a variety of approaches, particularly in connection with the polarization of quarks and hadrons~\cite{Liang:2004ph,Betz:2007kg,Gao:2007bc,Becattini:2007sr,Becattini:2013fla,Becattini:2015ska,Becattini:2016gvu,Florkowski:2018ahw,Pang:2016igs,Karpenko:2016jyx,Sun:2017xhx,Li:2017slc,Xia:2018tes,Wei:2018zfb,Fu:2020oxj,Liu:2021uhn,Ayala:2021xrn,Ayala:2020soy,Ayala:2023xyn,Ayala:2026wro,Becattini:2020ngo}. From the quantum-field-theoretical perspective, however, the effects of rotation extend beyond spin polarization. A rotating background breaks spatial isotropy and selects the direction of the angular velocity, thereby distinguishing the longitudinal and transverse sectors of the dynamics. Then, quantities that in vacuum satisfy the full Lorentz symmetry can acquire additional tensor structures associated with the rotation.

This structure has been explored for single-particle propagation in rotating systems, where the fermion and boson spectra, propagators, and self-energies are modified by the rotational background~\cite{Ayala:2021osy,Gaspar:2023nqk,Ayala:2026dmz,CastanoYepes:2025,Ambrus_PhysRevD.93.104014}. Similar modifications arise at the level of interaction vertices when an external background breaks the Lorentz symmetry of the vacuum. In particular, recent studies of the QED fermion-photon vertex in a constant magnetic field have shown that the vertex acquires distinct longitudinal and transverse contributions already at tree level, while radiative corrections generate additional tensor structures associated with the anomalous magnetic moment~\cite{Ayala:2026eja}. These results illustrate how a background field can modify the tensor structure of an elementary interaction and motivate the corresponding analysis for vortical backgrounds.

In this work, we address this question by computing the tree-level fermion-photon effective vertex for on-shell particles in a rigidly rotating background at zero temperature. We construct the fermion and photon modes in the rotating frame and use them to evaluate the corresponding matrix element. Owing to the tensor structure of the photon modes, the transverse part of the vertex is naturally expressed in a circular basis defined with respect to the rotation axis, while the longitudinal component is described in a Cartesian basis. The resulting vertex exhibits a qualitative difference from its vacuum counterpart: the energy and the longitudinal component of the momentum remain conserved, whereas transverse momentum is not conserved. The radial structure of the rotating modes instead leads to nontrivial support conditions for the transverse momentum magnitudes. The resulting tensor decomposition can be expressed in terms of 24 form-factor components, which encode the photon-mode components, the physical photon states, and the spin-dependent structures associated with the fermion coupling to the rotating background.

We consider on-shell fermions and photons whose wave functions incorporate the effects of the vortical motion of the medium. We show that, as a consequence of the tensor structure of the photon field, the vertex is naturally expressed in a circular basis for the transverse directions, defined with respect to the axis of rotation, together with a Cartesian component along the rotation axis. We demonstrate that, in contrast to the energy and the $z$-component of the momentum, the transverse momentum is not conserved in the interaction. Furthermore, this tensor decomposition leads to 24 form-factor components, which characterize the different photon-mode components, physical photon states, and spin-dependent structures associated with the coupling of the fermions to the rotating background.

The manuscript is organized as follows. In Sec.~\ref{secII}, we formulate the QED theory in rigidly rotating coordinates and introduce the corresponding metric, tetrad, spin connection, and covariant derivative. The action is then separated into a free part, which contains the full dependence on the rotating background, and the electromagnetic interaction. In Sec.~\ref{secIII}, we establish the perturbative expansion and define the tree-level fermion-photon vertex. Sec.~\ref{secIV} is devoted to the exact positive- and negative-energy fermion modes in the rotating frame and to the quantization of the fermion field, with the explicit derivation of the fermion modes and their normalization given in Appendix~\ref{app:RotatingFermionModes}. The construction and quantization of the physical photon modes are presented in Sec.~\ref{sec:photonmodes}, which summarizes the recent results presented in Ref.~\cite{CastanoYepes:2026PhotonRotating}. In Sec.~\ref{sec:treelevel}, these modes are used to construct the tree-level effective fermion-photon vertex and its decomposition into longitudinal and transverse structures. Sec.~\ref{sec:currentconservation} examines current conservation and the conditions under which the resulting tree-level matrix element is gauge invariant. In Sec.~\ref{sec:results}, we present the angular-momentum selection rules, the support of the transverse radial integrals, the resulting nonconservation of transverse momentum, and the associated kinematic constraints. We then express the rotating vertex in terms of form factors associated with independent Dirac structures. Finally, Sec.~\ref{sec:conclusions} summarizes the main results and discusses their implications.

\section{Rotating relativistic background}
\label{secII}

We construct the fermion-photon vertex in the presence of a uniformly
rotating background. The rotational effects are incorporated
nonperturbatively in the free fermion {and photon} dynamics, while the
electromagnetic interaction is treated perturbatively. We consider
the fermion-photon scattering process
\bea
f_\alpha+\gamma_\mu(q)\longrightarrow f_{\alpha'},
\label{VertexProcess}
\eea
Here, $\alpha=\left(l,h,p_z,|\boldsymbol{p}_\perp|\right)$ and $\alpha'$ distinguish the initial and final
fermion modes, respectively. The external photon is
characterized by its momentum $q$ and Lorentz index $\mu$, {and its angular momentum state}.

The starting point is the Dirac action in the rotating frame. We
introduce the tetrad $\mathrm{e}^\mu_{\ a}(x)$ through
\bea
g^{\mu\nu}(x)=\mathrm{e}^\mu_{\ a}(x)e^\nu_{\ b}(x)\eta^{ab},
\label{InverseMetrictetrad}
\eea
where $\eta^{ab}={\rm diag}(1,-1,-1,-1)$ is the Minkowski metric.
For rigid rotation around the $z$ axis, the nonvanishing components of
the tetrad are (see Appendix~\ref{app:RotatingFermionModes} for details)
\bea
&&\mathrm{e}^0_{\ 0}=1,\qquad
\mathrm{e}^1_{\ 1}=1,\qquad
\mathrm{e}^2_{\ 2}=1,\qquad
\mathrm{e}^3_{\ 3}=1,\qquad\nn\\
&&\mathrm{e}^1_{\ 0}=\Omega y,\qquad
\mathrm{e}^2_{\ 0}=-\Omega x.
\label{Rotatingtetrad}
\eea

The determinant of the tetrad is equal to one. The Dirac matrices
associated with the {non-inertial rotating system} are  obtained from~\cite{CastanoYepes:2025}
\bea
\gamma^\mu(x)=\mathrm{e}^\mu_{\ a}(x)\gamma^a,
\label{CurvedDiracMatrices}
\eea
where the matrices $\gamma^a$ satisfy the usual Minkowski-space
Clifford algebra,
\bea
\left\{\gamma^a,\gamma^b\right\}=2\eta^{ab},
\label{MinkowskiCliffordAlgebra}
\eea
while the matrices $\gamma^{\mu}(x)$ satisfy
\bea
\left\{\gamma^{\mu}(x),\gamma^{\nu}(x) \right\} = 2 g^{\mu\nu}(x).
\eea
For the tetrad in Eq.~\eqref{Rotatingtetrad}, Eq.~\eqref{CurvedDiracMatrices}
gives
\bea
&&\gamma^0(x)=\gamma^0,\qquad
\gamma^1(x)=\gamma^1+\Omega y\gamma^0,\qquad\nn\\
&&\gamma^2(x)=\gamma^2-\Omega x\gamma^0,\qquad
\gamma^3(x)=\gamma^3.
\label{RotatingGammaMatrices}
\eea
The covariant derivative acting on the fermion field is
\bea
{\cal D}_\mu
\equiv
\partial_\mu+\Gamma_\mu+\ii q_f A_\mu,
\label{GaugeCovariantDerivative}
\eea
where $q_f$ is the fermion electric charge and $\Gamma_\mu$ denotes
the spin connection. For the rotating tetrad considered here, its
nonvanishing components are
\bea
\Gamma_0=-\frac{\ii}{2}\Omega\Sigma_3,\qquad
\Gamma_1=\Gamma_2=\Gamma_3=0,
\label{RotatingSpinConnection}
\eea
{where $\Sigma_3 = i\gamma^1\gamma^2$ is the z-component of the spin operator.}
The minimally coupled Dirac action is therefore
\bea
S[\bar\psi,\psi,A]
&=&
\int d^4x\,\sqrt{-g}\;\bar\psi(x)\nn\\
&&\times
\left[
\ii\gamma^\mu(x){\cal D}_\mu-m_f
\right]\psi(x),
\label{MinimallyCoupledAction}
\eea
{where, for the particular tetrad system involved in the rotating frame, we have $\sqrt{-g} = 1$ (see Eq.~\eqref{PhotonRotatingMetric}).}
Using Eq.~\eqref{GaugeCovariantDerivative}, the Dirac operator can be
separated into a free contribution, which contains the full
dependence on the rotating background, and the electromagnetic
interaction~\cite{CastanoYepes:2025},
\bea
&&\ii\gamma^\mu(x){\cal D}_\mu-m_f\nn\\
&=&\ii\gamma^\mu(x)
\left(
\partial_\mu+\Gamma_\mu
\right)
-m_f-q_f\gamma^\mu(x)A_\mu.
\label{DiracOperatorDecomposition}
\eea

We first evaluate the free contribution. Substituting
Eqs.~\eqref{RotatingGammaMatrices} and
\eqref{RotatingSpinConnection} into
Eq.~\eqref{DiracOperatorDecomposition}, we obtain
\bea
&&\ii\gamma^\mu (x)
\left(
\partial_\mu+\Gamma_\mu
\right)-m_f\nn\\
&=&
\ii\gamma^0
\left(
\partial_t-\frac{\ii}{2}\Omega\Sigma_3
\right)
+\ii\left(\gamma^1+\Omega y\gamma^0\right)\partial_x\nn\\
&+&\ii\left(\gamma^2-\Omega x\gamma^0\right)\partial_y
+\ii\gamma^3\partial_z-m_f.\nn\\
\label{FreeOperatorFirstStep}
\eea

The terms proportional to $\Omega$ have two origins. The terms
involving the transverse coordinates arise from the spatial
dependence of the tetrad, whereas the spin-dependent contribution
originates from the spin connection. Collecting these terms gives~\cite{CastanoYepes:2025}
\bea
&&\ii\gamma^\mu(x)
\left(
\partial_\mu+\Gamma_\mu
\right)-m_f\nn\\
&=&
\ii\gamma^a\partial_a
+\ii\Omega\gamma^0
\left(
y\partial_x-x\partial_y
\right)
+\frac{\Omega}{2}\gamma^0\Sigma_z-m_f.\nn\\
\label{FreeOperatorSecondStep}
\eea

The orbital contribution is expressed in terms of the $z$ component
of the orbital angular momentum,
\bea
L_z=-\ii\left(x\partial_y-y\partial_x\right),
\label{OrbitalAngularMomentum}
\eea
and the total angular momentum is
\bea
J_z=L_z+\frac{\Sigma_z}{2}.
\label{AngularMomentumOperators}
\eea
Then, the free Dirac operator takes the compact form
\bea
\ii\gamma^\mu(x)
\left(
\partial_\mu+\Gamma_\mu
\right)-m_f
=
\ii\gamma^a\partial_a+\gamma^0\Omega J_z-m_f.
\label{FreeRotatingDiracOperator}
\eea

This expression is the central result for the free rotating theory.
The rotational background is therefore incorporated through the
coupling of the angular velocity to the total angular momentum. The
fermion modes used as external states and the corresponding free
propagator are defined with respect to this operator.

We next isolate the electromagnetic interaction. In the tetrad
formalism, the contraction of the electromagnetic field with the
coordinate-dependent Dirac matrices can be written as
\bea
\gamma^\mu(x)A_\mu
=
\mathrm{e}^\mu_{\ a}(x)\gamma^a A_\mu.
\label{PhotonCouplingtetrad}
\eea
Here, $\gamma^a$ denote the constant Dirac matrices in the local
Minkowski frame, {as defined in Eq.~\eqref{MinkowskiCliffordAlgebra}}. For the rotating tetrad used here, the contraction
reads
\bea
&&\gamma^\mu(x)A_\mu\nn\\
&=&
\gamma^0
\left(
A_0+\Omega y A_1-\Omega x A_2
\right)
+\gamma^1A_1+\gamma^2A_2+\gamma^3A_3.\nn\\
\label{PhotonCouplingRotatingFrame}
\eea
Equation~\eqref{PhotonCouplingRotatingFrame} shows explicitly how the
geometry of the rotating frame enters the electromagnetic coupling.
The terms proportional to $\Omega$ originate from the spatial
dependence of the tetrad and therefore do not represent a new
electromagnetic interaction. Rather, they arise from expressing the
same local gauge connection in the rotating coordinate basis.

It is therefore useful to retain the tetrad explicitly in the
interaction term rather than absorb its contribution into a
redefinition of the electromagnetic field. Using
$\sqrt{-g}=1$ for the {rotating tetrad system} considered here, the action
takes the form
\bea
S[\bar\psi,\psi,A]
&=&
\int d^4x\,\bar\psi(x)
\left[
\ii\gamma^a\partial_a+\gamma^0\Omega J_z-m_f
\right]\psi(x)\nn\\
&-&q_f\int d^4x\,
\bar\psi(x)\mathrm{e}^\mu_{\ a}(x)\gamma^a A_\mu(x)\psi(x).
\label{RotatingQEDAction}
\eea
This form displays the two ingredients of the theory explicitly.
The rotational background enters the free Dirac operator through
$\gamma^0\Omega J_z$. The electromagnetic field instead enters through
the gauge connection $\mathrm{e}^\mu_{\ a}(x)\gamma^a A_\mu(x)$. 

We can  separate the action into a free part and an
interaction part,
\bea
S[\bar\psi,\psi,A]
=
S_0[\bar\psi,\psi]
+
S_{\rm int}[\bar\psi,\psi,A],
\label{ActionDecomposition}
\eea
with
\bea
S_0[\bar\psi,\psi]
=
\int d^4x\,\bar\psi(x)
\left[
\ii\gamma^a\partial_a+\gamma^0\Omega J_z-m_f
\right]\psi(x),
\label{FreeRotatingAction}
\eea
and
\bea
S_{\rm int}[\bar\psi,\psi,A]
=
-q_f\int d^4x\,
\bar\psi(x)\mathrm{e}^\mu_{\ a}(x)\gamma^a A_\mu(x)\psi(x).
\label{InteractionAction}
\eea
This separation fixes the perturbative organization used below. The
rotational background is treated exactly through $S_0$, whereas the
expansion is performed in powers of the electromagnetic interaction
$S_{\rm int}$. Thus, the perturbative expansion is not an expansion in
the angular velocity $\Omega$. The latter remains present in the free
fermion modes and propagators, as well as in the matrix elements of
the interaction through the tetrad.

\section{Perturbative expansion and tree-level vertex}
\label{secIII}

The expansion is performed in powers of $q_f$, while the dependence on the angular velocity $\Omega$ is retained exactly in the free fermion dynamics and in the interaction through the tetrad.

The interaction-picture $\mathcal{S}$ matrix is
\bea
{\cal S}={\cal T}\exp\left(\ii S_{\rm int}\right) = 1+\sum_{n=1}^{\infty}{\cal S}^{(n)},
\label{SMatrixDefinition}
\eea
where ${\cal T}$ is the time ordering operator, and the terms of the $\mathcal{S}$-matrix perturbative expansion are defined by
\bea
{\cal S}^{(n)}=\frac{\ii^n}{n!}{\cal T}\left[S_{\rm int}^n\right].
\label{SMatrixExpansionTerms}
\eea
The interaction action is
\bea
S_{\rm int}=-q_f\int d^4x\,\bar\psi(x)\gamma^\mu(x)A_\mu(x)\psi(x),
\label{InteractionActionPerturbative}
\eea
where $\gamma^\mu(x)$ is the  Dirac matrix in the non-inertial rotating frame defined in Eq.~\eqref{CurvedDiracMatrices}. Using the tetrad representation, the interaction takes the form
\bea
S_{\rm int}=-q_f\int d^4x\,\bar\psi(x)\mathrm{e}^\mu_{\ a}(x)\gamma^a A_\mu(x)\psi(x).
\label{InteractionActiontetrad}
\eea
The rotational background therefore enters the electromagnetic interaction through the spacetime-dependent tetrad, in addition to its presence in the free fermion modes.

The fermion-photon matrix element is obtained by evaluating the $S$-matrix between the one-fermion and one-photon external states introduced below. At each order in the electromagnetic interaction, the dependence on the external photon is linear and can  be separated from the fermion matrix element. We therefore define the vertex amplitude with the photon Lorentz index uncontracted as
\bea
{\cal M}_{\alpha'\alpha}^{\mu}(p',p;q)=\sum_{n=0}^{\infty}{\cal M}_{\alpha'\alpha}^{\mu(n)}(p',p;q).
\label{VertexPerturbativeExpansion}
\eea
In particular,
\bea
{\cal M}_{\alpha'\alpha}^{\mu}(p',p;q)
&=&{\cal M}_{\alpha'\alpha}^{\mu(0)}(p',p;q)\nn\\
&+&{\cal M}_{\alpha'\alpha}^{\mu(1)}(p',p;q)+\cdots.
\label{VertexAmplitudeExpansion}
\eea
The superscripts $(0)$ and $(1)$ denote the leading-order and one-loop contributions, respectively. 
%


\section{Rotating fermion modes}\label{secIV}

We first specify the free-field representation used to construct the
external states. The free fermion modes used to construct the external states are solutions of the free rotating Dirac equation~\cite{Ayala:2021osy,CastanoYepes:2025},
\bea
\left[\ii\gamma^a\partial_a+\gamma^0\Omega J_z-m_f\right]U_{\alpha}(x)=0,
\label{PositiveEnergyRotatingMode}
\eea
together with the corresponding negative-energy solutions $V_{\alpha}(x)$. We recall that the collective label
$\alpha$ specifies the complete set of quantum numbers of the fermion mode,
\bea
\alpha=\left(l,h,p_z,|\boldsymbol{p}_\perp|\right),
\label{FermionModeQuantumNumbers}
\eea
where $l\in\mathbb{Z}$ is the orbital-angular-momentum quantum number, $h=\pm 1/2$ is the helicity, and $p_z$ and $|\boldsymbol{p}_\perp|$ are the continuous longitudinal and transverse momentum variables. The positive-energy rotating modes can be written as
\bea
U_{\alpha}(x)=\frac{1}{2\pi}e^{-\ii\widetilde E_jt}u_{\alpha}(x),
\label{FermionPositiveRotatingMode}
\eea
with
\bea
u_{\alpha}(x)=e^{\ii p_zz}\begin{pmatrix}\varphi_{\alpha}(\rho,\phi)\\\chi_{\alpha}(\rho,\phi)\end{pmatrix},
\label{FermionPositiveBispinor}
\eea
where the upper Pauli spinor is
\bea
\varphi_{\alpha}(\rho,\phi)=\frac{{\cal N}}{\sqrt{2}}\begin{pmatrix}N_{2h}J_l\left(|\boldsymbol{p}_\perp|\rho\right)e^{\ii l\phi}\\2\ii hN_{-2h}J_{l+1}\left(|\boldsymbol{p}_\perp|\rho\right)e^{\ii(l+1)\phi}\end{pmatrix},
\label{FermionUpperPauliSpinor}
\eea
and the lower Pauli spinor is
\bea
\chi_{\alpha}(\rho,\phi)=\frac{2h|\boldsymbol{p}|}{E+m_f}\varphi_{\alpha}(\rho,\phi).
\label{FermionLowerPauliSpinor}
\eea
Here, $J_l\left(|\boldsymbol{p}_\perp|\rho\right)$ denotes the Bessel function of the first kind, ${\cal N}$ is the normalization factor, 
\bea
{\cal N}=\sqrt{\frac{E+m_f}{2E}},
\label{AppNormalizationFactor_main}
\eea
and
\bea
N_{2h}=\sqrt{1+\frac{2hp_z}{|\boldsymbol{p}|}},
\label{FermionNpmDefinition}
\eea
with $|\boldsymbol{p}|$ denoting the magnitude of the three-momentum. The energy $E$ and the momentum variables are related by the dispersion relation given above, while the rotating-frame energy is
\bea
\widetilde E_{l+1/2}=E-\Omega(l+1/2).
\label{E-tilde-fermion}
\eea
The corresponding negative-energy modes are constructed analogously from the solutions $V_{\alpha}(x)$, with the explicit spinor structure and normalization given in Appendix~\ref{app:RotatingFermionModes}.

The use of helicity to label the fermion states deserves a brief comment. Since the rotational background is characterized by the angular velocity $\boldsymbol{\Omega}$, one might be tempted to label the states by the projection of the spin along the rotation axis, in analogy with the usual treatment of a fermion in a uniform magnetic field, where the spin projection along the field direction provides a natural quantum number. However, the analogy between a magnetic field and vorticity does not extend to the structure of the free fermion eigenstates in this way. In the present problem, the free Hamiltonian commutes with the helicity operator~\cite{CastanoYepes:2025},
\bea
\hat W_0=\frac{\hat{\boldsymbol{\Sigma}}\cdot\hat{\boldsymbol{p}}}{2|\boldsymbol{p}|},
\qquad [\hat H_0,\hat W_0]=0,
\label{FermionHelicityOperator}
\eea
so that the positive-energy modes can be chosen as eigenstates of helicity,
\bea
\hat W_0u_{\alpha}=h u_{\alpha},
\qquad h=\pm\frac12.
\label{FermionHelicityEigenvalue}
\eea

Thus, helicity is a natural label of the free Dirac solutions used here. By contrast, the angular velocity defines the axis associated with rotational symmetry through $J_z$, but it does not generate a Zeeman-like spin projection quantum number analogous to the spin projection along a magnetic-field direction. The distinction is important because the role of the rotation axis is encoded through the total angular momentum quantum number $j=l+1/2$, whereas the spinor structure of the free Dirac solutions is fixed by the helicity condition. The explicit derivation of these modes, including the angular dependence, the Bessel functions, the relative coefficients of the Pauli spinors, the normalization factors, and the corresponding negative-energy solutions, is given in Appendix~\ref{app:RotatingFermionModes}.

Upon quantization, the coefficients of the positive- and negative-energy modes become fermion annihilation and creation operators, respectively. The fermion field is therefore expanded as
\bea
\psi(x)=\sumint_{l,h}[dp]\left[\hat b_{\alpha}U_{\alpha}(x)+\hat d_{\alpha}^{\dagger}V_{\alpha}(x)\right],
\label{FermionFieldExpansion}
\eea
with the corresponding adjoint field
\bea
\bar\psi(x)=\sumint_{l,h}[dp]\left[\hat b_{\alpha}^{\dagger}\overline{U}_{\alpha}(x)+\hat d_{\alpha}\overline{V}_{\alpha}(x)\right].
\label{AdjointFermionFieldExpansion}
\eea

The measure symbol denotes the complete set of discrete and continuous mode labels,
\bea
\sumint_{l,h}[dp]\equiv\sum_{l=-\infty}^{\infty}\sum_{h=\pm1/2}\int_{-\infty}^{\infty}dp_z\int_0^\infty |\boldsymbol{p}_\perp|\,d|\boldsymbol{p}_\perp|.
\label{FermionModeMeasure}
\eea

The operators $\hat b_{\alpha}$ and $\hat d_{\alpha}$ annihilate fermion excitations associated with the positive- and negative-energy modes, respectively, and satisfy the canonical anticommutation relations
\bea
\{\hat b_{\alpha},\hat b_{\alpha' }^\dagger\}
&=&\delta_{ll'}\delta_{hh'}(2\pi)\delta(p_z-p_z')\nn\\
&&\times(2\pi)\frac{1}{|\boldsymbol{p}_\perp|}\delta\left(|\boldsymbol{p}_\perp|-|\boldsymbol{p}_\perp'|\right),
\label{FermionAnnihilationAnticommutator}
\eea
and
\bea
\{\hat d_{\alpha},\hat d_{\alpha' }^\dagger\}
&=&\delta_{ll'}\delta_{hh'}(2\pi)\delta(p_z-p_z')\nn\\
&&\times(2\pi)\frac{1}{|\boldsymbol{p}_\perp|}\delta\left(|\boldsymbol{p}_\perp|-|\boldsymbol{p}_\perp'|\right),
\label{FermionCreationAnticommutator}
\eea
with all other anticommutators vanishing. The corresponding fermion vacuum is defined by
\bea
\hat b_{\alpha}|0_\Omega\rangle_f=0,\qquad \hat d_{\alpha}|0_\Omega\rangle_f=0.
\label{RotatingFermionVacuum}
\eea
Here, $|0_\Omega\rangle_f$ denotes the vacuum associated with the free rotating fermion theory, whose mode decomposition is determined by the free rotating Hamiltonian.

\section{Rotating Photon Modes}\label{sec:photonmodes}

The photon field is described by the Maxwell action in the rotating
coordinates introduced above. After fixing the covariant gauge with
$\xi=1$, corresponding to the Feynman gauge, the equation of motion for
the vector potential becomes
\bea
&&\nabla_\mu\nabla^\mu A^\nu=0,
\label{PhotonFeynmanGaugeWaveEquation}
\eea
{
where we introduced the Levi-Civita covariant derivative $\nabla_\mu$, whose action on a vector field $V^\nu$ is defined by 
\bea
\nabla_\mu V^\nu=\partial_\mu V^\nu+\Gamma^\nu_{\mu\lambda}V^\lambda,
\label{covariantder}
\eea
with $\Gamma^\lambda_{\mu\nu}$ the Christoffel symbols determined by the metric $g^{\mu\nu}(x)$
\bea
\Gamma^{\lambda}_{\mu\nu} = \frac{1}{2}g^{\lambda\sigma}\left( \partial_{\mu}g_{\nu\sigma} + \partial_{\nu}g_{\sigma\mu} - \partial_{\sigma}g_{\mu\nu} \right).
\label{eq:Christoffel}
\eea
}
The corresponding solutions {of Eq.~\eqref{PhotonFeynmanGaugeWaveEquation}} are conveniently organized according to the
eigenvalue of the total angular momentum operator $\hat J_z$. We denote this eigenvalue by $j$, so that
\bea
&&\hat J_z A_j^\mu=j A_j^\mu.
\eea
With this convention, $j$ labels the total angular momentum of the mode,
rather than just the orbital angular momentum of each individual component.
The temporal and longitudinal components carry orbital angular momentum
$j$, while the circular transverse components carry orbital angular
momenta $j-1$ and $j+1$, respectively. The construction of these modes,
including the action of the vector wave operator, the Lorenz condition,
and the residual gauge freedom, is given in
Ref.~\cite{CastanoYepes:2026PhotonRotating}. For propagating modes with $\mathbf q\neq0$, the residual gauge freedom
can be used to eliminate the temporal component, and the two independent physical photon modes ($\beta=1,2$) are
 \bea
\widetilde A_{j,\beta}^\mu&=&\widetilde{\mathcal N}_\beta e^{-\ii \omega t}e^{\ii q_z z}\Big[\widetilde C_\beta^+J_{j-1}(|\mathbf q_\perp|\rho)e^{\ii(j-1)\phi}\boldsymbol{\epsilon}_+^\mu\nn\\
&+&\widetilde C_\beta^-J_{j+1}(|\mathbf q_\perp|\rho)e^{\ii(j+1)\phi}\boldsymbol{\epsilon}_-^\mu+\widetilde C_\beta^zJ_j(|\mathbf q_\perp|\rho)e^{\ii j\phi}\delta^\mu_z\Big]\nn\\
\label{PhotonSymmetricPhysicalModes}
\eea  
where 
\bea
\widetilde C_1^+&=&\widetilde C_2^-=\frac{1}{2}\left(1+\frac{q_z}{|\mathbf q|}\right),\nn\\
\widetilde C_1^-&=&\widetilde C_2^+=\frac{1}{2}\left(1-\frac{q_z}{|\mathbf q|}\right),\nn\\
\widetilde C_1^z&=&- \widetilde C_2^z=-\frac{\ii}{\sqrt{2}}\frac{|\mathbf q_\perp|}{|\mathbf q|},
\eea
the rotating-frame energy is
\bea
\omega=\sqrt{|\mathbf q_\perp|^2+q_z^2}-\Omega j=|\mathbf{q}|-\Omega j,
\label{PhotonRotatingEnergy}
\eea
and the normalization constants are
\bea
\widetilde{\mathcal N}_1=\widetilde{\mathcal N}_2=\frac{1}{2\pi\sqrt{2|\mathbf{q}|}}.
\label{PhotonSymmetricNormalization}
\eea

Moreover, the circular basis vectors are defined by
\bea
\boldsymbol{\epsilon}_\pm^\mu=\frac{1}{\sqrt{2}}\left(0,1,\pm\ii,0\right),
\label{PhotonCircularPolarizations}
\eea
and satisfy
\bea
\hat S_z\boldsymbol{\epsilon}_\pm^\mu=\pm\boldsymbol{\epsilon}_\pm^\mu,
\eea
where $\hat S_z$ is the spin-one generator associated with rotations
around the $z$ axis. The orbital factors appearing in the two
transverse components combine with their spin contributions such that
each term has the same total angular momentum $j$  under
$\hat J_z=\hat L_z+\hat S_z$.

The vector potential can then be expanded in the basis of these physical
solutions as
\bea
&&A^\mu(x)=\sumint_{j,\beta}\left[ dq \right]\left[\hat a_{j,\beta}(\boldsymbol q)A_{j,\beta}^\mu(x)+\hat a_{j,\beta}^\dagger(\boldsymbol q)A_{j,\beta}^{\mu *}(x)\right],\nn\\
\label{PhotonFieldExpansion}
\eea
where $\alpha=1,2$ labels the two independent physical photon modes and
the momentum measure is chosen as
\bea
\sumint_{j,\beta} \left[ dq \right]\equiv\sum_{j=-\infty}^{\infty}\sum_{\beta=1,2}\int_{-\infty}^{\infty}dq_z\int_0^\infty |\boldsymbol{q}_\perp|\,d|\boldsymbol{q}_\perp|.
\label{PhotonModeMeasure}
\eea

The normalization factors $\widetilde{\mathcal N}_\beta$ specify the
normalization of the mode functions and are understood to be chosen
consistently with the normalization of the creation and annihilation
operators. With the above convention, these operators satisfy
\bea
&&[\hat a_{j,\beta}(\boldsymbol{q}),\hat a_{j',\alpha'}^\dagger(\boldsymbol{q}')]\nn\\
&=&\delta_{jj'}\delta_{\alpha\alpha'}\frac{1}{|\boldsymbol{q}_\perp|}\delta(|\boldsymbol{q}_\perp|-|\boldsymbol{q}_\perp'|)\delta(q_z-q_z').
\label{PhotonCommutationRelations}
\eea
with all other commutators vanishing. The photon vacuum is defined by
\bea
&&\hat a_{j,\beta}(\boldsymbol q)|0_\Omega\rangle_\gamma=0.
\label{PhotonVacuum}
\eea

\section{Tree-level effective vertex}\label{sec:treelevel}
The vacuum of the complete free theory can then be written as the tensor product
\bea
|0\rangle_\Omega=|0_\Omega\rangle_f\otimes|0_\Omega\rangle_\gamma.
\label{FullFreeVacuum}
\eea
This choice specifies the perturbative representation used below: the fermion and photon sectors are constructed from the exact modes of the rotating free Hamiltonian.

We consider the matrix element containing two external fermion states and one external photon. For photon absorption, with the external photon prepared in one of the physical modes $\beta=1,2$. The initial and final states are
\bea
|i\rangle=\hat a_{j,\beta}^\dagger(q)\hat b_{\alpha}^\dagger(p)|0\rangle_\Omega,\qquad |f\rangle=\hat b_{\alpha'}^\dagger(p')|0\rangle_\Omega.
\label{VertexExternalStates}
\eea
{Here, $j$} denotes the total angular momentum of the photon mode, while $\beta$ labels the two independent physical photon modes.

The vertex amplitude is organized according to the number of electromagnetic interactions as shown in Eq.~\eqref{VertexAmplitudeExpansion}.

Using Eq.~\eqref{InteractionActionPerturbative} and the external states defined above, the first-order matrix element is
\bea
&&{\cal M}_{\alpha'\alpha;j\beta}^{(0)}(p',p;q)\nn\\
&=&-\ii q_f\left\langle f\left|\int d^4x\,\bar\psi(x)\gamma^\mu(x)A_\mu(x)\psi(x)\right|i\right\rangle.
\label{TreeLevelAmplitudeFirstStep}
\eea
Upon substituting the free-field expansions, the annihilation operator associated with the external photon selects the corresponding positive-frequency mode $A_{j,\beta}^\mu(x;q)$, while the fermion matrix element selects the positive-energy rotating modes $U_{\alpha}(x;p)$ and $U_{\alpha'}(x;p')$. Using $\gamma^\mu(x)=\mathrm{e}^\mu_{\ a}(x)\gamma^a$, one obtains the tree-level vertex in coordinate space,
\bea
&&{\cal M}_{\alpha'\alpha;j\beta}^{(0)}(p',p;q)\nn\\
&=&-\ii q_f\int d^4x\,\overline{U}_{\alpha'}(x;p')\gamma^\mu(x)U_{\alpha}(x;p)A_{j,\beta\,\mu}(x;q)\nn\\
&=&-\ii q_f\int d^4x\,\overline{U}_{\alpha'}(x;p')\gamma^\mu(x)U_{\alpha}(x;p)g_{\mu\nu}(x)A_{j,\beta}^\nu(x;q),\nn\\
\label{TreeLevelAmplitudetetrad}
\eea
where the inverse metric $g_{\mu\nu}(x)$ is given by~\cite{CastanoYepes:2026PhotonRotating}
\bea
g_{\mu\nu}(x)&=&\eta_{ab}\,e^a_{\ \mu}(x)e^b_{\ \nu}(x)\nn\\
&=&
\begin{pmatrix}
1-\Omega^2(x^2+y^2)&\Omega y&-\Omega x&0\\
\Omega y&-1&0&0\\
-\Omega x&0&-1&0\\
0&0&0&-1
\end{pmatrix},
\label{PhotonRotatingMetric}
\eea
and $\eta_{ab}=\mathrm{diag}(1,-1,-1,-1)$.

The rotating fermion modes and their spinor structure are those specified by Eqs.~\eqref{FermionPositiveRotatingMode}--\eqref{FermionNpmDefinition}, with
\bea
\overline{U}_{\alpha}(x)=U_{\alpha}^{\dagger}(x)\gamma^0.
\label{TreeLevelRotatingAdjoint}
\eea
The external photon mode $A_{j,\beta}^\mu(x;q)$ is given by Eq.~\eqref{PhotonSymmetricPhysicalModes}, with $\beta=1,2$ denoting the two independent physical photon modes.

To separate the spatial dependence of the rotating fermion modes from
their spinor structure, we write the positive-energy modes in a
factorized form,
\bea
u_{\alpha}(x)=\mathcal R_{l,\boldsymbol p}(x)u(p,h),
\label{RotatingModeFactorization}
\eea
where $u(p,h)$ is a coordinate-independent Dirac spinor containing the
kinematic normalization, while $\mathcal R_{l,\boldsymbol p}(x)$
contains the complete spatial dependence of the mode. From
Eqs.~\eqref{FermionPositiveBispinor}--\eqref{FermionLowerPauliSpinor},
the spinor can be written as
\bea
u(p,h)=\sqrt{\frac{E+m_f}{4E}}\begin{pmatrix}\sqrt{1+\dfrac{2hp_z}{|\boldsymbol p|}}\\2\ii h\sqrt{1-\dfrac{2hp_z}{|\boldsymbol p|}}\\\dfrac{2h|\boldsymbol p|}{E+m_f}\sqrt{1+\dfrac{2hp_z}{|\boldsymbol p|}}\\\dfrac{2h|\boldsymbol p|}{E+m_f}2\ii h\sqrt{1-\dfrac{2hp_z}{|\boldsymbol p|}}\end{pmatrix}.
\label{RotatingConstantSpinor}
\eea
and the spatial matrix is then given by
\bea
&&\mathcal R_{l,\boldsymbol p}(\mathbf{x}_\perp)\nn\\
&\equiv&\mathrm{diag}\left[\mathcal F_l(\mathbf{x}_\perp),\mathcal F_{l+1}(\mathbf{x}_\perp),\mathcal F_l(\mathbf{x}_\perp),\mathcal F_{l+1}(\mathbf{x}_\perp)\right].\nn\\
\label{RotatingSpatialMatrix}
\eea
where
\bea
\mathcal F_l(\mathbf{x}_\perp)&\equiv&J_l(|\boldsymbol p_\perp|\rho)e^{\ii l\phi},\nn\\
\mathcal F_{l+1}(\mathbf{x}_\perp)&\equiv& J_{l+1}(|\boldsymbol p_\perp|\rho)e^{\ii(l+1)\phi}.
\eea

With these definitions, the positive-energy rotating modes take the form
\bea
U_{\alpha}(x)=\frac{1}{2\pi}e^{-\ii\widetilde E_jt}e^{\ii p_zz}\mathcal R_{l,\boldsymbol p}(\mathbf{x}_\perp)u(p,h).
\label{RotatingModeFactorized}
\eea

Using the factorized form of the rotating fermion modes introduced above, together with the decomposition of the physical photon modes in Eq.~\eqref{PhotonSymmetricPhysicalModes}, the coordinate dependence of the tree-level matrix element can be isolated from the spinor structure, and the spinors $u(p,h)$ and $\overline u(p',h')$ can be taken outside the coordinate integral, yielding
\bea
{\cal M}_{\alpha'\alpha;j\beta}^{(0)}=-\frac{\ii q_f}{(2\pi)^2}\overline u(p',h')\Lambda_{l,j,\beta}(\mathbf{p},\mathbf{p}',q) u(p,h),\nn\\
\label{FactorizedTreeLevelAmplitude}
\eea
where
\bea
&&\Lambda_{l,j,\beta}(\mathbf{p},\mathbf{p}',q)=\int d^4x\Big[e^{\ii(\widetilde E_{j'}-\widetilde E_j)t}e^{\ii(p_z-p_z')z}\nn\\
&\times&\overline{\mathcal R}_{l',\boldsymbol p'}(\mathbf{x}_\perp)\gamma^\mu(x)\mathcal R_{l,\boldsymbol p}(\mathbf{x}_\perp)g_{\mu\nu}(x)A_{j,\beta}^\nu(x;q)\Big],\nn\\
\eea
and
\bea
\overline{\mathcal R}_{l,\boldsymbol p}(\mathbf{x}_\perp)=\gamma^0\mathcal R_{l,\boldsymbol p}^{\dagger}(\mathbf{x}_\perp)\gamma^0.
\eea

Since the photon mode is itself a superposition of the circular polarization vectors $\boldsymbol{\epsilon}_\pm^\mu$ and the longitudinal vector $\delta^\mu_z$, its polarization structure can be separated from the corresponding spatial mode functions. We therefore write
\bea
A_{j,\beta}^{\mu}(x;q)=e^{-\ii \omega t+\ii q_zz}\sum_{s=\pm,z}\mathcal A_{j,\beta}^{s}(\mathbf{x}_\perp;\mathbf{q}_\perp)\boldsymbol{\epsilon}_s^\mu,
\label{PhotonModePolarizationDecomposition}
\eea
where $\boldsymbol{\epsilon}_z^\mu\equiv\delta^\mu_z$, while $\mathcal A_{j,\beta}^{s}$ contains the corresponding Bessel function and azimuthal dependence of each component, namely
\bea
\mathcal A_{j,\beta}^{s}(\mathbf{x}_\perp;\mathbf q_\perp)&=&\widetilde{\mathcal N}_{\beta}\widetilde C_\beta^sJ_{j+\lambda_s}(|\mathbf q_\perp|\rho)e^{\ii(j+\lambda_s)\phi},
\eea
where
\bea
\lambda_{+}=-1,\quad\lambda_{-}=+1,\quad\lambda_z=0
\eea

Then, after using Eqs.~\eqref{E-tilde-fermion} and~\eqref{PhotonRotatingEnergy}, the matrix element can then be expressed as
\bea
&&{\cal M}^{(0)}(p',p;q)\nn\\
&=&-\ii q_f\,\delta\left[E'-E-|\mathbf{q}|+\Omega(j-(l'-l))\right]\delta(p_z-p_z'+q_z)\nn\\
&\times&\overline u(p',h')\left[\sum_{s=\pm,z}\Gamma^s_{\nu}\boldsymbol{\epsilon}^{\nu}_s\right]u(p,h),
\label{EffectiveVertexAmplitude}
\eea
where
\bea
\Gamma_{\nu}^s=\int d\mathbf{x}_\perp\,\mathcal A_{j}^{s}(\mathbf{x}_\perp;\mathbf{q}_\perp)\overline{\mathcal R}_{l',\boldsymbol p'}(\mathbf{x}_\perp)\gamma_\nu(x)\mathcal R_{l,\boldsymbol p}(\mathbf{x}_\perp),\nn\\
\label{EffectiveVertexDefinition}
\eea  
with
\bea
\gamma_\nu(x)=g_{\mu\nu}(x)\gamma^\mu(x)=g_{\mu\nu}(x)\mathrm{e}^\mu_{\ a}(x)\gamma^a,
\label{eq:gamma_abajo}
\eea
and for compactness, we have suppressed the explicit mode labels $\alpha'\alpha;j\beta$ in ${\cal M}^{(0)}(p',p;q)$ and in $\Gamma^{s(0)}_{\nu}$, with their dependence understood through the corresponding fermion and photon mode functions. Moreover, we suppress the superscript $(0)$ for notational convenience, as all vertices considered below correspond to the tree-level contribution.

Let us define the matrix containing the spatial and Dirac structure of the
fermionic modes as
\bea
\widetilde{\mathcal{R}}_\nu(\mathbf{x}_\perp)\equiv\overline{\mathcal R}_{l',\boldsymbol p'}(\mathbf{x}_\perp)\gamma_\nu(x)\mathcal R_{l,\boldsymbol p}(\mathbf{x}_\perp).
\eea
Since $\mathcal R_{l,\boldsymbol p}$ is diagonal, it is convenient to introduce
\bea
\mathcal S_{l',l}(\mathbf{x}_\perp)\equiv\mathcal F_{l'}^{\prime *}(\mathbf{x}_\perp)\mathcal F_l(\mathbf{x}_\perp),
\eea
so that the product of the spatial matrices can be written as
\bea
&&\overline{\mathcal R}_{l',\boldsymbol p'}(\mathbf{x}_\perp)\mathcal R_{l,\boldsymbol p}(\mathbf{x}_\perp)\nn\\
&=&\operatorname{diag}\left(\mathcal S_{l',l},\mathcal S_{l'+1,l+1},\mathcal S_{l',l},\mathcal S_{l'+1,l+1}\right).
\eea

This diagonal matrix can be decomposed in terms of the identity and the spin matrix $\Sigma_3=\ii\gamma^1\gamma^2$ as
\bea
\overline{\mathcal R}_{l',\boldsymbol p'}\mathcal R_{l,\boldsymbol p}=\mathcal P_{l',l}^{+}\mathbb 1+\mathcal P_{l',l}^{-}\Sigma_3,
\eea
where
\bea
\mathcal P_{l',l}^\pm(\mathbf{x}_\perp)=\frac{\mathcal S_{l',l}(\mathbf{x}_\perp)\pm\mathcal S_{l'+1,l+1}(\mathbf{x}_\perp)}{2}.
\eea
and in the standard Dirac representation,
\bea
\Sigma_3=\ii\gamma^1\gamma^2=\begin{pmatrix}1&0&0&0\\0&-1&0&0\\0&0&1&0\\0&0&0&-1\end{pmatrix}.
\eea

For the present mode structure, the spatial matrices commute with the longitudinal Dirac matrices $\gamma^0$ and $\gamma^3$. Thus, for $a=0,3$,
\bea
\overline{\mathcal R}_{l',\boldsymbol p'}\gamma^a\mathcal R_{l,\boldsymbol p}=\left(\overline{\mathcal R}_{l',\boldsymbol p'}\mathcal R_{l,\boldsymbol p}\right)\gamma^a.
\eea

On the other hand, the spatial matrices do not commute with $\gamma^1$ and $\gamma^2$, and the corresponding contractions must therefore be evaluated directly. Introducing the crossed spatial products
\bea
\mathcal Q_{l',l}^{\pm}(\mathbf{x}_\perp)&\equiv&\frac{1}{2}\left(\mathcal F_{l'}^{\prime *}\mathcal F_{l+1}\pm\mathcal F_{l'+1}^{\prime *}\mathcal F_l\right),
\eea
the transverse contractions can be written as
\bea
\overline{\mathcal R}_{l',\boldsymbol p'}\gamma^1\mathcal R_{l,\boldsymbol p}&=&\mathcal Q_{l',l}^{+}\gamma^1+\ii\mathcal Q_{l',l}^{-}\gamma^2,\nn\\
\overline{\mathcal R}_{l',\boldsymbol p'}\gamma^2\mathcal R_{l,\boldsymbol p}&=&\mathcal Q_{l',l}^{+}\gamma^2-\ii\mathcal Q_{l',l}^{-}\gamma^1.
\eea

Moreover, using the rotating metric and the tetrad introduced above, Eq.~\eqref{eq:gamma_abajo} takes the form
\bea
\gamma_{\nu}(x)&=&g_{\mu\nu}(x)\mathrm{e}^\mu_{\ a}(x)\gamma^a\nn\\
&=&
\begin{pmatrix}
1 & \Omega y & -\Omega x & 0\\
0 & -1 & 0 & 0\\
0 & 0 & -1 & 0\\
0 & 0 & 0 & -1
\end{pmatrix}\gamma^a.
\eea
whose non-vanishing components are
\bea
\gamma_0(x)&=&\gamma^0+\Omega y\,\gamma^1-\Omega x\,\gamma^2,\nn\\
\gamma_1(x)&=&-\gamma^1,\nn\\
\gamma_2(x)&=&-\gamma^2,\nn\\
\gamma_3(x)&=&-\gamma^3,
\eea
or
\bea
\gamma_\nu(x)&=&\delta_{\nu0}\left(\gamma^0+\Omega y\,\gamma^1-\Omega x\,\gamma^2\right)-\sum_{i=1}^{3}\delta_{\nu i}\gamma^i.
\eea

Combining these relations,
\begin{widetext}
\bea
\Gamma_{\nu}^s\left( \mathbf{q}_{\perp} \right)  &=&\int d\mathbf{x}_\perp\,\mathcal A_{j}^{s}(\mathbf{x}_\perp;\mathbf{q}_\perp)
\Bigg\{
\left(\mathcal P_{l',l}^{+}(\mathbf{x}_\perp)\mathbb 1+\mathcal P_{l',l}^{-}(\mathbf{x}_\perp)\Sigma_3\right)
\left(\gamma^0\delta_{\nu0}-\gamma^3\delta_{\nu3}\right)\nn\\
&+&\left(\mathcal Q_{l',l}^{+}(\mathbf{x}_\perp)\gamma^1+\ii\mathcal Q_{l',l}^{-}(\mathbf{x}_\perp)\gamma^2\right)
\left(\Omega y\,\delta_{\nu0}-\delta_{\nu1}\right)+\left(\mathcal Q_{l',l}^{+}(\mathbf{x}_\perp)\gamma^2-\ii\mathcal Q_{l',l}^{-}(\mathbf{x}_\perp)\gamma^1\right)
\left(-\Omega x\,\delta_{\nu0}-\delta_{\nu2}\right)
\Bigg\}.
\label{Gammanu-general}
\eea
\end{widetext}

Note that, although $\Gamma_\nu^s$ is defined as a four-vector with respect to the Lorentz index $\nu$, the contraction entering Eq.~\eqref{EffectiveVertexAmplitude} receives no contribution from its $\nu=0$ component, since the physical polarization vectors satisfy $\epsilon_\pm^0=\delta_z^0=0$. For the transverse photon modes, this contraction naturally selects the circular combinations of the transverse components of the effective vertex. In fact, since $\boldsymbol{\epsilon}_\pm^\mu=(0,1,\pm\ii,0)/\sqrt{2}$, we define
\bea
\Gamma_\pm^s\equiv\frac{1}{\sqrt{2}}\left(\Gamma_1^s\pm\ii\Gamma_2^s\right),
\eea
such that
\bea
\Gamma_\nu^s\epsilon_\pm^\nu=\frac{1}{\sqrt{2}}\left(\Gamma_1^s\pm\ii\Gamma_2^s\right)=\Gamma_\pm^s.
\eea

This definition allows us to express the transverse contribution in the same circular basis as the physical photon modes. To obtain these combinations explicitly, we first extract the $\nu=1$ and $\nu=2$ components from Eq.~\eqref{Gammanu-general},
    \bea
\Gamma_1^s&=&-\int d\mathbf{x}_\perp\,\mathcal A_{j}^{s}(\mathbf{x}_\perp;\mathbf{q}_\perp)\left(\mathcal Q_{l',l}^{+}(\mathbf{x}_\perp)\gamma^1+\ii\mathcal Q_{l',l}^{-}(\mathbf{x}_\perp)\gamma^2\right),\nn\\
\Gamma_2^s&=&-\int d\mathbf{x}_\perp\,\mathcal A_{j}^{s}(\mathbf{x}_\perp;\mathbf{q}_\perp)\left(\mathcal Q_{l',l}^{+}(\mathbf{x}_\perp)\gamma^2-\ii\mathcal Q_{l',l}^{-}(\mathbf{x}_\perp)\gamma^1\right).\nn\\
\eea

Combining these components according to the circular basis gives
\bea
\Gamma_\pm^s&=&-\frac{1}{\sqrt{2}}\left(\gamma^1\pm\ii\gamma^2\right)\nn\\
&\times&\int d\mathbf{x}_\perp\,\mathcal A_{j}^{s}(\mathbf{x}_\perp;\mathbf{q}_\perp)\left(\mathcal Q_{l',l}^{+}(\mathbf{x}_\perp)\pm\mathcal Q_{l',l}^{-}(\mathbf{x}_\perp)\right).\nn\\
\eea

The combinations of $\mathcal Q_{l',l}^{\pm}$ appearing in the transverse components are not independent structures. Rather, they reorganize the crossed products of the radial mode functions associated with the two neighboring angular momentum components of the fermion modes. In the circular basis, these combinations therefore select a definite pairing of the $l$ and $l+1$ components, with the two circular sectors corresponding to the two possible shifts of the fermionic angular structure. Explicitly, from the definition of $\mathcal Q_{l',l}^{\pm}$, one finds
\bea
\mathcal Q_{l',l}^{+}+\mathcal Q_{l',l}^{-}&=&\mathcal F_{l'}^{\prime *}\mathcal F_{l+1},\nn\\
\mathcal Q_{l',l}^{+}-\mathcal Q_{l',l}^{-}&=&\mathcal F_{l'+1}^{\prime *}\mathcal F_l.
\eea
Then, the circular components of the effective vertex isolate these two transverse mode combinations according to
\bea
\Gamma_+^s&=&-\frac{1}{\sqrt{2}}\left(\gamma^1+\ii\gamma^2\right)\nn\\
&\times&\int d\mathbf{x}_\perp\,\mathcal A_{j}^{s}(\mathbf{x}_\perp;\mathbf{q}_\perp)
\mathcal F_{l'}^{\prime *}(\mathbf{x}_\perp)\mathcal F_{l+1}(\mathbf{x}_\perp),
\label{eq:Gamma+}
\eea
\bea
\Gamma_-^s&=&-\frac{1}{\sqrt{2}}\left(\gamma^1-\ii\gamma^2\right)\nn\\
&\times&\int d\mathbf{x}_\perp\,\mathcal A_{j}^{s}(\mathbf{x}_\perp;\mathbf{q}_\perp)
\mathcal F_{l'+1}^{\prime *}(\mathbf{x}_\perp)\mathcal F_l(\mathbf{x}_\perp).
\label{eq:Gamma-}
\eea

The longitudinal component is obtained by projecting the effective vertex onto the $z$ direction, $\Gamma_z^s=\Gamma_\nu^s\delta^\nu_z=\Gamma_3^s$. In this case, the tetrad contribution is restricted to $e_{33}=-1$, while the transverse tetrad components do not contribute. Since $\gamma^3$ commutes with $\Sigma_3$, the same decomposition of the fermionic spatial structure in terms of $\mathcal P_{l',l}^{\pm}$ can be used directly, yielding
\bea
\Gamma_z^s=-\int d\mathbf{x}_\perp\,\mathcal A_{j}^{s}(\mathbf{x}_\perp;\mathbf{q}_\perp)\left[\mathcal P_{l',l}^{+}(\mathbf{x}_\perp)+\mathcal P_{l',l}^{-}(\mathbf{x}_\perp)\Sigma_3\right]\gamma^3.\nn\\
\label{eq:Gammaz}
\eea

\section{Current conservation}\label{sec:currentconservation}

It is important to verify that the fermion current associated with the vertex satisfies the appropriate conservation law.

The fermion current is
\bea
j^\mu(x)=\bar\psi(x)\gamma^\mu(x)\psi(x).
\label{FermionCurrent}
\eea

Since $j^\mu$ is a spacetime vector with no free spinor indices, its conservation law is expressed through its spacetime covariant divergence. Using the identity
\bea
\nabla_\mu\gamma^\nu\equiv\partial_\mu\gamma^\nu+\Gamma^\nu_{\mu\lambda}\gamma^\lambda+[\Gamma_\mu,\gamma^\nu]=0,
\label{GammaCompatibility}
\eea
one obtains
\bea
\nabla_\mu j^\mu=\left(\bar\psi\overleftarrow D_\mu\right)\gamma^\mu\psi+\bar\psi\gamma^\mu D_\mu\psi,
\label{currentdiv}
\eea
{
where we remark that the covariant derivative acting on the adjoint spinor from the left leads to the expression
\bea
\bar\psi\overleftarrow D_\mu\equiv\partial_\mu\bar\psi-\bar\psi\Gamma_\mu-iq_fA_\mu\bar\psi.
\label{AdjointGaugeSpinCovariantDerivative}
\eea
}
The Dirac equation and its adjoint are
\bea
\left(i\gamma^\mu D_\mu-m_f\right)\psi&=&0,\nn\\
i\left(\bar\psi\overleftarrow D_\mu\right)\gamma^\mu+m_f\bar\psi&=&0,
\label{DiracEquationsCurrent}
\eea
so that multiplying the first equation from the left by $\bar\psi$ and using the second one in Eq.~\eqref{currentdiv}, the mass terms cancel and one obtains
\bea
\nabla_\mu j^\mu=0.
\label{currcons}
\eea

Finally, since the divergence of a vector can be written as
\bea
\nabla_\mu j^\mu=\frac{1}{\sqrt{-g}}\partial_\mu\left(\sqrt{-g}\,j^\mu\right),
\label{VectorDivergence}
\eea
and $\sqrt{-g}=1$ for the  {rotating frame metric} considered here, Eq.~\eqref{currcons} reduces to
\bea
\partial_\mu j^\mu=0.
\label{currconsCartesian}
\eea

The same argument applies to the transition current constructed from two on-shell fermion modes,
\bea
j^\mu_{fi}(x)=\overline U_f(x)\gamma^\mu(x)U_i(x),
\label{TransitionCurrent}
\eea
provided that $U_i$ and $U_f$ satisfy the same Dirac equation with the same fermion mass. The interaction matrix element can then be written as
\bea
{\mathcal M}_{fi}=-iq_f\int_{\mathcal V}d^4x\,j^\mu_{fi}(x)A_\mu(x),
\label{InteractionMatrixElement}
\eea
where ${\mathcal V}$ denotes the spacetime integration region. Under a gauge transformation
\bea
A_\mu\longrightarrow A_\mu+\partial_\mu\Lambda,
\eea
the corresponding variation of the matrix element is
\bea
\Delta{\mathcal M}_{fi}=-iq_f\int_{\mathcal V}d^4x\,j^\mu_{fi}\partial_\mu\Lambda.
\label{GaugeVariation}
\eea
Using integration by parts and Eq.~\eqref{currconsCartesian}, this becomes
\bea
\Delta{\mathcal M}_{fi}&=&iq_f\int_{\mathcal V}d^4x\,\Lambda\,\partial_\mu j^\mu_{fi}-iq_f\int_{\partial\mathcal V}d\Sigma_\mu\,\Lambda j^\mu_{fi}\nn\\
&=&-iq_f\int_{\partial\mathcal V}d\Sigma_\mu\,\Lambda j^\mu_{fi}.
\label{GaugeVariationSurface}
\eea
Thus, local current conservation eliminates the volume contribution, while gauge invariance of the matrix element additionally requires the surface contribution to vanish.

In the present treatment, the photon modes are taken to satisfy the boundary condition that suppresses the field at the outer boundary. We may therefore choose the spatial integration region to be a cylinder of radius $R$ satisfying $R\Omega<1$, so that the entire cylindrical boundary remains inside the light cylinder. For a rigidly rotating fermion configuration with fixed orbital radius $R$, the transition current is tangent to the cylindrical boundary,
\bea
j^\mu_{fi}d\Sigma_\mu=0,
\eea
and the contribution {normal to the cylindrical surface in} Eq.~\eqref{GaugeVariationSurface} vanishes. The remaining temporal boundary contributions are treated in the usual scattering prescription, where the initial and final states are specified asymptotically. Under these conditions, the gauge variation of the matrix element vanishes and the tree-level interaction is gauge invariant. This same rigid-rotation approximation has been employed in the derivation of the fermion propagator in a rotating environment in Refs.~\cite{Ayala:2021osy,Ayala:2026dmz,CastanoYepes2026FiniteSize}.

\section{Results and discussion}\label{sec:results}

\subsection{Selection rules and master integrals}\label{sec:master-integrals}

The contribution of each polarization sector to the tree-level fermion-photon vertex, given in Eqs.~\eqref{eq:Gamma+}-\eqref{eq:Gammaz}, is determined by a set of transverse integrals involving products of the fermionic and photon mode functions. The azimuthal dependence of these products can be integrated explicitly, yielding selection rules that relate the orbital angular momentum labels of the fermions and the photon to the spin projection $\lambda_s$ of the latter. 

For the crossed products entering the transverse components of the vertex, one obtains
\bea
&&\int d\mathbf{x}_\perp\,\mathcal A_{j}^{s}(\mathbf{x}_\perp;\mathbf{q}_\perp)\mathcal F_{l'}^{\prime *}(\mathbf{x}_\perp)\mathcal F_{l+1}(\mathbf{x}_\perp)\nn\\
&=&2\pi\delta_{l'-l,1+j+\lambda_s}\nn\\
&\times&\int_0^\infty d\rho\,\rho J_{j+\lambda_s}(|\mathbf q_\perp|\rho)J_{l'}(|\boldsymbol p_\perp'|\rho)J_{l+1}(|\boldsymbol p_\perp|\rho),\nn\\
\eea
\bea
&&\int d\mathbf{x}_\perp\,\mathcal A_{j}^{s}(\mathbf{x}_\perp;\mathbf{q}_\perp)\mathcal F_{l'+1}^{\prime *}(\mathbf{x}_\perp)\mathcal F_l(\mathbf{x}_\perp)\nn\\
&=&2\pi\delta_{l'-l,j+\lambda_s-1}\nn\\
&\times&\int_0^\infty d\rho\,\rho J_{j+\lambda_s}(|\mathbf q_\perp|\rho)J_{l'+1}(|\boldsymbol p_\perp'|\rho)J_l(|\boldsymbol p_\perp|\rho),\nn\\
\eea
while the longitudinal and temporal structures involve the combinations $\mathcal P_{l',l}^{\pm}$. For the longitudinal component, the azimuthal dependence of both terms in $\mathcal P_{l',l}^{\pm}$ is identical, and one obtains
\bea
&&\int d\mathbf{x}_\perp\,\mathcal A_{j}^{s}(\mathbf{x}_\perp;\mathbf{q}_\perp)\mathcal P_{l',l}^{\pm}(\mathbf{x}_\perp)\nn\\
&=&\pi\delta_{l'-l,j+\lambda_s}\int_0^\infty d\rho\,\rho J_{j+\lambda_s}(|\mathbf q_\perp|\rho)\nn\\
&\times&\Big[J_{l'}(|\boldsymbol p_\perp'|\rho)J_l(|\boldsymbol p_\perp|\rho)\pm J_{l'+1}(|\boldsymbol p_\perp'|\rho)J_{l+1}(|\boldsymbol p_\perp|\rho)\Big].\nn\\
\label{eq:int-with-P-longitudinal}
\eea

The azimuthal integration therefore gives three distinct selection rules,
\bea
j+\lambda_s&=&l'-l\pm1\nn\\
j+\lambda_s&=&l'-l,
\eea
which correspond to the three different angular structures appearing in the effective vertex. These relations express the balance between the angular momentum carried by the photon mode, through its orbital label $j$ and polarization-dependent spin projection $\lambda_s$, and the change in the fermionic orbital quantum number. Using these selection rules, the independent radial structures can be organized into the following master integrals:
\bea
\mathcal{I}_{l'l}&\equiv&\int_0^\infty d\rho\,\rho J_{l'}(|\boldsymbol p_\perp'|\rho)J_{l+1}(|\boldsymbol p_\perp|\rho)J_{l'-l-1}(|\mathbf q_\perp|\rho),\nn\\
\label{eq:I-def}
\eea
\bea
\mathcal{J}_{l'l}&\equiv&\int_0^\infty d\rho\,\rho J_{l'+1}(|\boldsymbol p_\perp'|\rho)J_l(|\boldsymbol p_\perp|\rho)J_{l'-l+1}(|\mathbf q_\perp|\rho),\nn\\
\label{eq:J-def}
\eea
\bea
\mathcal{K}_{l'l}&\equiv&\int_0^\infty d\rho\,\rho J_{l'}(|\boldsymbol p_\perp'|\rho)J_l(|\boldsymbol p_\perp|\rho)J_{l'-l}(|\mathbf q_\perp|\rho),\nn\\
\label{eq:K-def}
\eea

The three integrals encountered above belong to the class of triple-Bessel integrals for which closed analytical expressions can be obtained when the orders satisfy an appropriate integer relation. Following Gervois and Navelet~\cite{GervoisNavelet1984}, we consider
\bea
{\cal T}_{m,n}(a,b,c)=\int_0^\infty\rho\,d\rho\,J_m(a\rho)J_n(b\rho)J_{m+n}(c\rho),
\label{GeneralTripleBesselIntegral}
\eea
with $a$, $b$, and $c$ positive. The analytical form of this integral depends on the relative magnitudes of $a$, $b$, and $c$, leading to three distinct kinematic regions. In the non-degenerate region,
\bea
|a-b|<c<a+b,
\label{TripleBesselTriangleRegion}
\eea
the three magnitudes can be interpreted as the side lengths of an auxiliary triangle. Denoting by $A$, $B$, and $C$ the angles opposite to the sides $a$, $b$, and $c$, respectively, these angles are determined by
\bea
\cos A&=&\frac{b^2+c^2-a^2}{2bc},\nn\\
\cos B&=&\frac{a^2+c^2-b^2}{2ac},\nn\\
\cos C&=&\frac{a^2+b^2-c^2}{2ab},
\label{TripleBesselAngles}
\eea
with $A+B+C=\pi$. In this region, the integral is given by
\bea
{\cal T}_{m,n}(a,b,c)=\frac{\cos(nB-mA)}{\pi ab\sin C}.
\label{GeneralTripleBesselResult}
\eea
The remaining two regions correspond to configurations in which either $a$ or $b$ is larger than the sum of the other two magnitudes, while the region $c>a+b$ gives
\bea
{\cal T}_{m,n}(a,b,c)=0.
\label{TripleBesselRegionC}
\eea

When $a>b+c$ or $b>a+c$, the ordinary triangle angles are replaced by the corresponding hyperbolic parameters. For $a>b+c$, we define $u_a$ through
\bea
a^2=b^2+c^2+2bc\cosh u_a,
\label{TripleBesselUa}
\eea
and the integral takes the form
\bea
{\cal T}_{m,n}(a,b,c)
=-\frac{\sin(\pi m)}{\pi bc\sinh u_a}
e^{-m u_c-n u_b},
\label{TripleBesselRegionA}
\eea
where $u_b$ and $u_c$ are the corresponding hyperbolic parameters associated with the remaining magnitudes,
\bea
b^2&=&c^2+a^2+2ac\cosh u_b,\nn\\
c^2&=&a^2+b^2+2ab\cosh u_c.
\label{TripleBesselUbUc}
\eea

Similarly, when $b>a+c$, we introduce
\bea
b^2=a^2+c^2+2ac\cosh u_b,
\label{TripleBesselUb}
\eea
and obtain
\bea
{\cal T}_{m,n}(a,b,c)
=-\frac{\sin(\pi n)}{\pi ac\sinh u_b}
e^{-n u_c-m u_a}.
\label{TripleBesselRegionB}
\eea

The master integrals introduced above can be mapped onto the general form in Eq.~\eqref{GeneralTripleBesselIntegral} by using the relation
\bea
J_{-n}(x)=(-1)^nJ_n(x),
\label{eq:BesselIdentity}
\eea
for integer $n$. 

We identify the indices of ${\cal T}_{m,n}$ separately for each master integral. For $\mathcal{I}_{l'l}$, we choose
\bea
m&=&l',\qquad n=-(l+1),
\eea
so that
\bea
m+n&=&l'-l-1.
\eea
Using Eq.~\eqref{eq:BesselIdentity} we therefore obtain
\bea
\mathcal{I}_{l'l}&=&(-1)^{l+1}{\cal T}_{l',-(l+1)}\left(|\boldsymbol p_\perp'|,|\boldsymbol p_\perp|,|\mathbf q_\perp|\right).
\eea

For $\mathcal{J}_{l'l}$, we instead choose
\bea
m&=&l'+1,\qquad n=-l,
\eea
which gives
\bea
m+n&=&l'-l+1.
\eea
Again for Eq~\eqref{eq:BesselIdentity}, the corresponding identification is
\bea
\mathcal{J}_{l'l}&=&(-1)^l{\cal T}_{l'+1,-l}\left(|\boldsymbol p_\perp'|,|\boldsymbol p_\perp|,|\mathbf q_\perp|\right).
\eea

Finally, for $\mathcal{K}_{l'l}$, we take
\bea
m&=&l',\qquad n=-l,
\eea
so that
\bea
m+n&=&l'-l,
\eea
and we find
\bea
\mathcal{K}_{l'l}&=&(-1)^l{\cal T}_{l',-l}\left(|\boldsymbol p_\perp'|,|\boldsymbol p_\perp|,|\mathbf q_\perp|\right).
\eea
Thus, although the three master integrals correspond to different pairs of Bessel orders, they are all instances of the same triple-Bessel integral with the common arguments
$(a,b,c)=(|\boldsymbol p_\perp'|,|\boldsymbol p_\perp|,|\mathbf q_\perp|)$. Their kinematic support is therefore governed by the same conditions on the three transverse momentum magnitudes.

An alternative form of Eq.~\eqref{GeneralTripleBesselResult} that avoids the explicit inversion of the cosine functions can be obtained from Euler's identity. Writing
\bea
\cos(nB-mA)=\operatorname{Re}\left[e^{\ii nB}e^{-\ii mA}\right],
\label{TripleBesselEuler}
\eea
we can express the angular dependence directly in terms of the cosines and sines of $A$ and $B$. 

From Eq.~\eqref{TripleBesselAngles}, the sine of $A$ is given by
\bea
\sin A&=&\sqrt{1-\cos^2A}\nn\\
&=&\frac{\sqrt{4b^2c^2-(b^2+c^2-a^2)^2}}{2bc},
\label{TripleBesselSinA}
\eea
where the positive branch is chosen since $0<A<\pi$ in the non-degenerate triangular region. Similarly, for $B$,
\bea
\sin B&=&\sqrt{1-\cos^2B}\nn\\
&=&\frac{\sqrt{4a^2c^2-(a^2+c^2-b^2)^2}}{2ac},
\label{TripleBesselSinB}
\eea
with the positive branch again corresponding to $0<B<\pi$. Therefore,
\bea
e^{\ii A}&=&\cos A+\ii\sin A\nn\\
&=&\frac{b^2+c^2-a^2+\ii\sqrt{4b^2c^2-(b^2+c^2-a^2)^2}}{2bc},\nn\\
e^{\ii B}&=&\cos B+\ii\sin B\nn\\
&=&\frac{a^2+c^2-b^2+\ii\sqrt{4a^2c^2-(a^2+c^2-b^2)^2}}{2ac}.\nn\\
\label{TripleBesselComplexAngles}
\eea

Defining
\bea
Z_A&\equiv&\frac{b^2+c^2-a^2+\ii\sqrt{4b^2c^2-(b^2+c^2-a^2)^2}}{2bc},\nn\\
Z_B&\equiv&\frac{a^2+c^2-b^2+\ii\sqrt{4a^2c^2-(a^2+c^2-b^2)^2}}{2ac},\nn\\
\label{TripleBesselZA-ZB}
\eea
the numerator of Eq.~\eqref{GeneralTripleBesselResult} can then be written as
\bea
\cos(nB-mA)&=&\operatorname{Re}\left[Z_B^nZ_A^{-m}\right].
\label{TripleBesselNumerator}
\eea
The remaining factor in the denominator can be expressed directly in terms of the three magnitudes. From Eq.~\eqref{TripleBesselAngles},
\bea
\sin C&=&\sqrt{1-\cos^2C}\nn\\
&=&\frac{\sqrt{4a^2b^2-(a^2+b^2-c^2)^2}}{2ab},
\label{TripleBesselSinC}
\eea
where the positive branch is selected in the non-degenerate triangular region. Combining Eqs.~\eqref{TripleBesselNumerator} and \eqref{TripleBesselSinC}, we obtain
\bea
{\cal T}_{m,n}(a,b,c)&=&\frac{2\operatorname{Re}\left[Z_B^nZ_A^{-m}\right]}{\pi\sqrt{4a^2b^2-(a^2+b^2-c^2)^2}}.
\label{GeneralTripleBesselComplexResult}
\eea
This representation avoids the explicit inversion of the cosine functions and expresses the triple-Bessel integral entirely in terms of the three positive magnitudes $a$, $b$, and $c$ {when they satisfy the inequality~\eqref{TripleBesselTriangleRegion}.}

\subsection{Nonconservation of transverse momentum}\label{sec:nonconservation}
In order to analyze the nonconservation of the transverse momentum, we choose the azimuthal direction of the photon momentum $\boldsymbol q$ as the reference direction and set $\phi_q=0$. Since the rotating background is invariant under rotations about the $z$ axis, defined by $\boldsymbol{\Omega}=\Omega\hat{\boldsymbol z}$, this choice fixes the overall azimuthal orientation without restricting the relative orientation of the three momenta. The spatial momenta are therefore parametrized as
\bea
\boldsymbol p&=&|\boldsymbol p|\left(\sin\theta_p\cos\phi_p,\sin\theta_p\sin\phi_p,\cos\theta_p\right)\nn\\
\boldsymbol p'&=&|\boldsymbol p'|\left(\sin\theta_{p'}\cos\phi_{p'},\sin\theta_{p'}\sin\phi_{p'},\cos\theta_{p'}\right)\nn\\
\boldsymbol q&=&|\boldsymbol q|\left(\sin\theta_q,0,\cos\theta_q\right),
\label{KinematicMomentumParametrization}
\eea
with
\bea
|\boldsymbol p_\perp|=|\boldsymbol p|\sin\theta_p,&&p_z=|\boldsymbol p|\cos\theta_p,\nn\\
|\boldsymbol p_\perp'|=|\boldsymbol p'|\sin\theta_{p'},&&p_z'=|\boldsymbol p'|\cos\theta_{p'},\nn\\
|\boldsymbol q_\perp|=|\boldsymbol q|\sin\theta_q,&&q_z=|\boldsymbol q|\cos\theta_q.
\label{KinematicMomentumComponents}
\eea
The translational invariance along the rotation axis gives the longitudinal momentum-conservation condition
\bea
p_z'&=&p_z+q_z,\nn\\
|\boldsymbol p'|\cos\theta_{p'}&=&|\boldsymbol p|\cos\theta_p+|\boldsymbol q|\cos\theta_q.
\label{KinematicLongitudinalConservation}
\eea
In contrast, translational invariance in the transverse plane is broken by the rotating background, i.e., the transverse integration does not generate an independent two-dimensional momentum-conservation delta function. Instead, the different structures arising from the fermion and photon modes reduce, after the azimuthal integration, to radial triple-Bessel integrals. The relevant structures are represented by Eqs.~\eqref{eq:I-def}, \eqref{eq:J-def}, and \eqref{eq:K-def}. Although these integrals differ in their Bessel orders, they involve the same three transverse momentum magnitudes, $|\boldsymbol p_\perp'|$, $|\boldsymbol p_\perp|$, and $|\boldsymbol q_\perp|$, and therefore probe the same underlying transverse momentum configuration.

The general structure discussed above can now be applied directly to the three radial integrals in Eqs.~\eqref{eq:I-def}, \eqref{eq:J-def}, and \eqref{eq:K-def}. In each case, the Bessel orders satisfy the integer relation required to identify the integral with Eq.~\eqref{GeneralTripleBesselIntegral}, while the three arguments are given by the same set of transverse momentum magnitudes, $|\boldsymbol p_\perp'|$, $|\boldsymbol p_\perp|$, and $|\boldsymbol q_\perp|$. Therefore, the regions identified in Eqs.~\eqref{TripleBesselTriangleRegion}, \eqref{TripleBesselRegionC}, \eqref{TripleBesselRegionA}, and \eqref{TripleBesselRegionB} translate into the following configurations:
\begin{subequations}
\bea
{\cal R}_1&:&\Big||\boldsymbol p_\perp'|-|\boldsymbol p_\perp|\Big|<|\boldsymbol q_\perp|<|\boldsymbol p_\perp'|+|\boldsymbol p_\perp|,
\label{TripleBesselKinematicRegion1}
\eea
\bea
{\cal R}_2&:&|\boldsymbol p_\perp'|>|\boldsymbol p_\perp|+|\boldsymbol q_\perp|,
\label{TripleBesselKinematicRegion2}
\eea
\bea
{\cal R}_3&:&|\boldsymbol p_\perp|>|\boldsymbol p_\perp'|+|\boldsymbol q_\perp|,
\label{TripleBesselKinematicRegion3}
\eea
\bea
{\cal R}_4&:&|\boldsymbol q_\perp|>|\boldsymbol p_\perp'|+|\boldsymbol p_\perp|.
\label{TripleBesselKinematicRegion4}
\eea
\end{subequations}
Here, ${\cal R}_1$ is the direct realization of the triangle region in Eq.~\eqref{TripleBesselTriangleRegion}, while ${\cal R}_2$ and ${\cal R}_3$ correspond to the regions $a>b+c$ and $b>a+c$ in Eqs.~\eqref{TripleBesselRegionA} and \eqref{TripleBesselRegionB}, respectively. The remaining possibility, ${\cal R}_4$, corresponds to the region $c>a+b$ for which the general integral vanishes according to Eq.~\eqref{TripleBesselRegionC}. Since the orders entering the present integrals are integers, the factors $\sin(\pi m)$ and $\sin(\pi n)$ appearing in Eqs.~\eqref{TripleBesselRegionA} and \eqref{TripleBesselRegionB} vanish. Thus, ${\cal R}_2$ and ${\cal R}_3$ do not contribute either, leaving ${\cal R}_1$ as the only nonvanishing kinematic region for the integrals considered here.

At this stage, it is useful to characterize ${\cal R}_1$ solely in terms of the three transverse momentum magnitudes, without evaluating the radial integrals themselves. The auxiliary triangle introduced in Eq.~\eqref{TripleBesselTriangleRegion} has side lengths $|\boldsymbol p_\perp'|$, $|\boldsymbol p_\perp|$, and $|\boldsymbol q_\perp|$. Denoting by $A$, $B$, and $C$ the angles opposite to $|\boldsymbol p_\perp'|$, $|\boldsymbol p_\perp|$, and $|\boldsymbol q_\perp|$, respectively, Eq.~\eqref{TripleBesselAngles} gives
\bea
\cos A&=&\frac{|\boldsymbol p_\perp|^2+|\boldsymbol q_\perp|^2-|\boldsymbol p_\perp'|^2}{2|\boldsymbol p_\perp||\boldsymbol q_\perp|},\nn\\
\cos B&=&\frac{|\boldsymbol p_\perp'|^2+|\boldsymbol q_\perp|^2-|\boldsymbol p_\perp|^2}{2|\boldsymbol p_\perp'||\boldsymbol q_\perp|},\nn\\
\cos C&=&\frac{|\boldsymbol p_\perp'|^2+|\boldsymbol p_\perp|^2-|\boldsymbol q_\perp|^2}{2|\boldsymbol p_\perp'||\boldsymbol p_\perp|},
\label{TripleBesselKinematicAngles}
\eea
with $A+B+C=\pi$. The common denominator appearing in the analytical result Eq.~\eqref{GeneralTripleBesselResult} can then be written as
\bea
&&|\boldsymbol p_\perp'||\boldsymbol p_\perp|\sin C\nn\\
&=&\frac{1}{2}\sqrt{4|\boldsymbol p_\perp'|^2|\boldsymbol p_\perp|^2-\left(|\boldsymbol p_\perp'|^2+|\boldsymbol p_\perp|^2-|\boldsymbol q_\perp|^2\right)^2}.\nn\\
\label{TripleBesselKinematicDenominator}
\eea
This expression makes explicit that the same geometric combination of transverse momentum magnitudes enters the analytical expressions obtained from Eq.~\eqref{GeneralTripleBesselResult}, although the different Bessel orders in Eqs.~\eqref{eq:I-def}, \eqref{eq:J-def}, and \eqref{eq:K-def} lead to different angular factors through the corresponding values of $m$ and $n$.

It is important to distinguish these geometric conditions from transverse momentum conservation. The condition defining ${\cal R}_1$ does not imply the vector relation $\boldsymbol p_\perp'=\boldsymbol p_\perp+\boldsymbol q_\perp$. Rather, it specifies the range of transverse momentum magnitudes for which the radial triple-Bessel integrals are nonvanishing. This distinction can be made explicit by writing the physical transverse momenta as
\bea
\boldsymbol p_\perp&=&|\boldsymbol p_\perp|(\cos\phi_p,\sin\phi_p)\nn\\
\boldsymbol p_\perp'&=&|\boldsymbol p_\perp'|(\cos\phi_{p'},\sin\phi_{p'})\nn\\
\boldsymbol q_\perp&=&|\boldsymbol q_\perp|(\cos\phi_q,\sin\phi_q).
\label{TransverseVectorsPolar}
\eea
With the choice $\phi_q=0$ adopted above, $\phi_p$ and $\phi_{p'}$ specify the relative azimuthal orientations of $\boldsymbol p_\perp$ and $\boldsymbol p_\perp'$ with respect to $\boldsymbol q_\perp$. These angles do not enter the support conditions in Eqs.~\eqref{TripleBesselKinematicRegion1}--\eqref{TripleBesselKinematicRegion4}. Then, different azimuthal configurations with the same three transverse momentum magnitudes have identical radial support. { The following relations are clearly satisfied 
\begin{eqnarray}
\cos A &=& \cos\phi_p\cos\phi_q + \sin\phi_p\sin\phi_q = \cos(\phi_p - \phi_q)\nonumber\\
\cos B &=& \cos\phi_{p'}\cos\phi_q + \sin\phi_{p'}\sin\phi_q = \cos(\phi_{q} - \phi_{p'})\nonumber\\
\cos C &=& \cos\phi_{p'}\cos\phi_p + \sin\phi_{p'}\sin\phi_p = \cos(\phi_{p'} - \phi_p).\nonumber\\
\end{eqnarray}}
In particular, one may vary the three azimuthal angles according to 
\bea
\phi_p&\rightarrow&\phi_p+\delta\phi_p,\nn\\
\phi_{p'}&\rightarrow&\phi_{p'}+\delta\phi_{p'},\nn\\
\phi_q&\rightarrow&\phi_q+\delta\phi_q,
\label{IndependentAzimuthalVariations}
\eea
{
However, the constraint $A+B+C = \pi$ implies that the variations 
\begin{eqnarray}
&&\delta A + \delta B + \delta C = 0\nonumber\\
&&= \delta(\phi_p - \phi_q) + \delta(\phi_q - \phi_{p'}) + \delta (\phi_{p'} - \phi_p),
\end{eqnarray}
which is clearly satisfied for arbitrary and independent variations $\delta\phi_p$, $\delta\phi_{p'}$, and $\delta\phi_q$.} 
Such variations leave the transverse momenta magnitudes unchanged,
\bea
|\boldsymbol p_\perp|\rightarrow|\boldsymbol p_\perp|,\qquad|\boldsymbol p_\perp'|\rightarrow|\boldsymbol p_\perp'|,\qquad|\boldsymbol q_\perp|\rightarrow|\boldsymbol q_\perp|,
\label{AzimuthalMagnitudeInvariance}
\eea
and therefore leave the regions ${\cal R}_1,\ldots,{\cal R}_4$ unchanged. Unlike a common rigid rotation, these independent variations modify the relative azimuthal orientations of the physical transverse momenta. Hence, the support of the radial Bessel integrals cannot determine the relative orientations of the physical vectors.

\begin{figure*}
    \centering
    \includegraphics[width=0.2\textwidth]{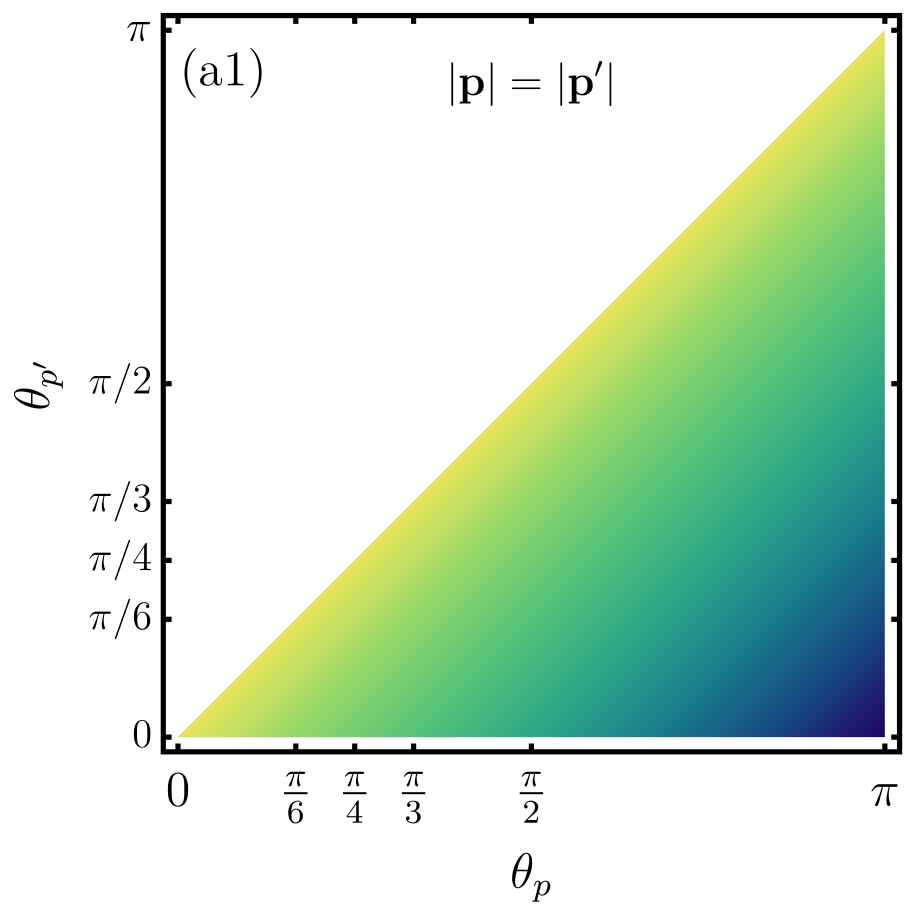}\hspace{-0.05cm}
    \includegraphics[width=0.2\textwidth]{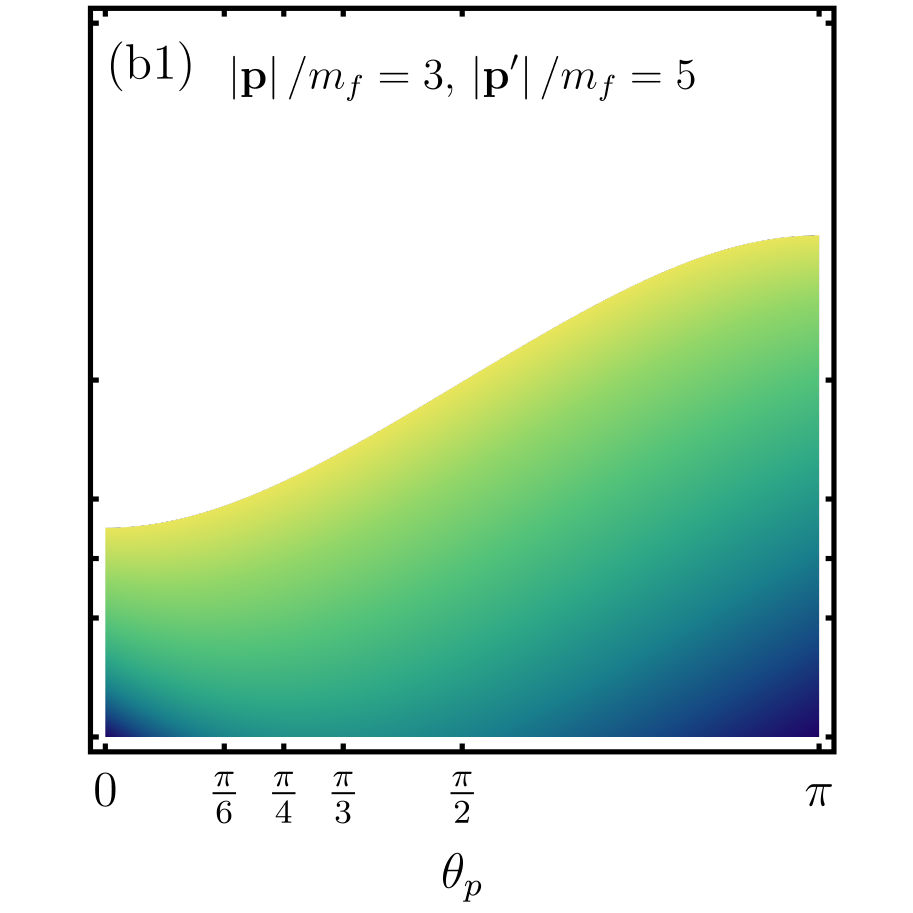}\hspace{-0.3cm}
    \includegraphics[width=0.2\textwidth]{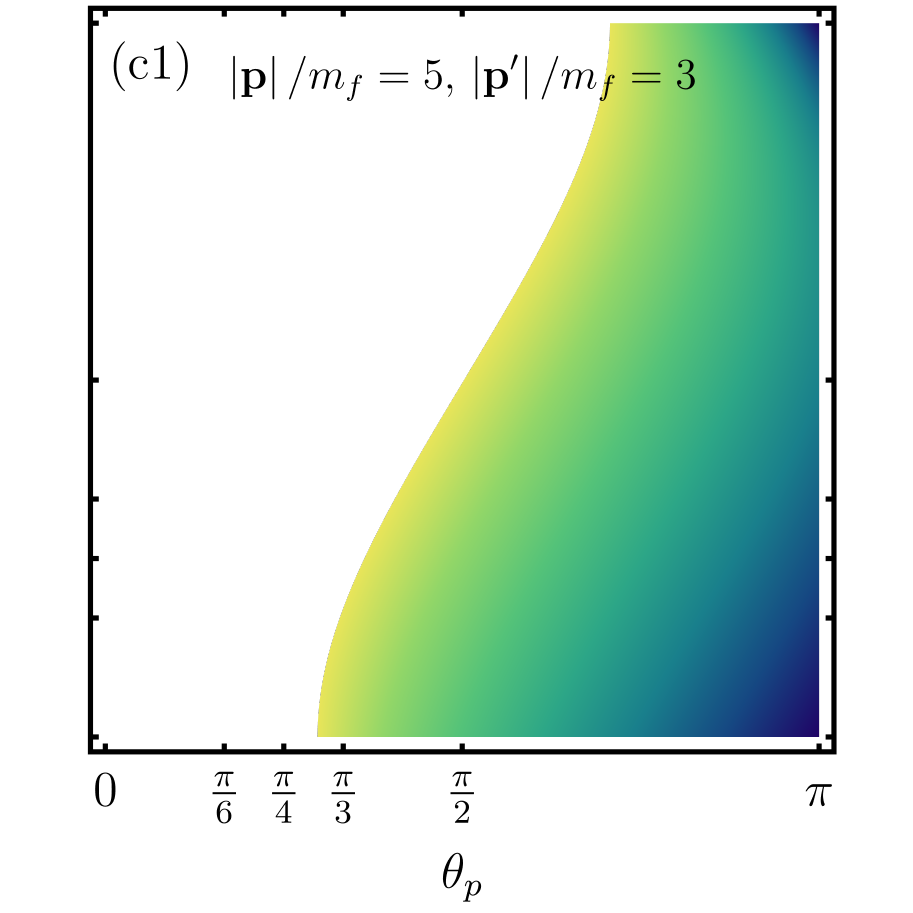}\hspace{-0.3cm}
    \includegraphics[width=0.2\textwidth]{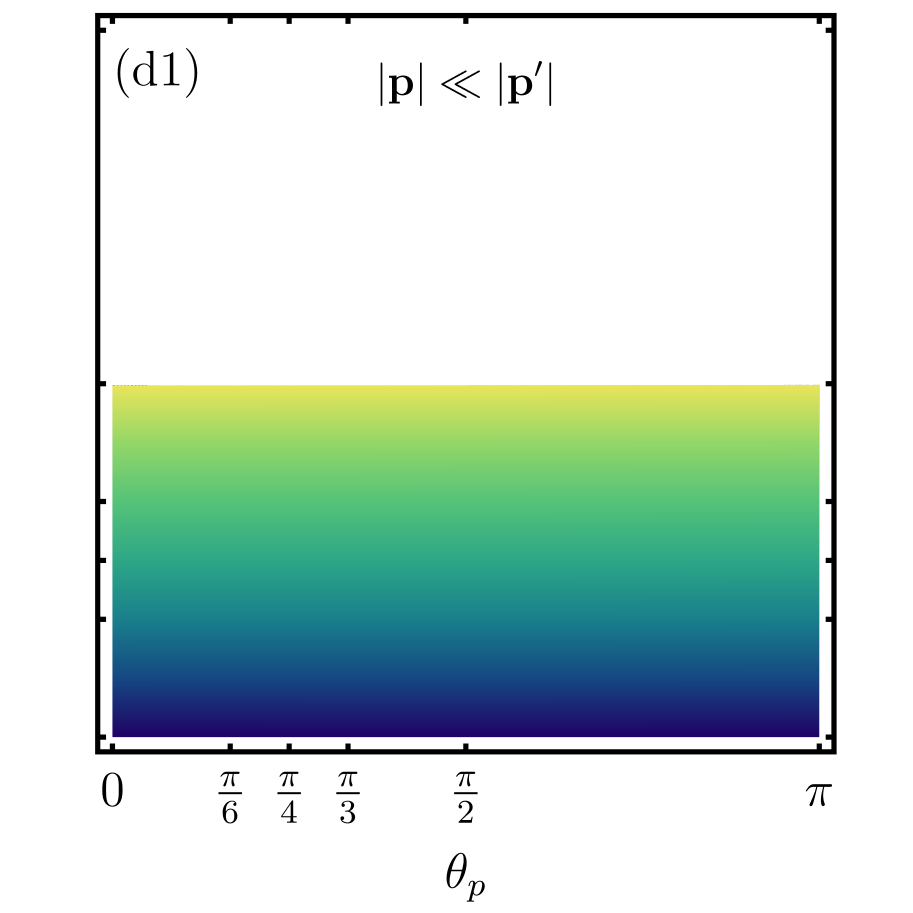}\hspace{-0.3cm}
    \includegraphics[width=0.2\textwidth]{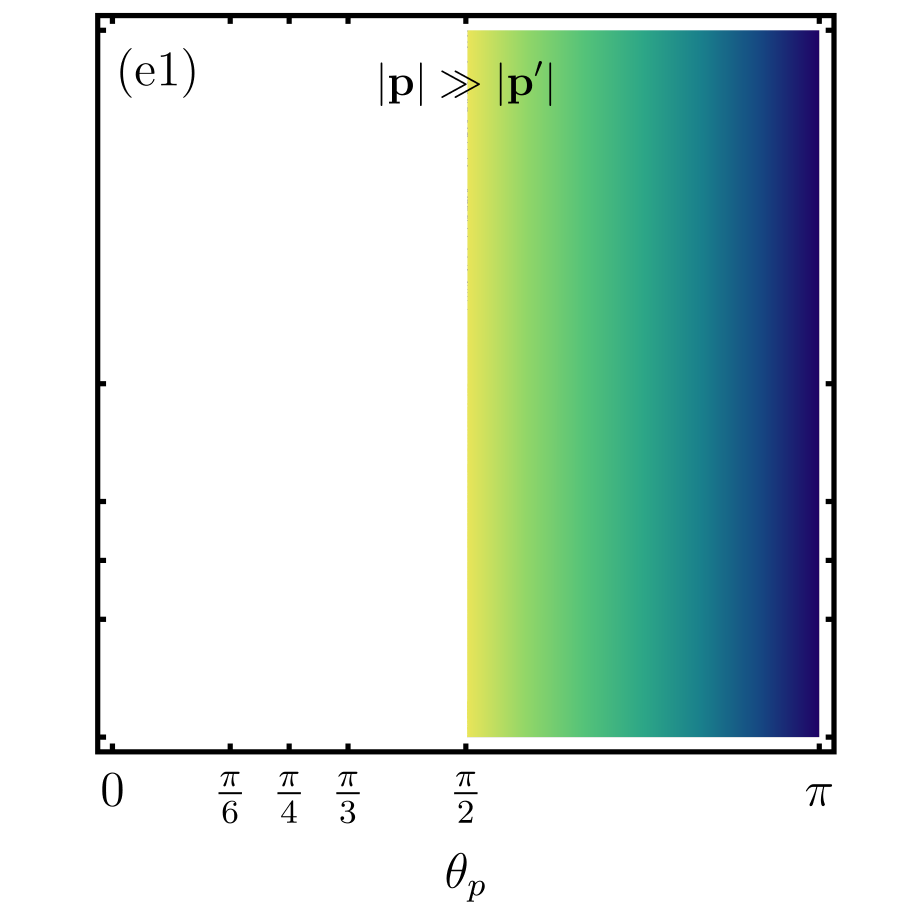}\\
    \vspace{0.4cm}
    \includegraphics[width=0.2\textwidth]{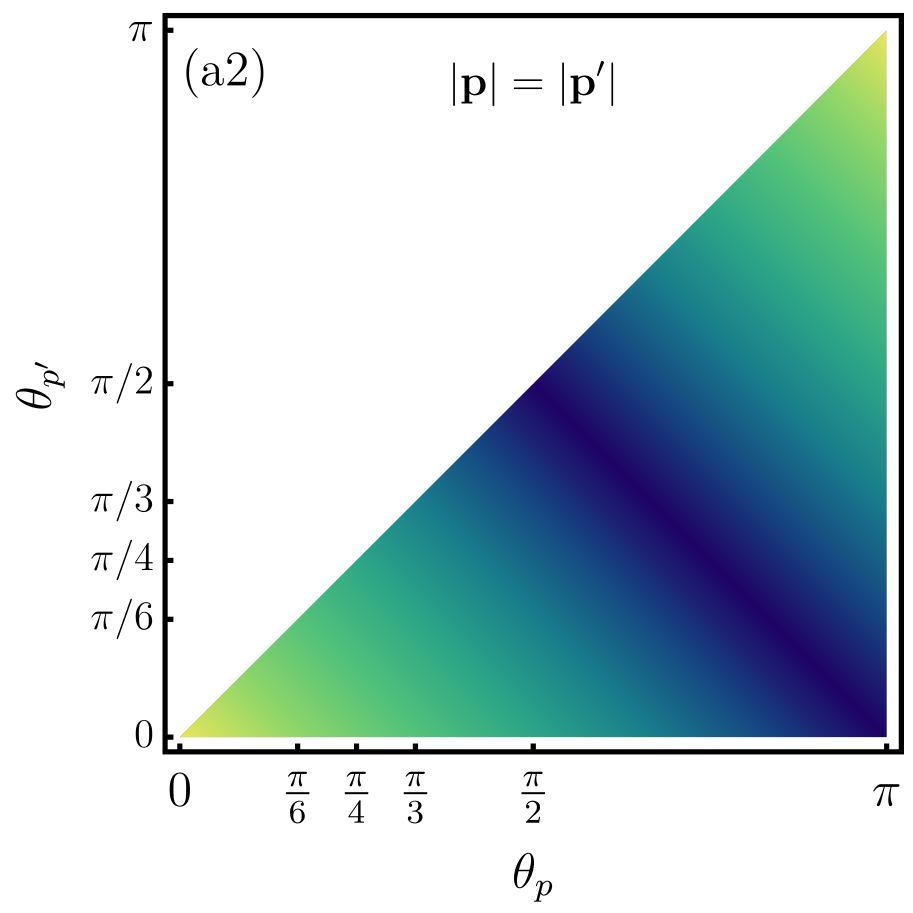}\hspace{-0.05cm}
    \includegraphics[width=0.2\textwidth]{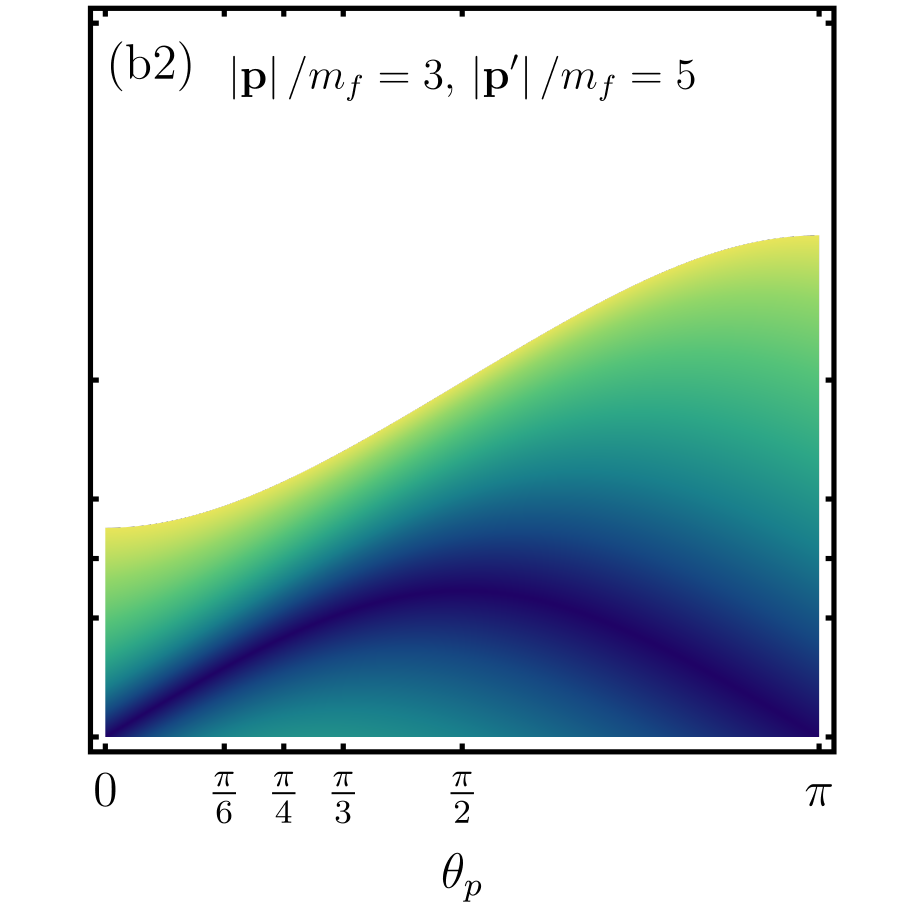}\hspace{-0.3cm}
    \includegraphics[width=0.2\textwidth]{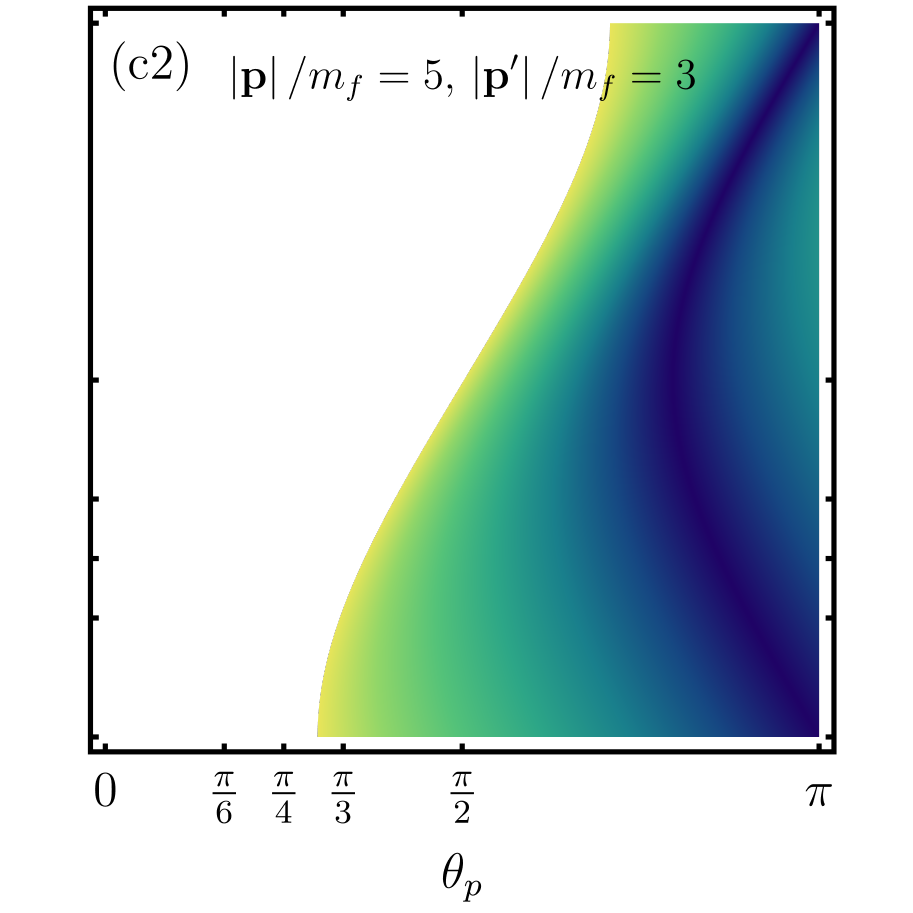}\hspace{-0.3cm}
    \includegraphics[width=0.2\textwidth]{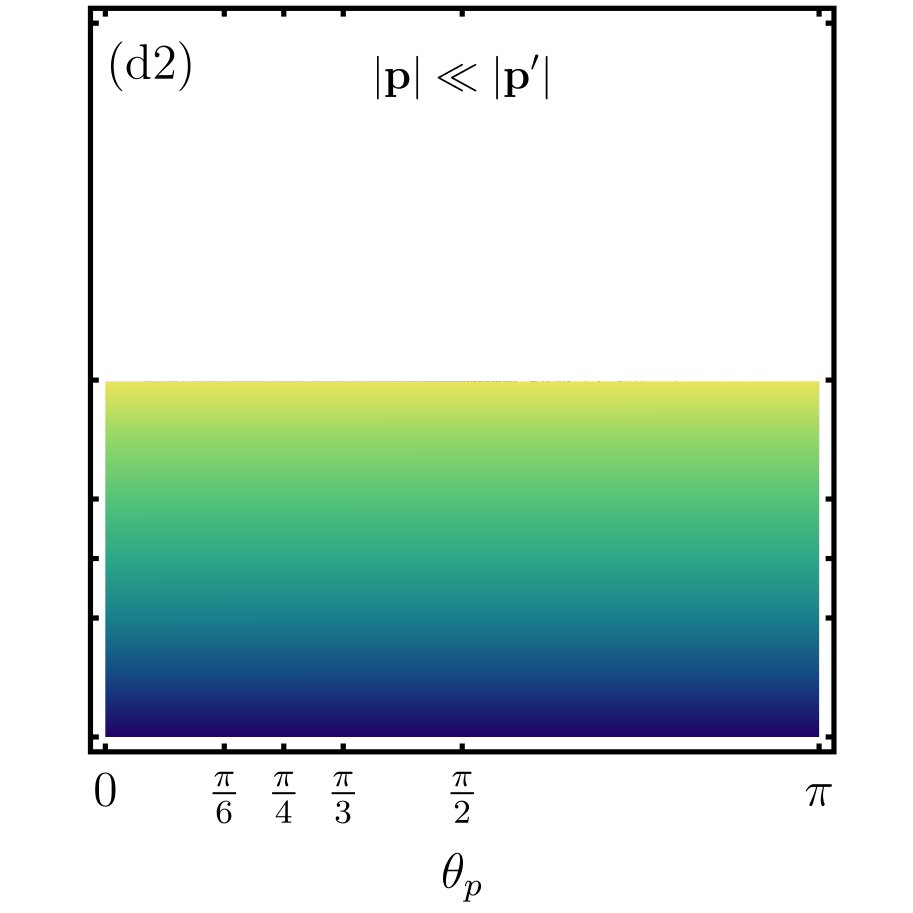}\hspace{-0.3cm}
    \includegraphics[width=0.2\textwidth]{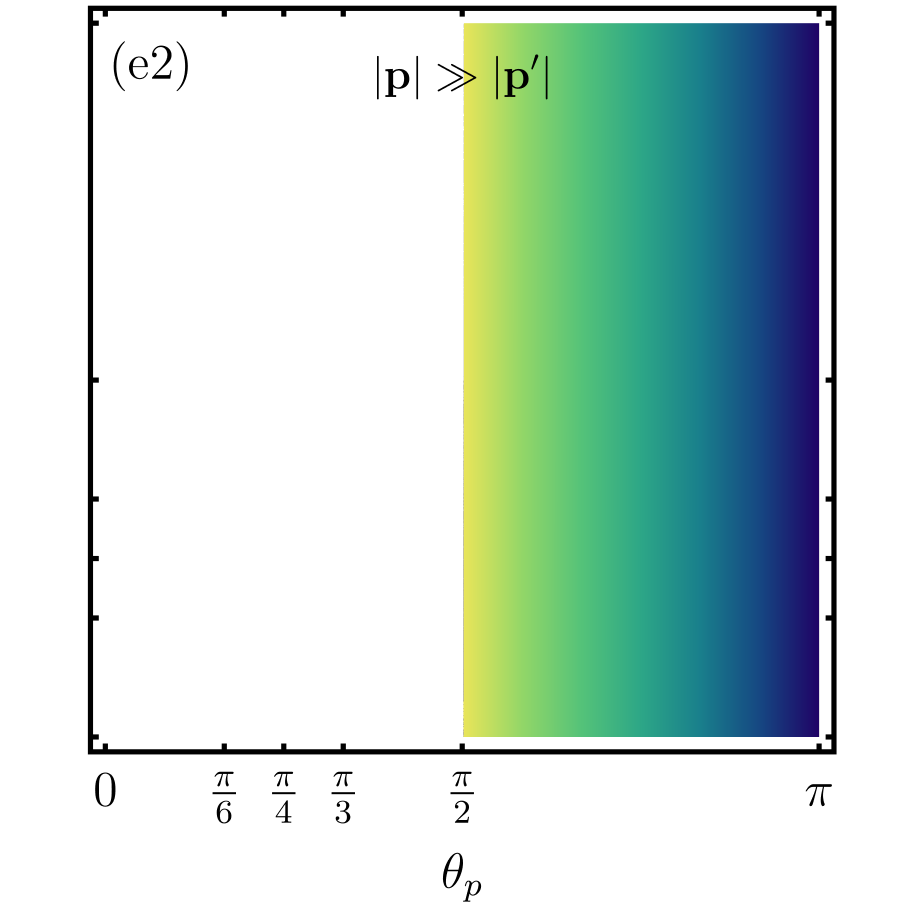}\\
    \vspace{0.4cm}
    \includegraphics[scale=0.4]{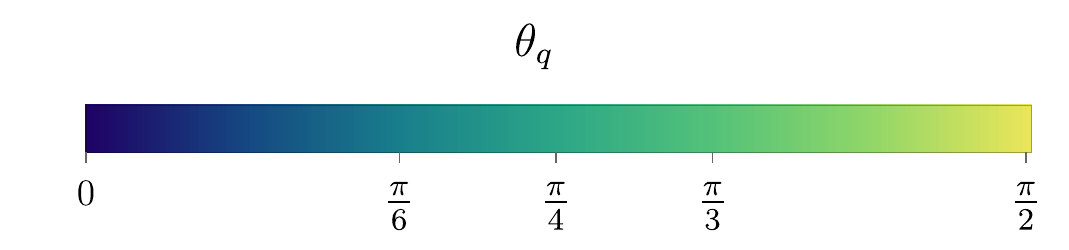}
    \caption{Allowed range of the polar angle $\theta_q$ as a function of the polar angles $\theta_p$ and $\theta_{p'}$ for different choices of the transverse momentum magnitudes, obtained from Eq.~\eqref{KinematicMainTanThetaQ} for the branch $q_z>0$. The upper row shows the upper bound $\theta_{q}^{\max}$, while the lower row shows the lower bound $\theta_{q}^{\min}$. The white regions correspond to kinematic configurations outside the non-degenerate transverse support of the three-Bessel-function integrals.}
    \label{fig:Tqmax}
\end{figure*}

This differs from imposing the vector relation
\bea
\boldsymbol p_\perp'=\boldsymbol p_\perp+\boldsymbol q_\perp,
\label{TransverseMomentumConservation}
\eea
which, for fixed transverse momentum magnitudes, imposes a definite relation between their relative azimuthal orientations. With the choice $\phi_q=0$, for example, Eq.~\eqref{TransverseMomentumConservation} gives
\bea
|\boldsymbol p_\perp'|^2=|\boldsymbol p_\perp|^2+|\boldsymbol q_\perp|^2+2|\boldsymbol p_\perp||\boldsymbol q_\perp|\cos\phi_p.
\label{PhysicalTransverseRelation}
\eea
Thus, the scalar conditions defining ${\cal R}_1$ cannot be identified with vectorial transverse-momentum conservation. The auxiliary triangle introduced in Eq.~\eqref{TripleBesselTriangleRegion} is instead a geometrical representation of the three transverse momentum magnitudes. Once these magnitudes are fixed, the corresponding angles $A$, $B$, and $C$ are fixed, whereas the physical transverse vectors can still have different relative azimuthal orientations. Therefore, $A$, $B$, and $C$ cannot, in general, be identified with the physical angles between $\boldsymbol p_\perp$, $\boldsymbol p_\perp'$, and $\boldsymbol q_\perp$.

For comparison, in a translationally invariant transverse plane, the transverse spatial integration would yield a two-dimensional vector Dirac delta function,
\bea
\int d^2x_\perp\,e^{\ii(\boldsymbol p_\perp+\boldsymbol q_\perp-\boldsymbol p_\perp')\cdot\boldsymbol x_\perp}\propto\delta^{(2)}\left(\boldsymbol p_\perp+\boldsymbol q_\perp-\boldsymbol p_\perp'\right).
\label{TransverseVectorDelta}
\eea
Such a delta function contains the complete vector information and, for fixed transverse momentum magnitudes, determines the relative azimuthal orientations required by transverse momentum conservation. No analogous transverse vector delta function is generated by the present rotating background. Instead, the transverse integration produces the scalar radial integrals in Eqs.~\eqref{eq:I-def}, \eqref{eq:J-def}, and \eqref{eq:K-def}, whose kinematic support follows from the triple-Bessel regions discussed above. The resulting constraints therefore concern the allowed combinations of transverse momentum magnitudes rather than a transverse vector conservation law.

The kinematic consequences of the support condition are illustrated in Fig.~\ref{fig:Tqmax}. The transverse momentum magnitudes entering the Bessel support are
\bea
|\boldsymbol p_\perp|&=&|\boldsymbol p|\sin\theta_p,\nn\\
|\boldsymbol p_\perp'|&=&|\boldsymbol p'|\sin\theta_{p'}\nn\\
|\boldsymbol q_\perp|&=&|\boldsymbol q|\sin\theta_q,
\label{TransverseMagnitudeDefinitions}
\eea
and therefore, the support condition becomes
{\small
\bea
\frac{\Big||\boldsymbol p'|\sin\theta_{p'}-|\boldsymbol p|\sin\theta_p\Big|}{|\boldsymbol q|}<\sin\theta_q<\frac{|\boldsymbol p'|\sin\theta_{p'}+|\boldsymbol p|\sin\theta_p}{|\boldsymbol q|}.\nn\\
\label{KinematicMainSupport}
\eea
}
This condition constrains the transverse magnitude $|\boldsymbol q_\perp|$ for a given pair $(\theta_p,\theta_{p'})$. It should therefore be interpreted as a condition on the magnitudes of the transverse projections and not as a condition on their orientations within the transverse plane. In particular, the azimuthal angles $\phi_p$ and $\phi_{p'}$ do not appear in Eq.~\eqref{KinematicMainSupport}.

The longitudinal direction is constrained independently. Using longitudinal momentum conservation, $|\boldsymbol q|$ can be eliminated in favor of the polar angles according to
\bea
|\boldsymbol q|=\frac{|\boldsymbol p'|\cos\theta_{p'}-|\boldsymbol p|\cos\theta_p}{\cos\theta_q}.
\label{KinematicMainQMagnitude}
\eea
Substitution into the transverse support condition yields
{\small
\bea
\frac{\Big||\boldsymbol p'|\sin\theta_{p'}-|\boldsymbol p|\sin\theta_p\Big|}{|\boldsymbol p'|\cos\theta_{p'}-|\boldsymbol p|\cos\theta_p}<\tan\theta_q<\frac{|\boldsymbol p'|\sin\theta_{p'}+|\boldsymbol p|\sin\theta_p}{|\boldsymbol p'|\cos\theta_{p'}-|\boldsymbol p|\cos\theta_p}.\nn\\
\label{KinematicMainTanThetaQ}
\eea
}
for the branch $q_z>0$, corresponding to $0\leq\theta_q<\pi/2$. For $q_z<0$, corresponding to $\pi/2<\theta_q\leq\pi$, the division by the negative longitudinal component reverses the inequalities, giving
{\small
\bea
\frac{|\boldsymbol p'|\sin\theta_{p'}+|\boldsymbol p|\sin\theta_p}{|\boldsymbol p'|\cos\theta_{p'}-|\boldsymbol p|\cos\theta_p}<\tan\theta_q<\frac{\Big||\boldsymbol p'|\sin\theta_{p'}-|\boldsymbol p|\sin\theta_p\Big|}{|\boldsymbol p'|\cos\theta_{p'}-|\boldsymbol p|\cos\theta_p}.\nn\\
\label{KinematicMainTanThetaQNegative}
\eea
}

For the $q_z>0$ branch, the upper and lower bounds of Eq.~\eqref{KinematicMainTanThetaQ} determine $\theta_q^{\max}$ and $\theta_q^{\min}$, respectively. The upper and lower rows of Fig.~\ref{fig:Tqmax} display these two bounds as functions of $(\theta_p,\theta_{p'})$, while their separation characterizes the angular freedom available to $\boldsymbol q$ for a given configuration. The white regions correspond to configurations outside the $q_z>0$ branch, for which $|\boldsymbol p'|\cos\theta_{p'}-|\boldsymbol p|\cos\theta_p<0$.

\begin{figure}
  \centering
  \subfigure[$|\mathbf{p}|/m_f=5$, $|\mathbf{p}'|/m_f=3$, $\theta_p=\pi/2$, $\theta_{p'}=\pi/6$ ]
  {
    \includegraphics[scale=0.195]{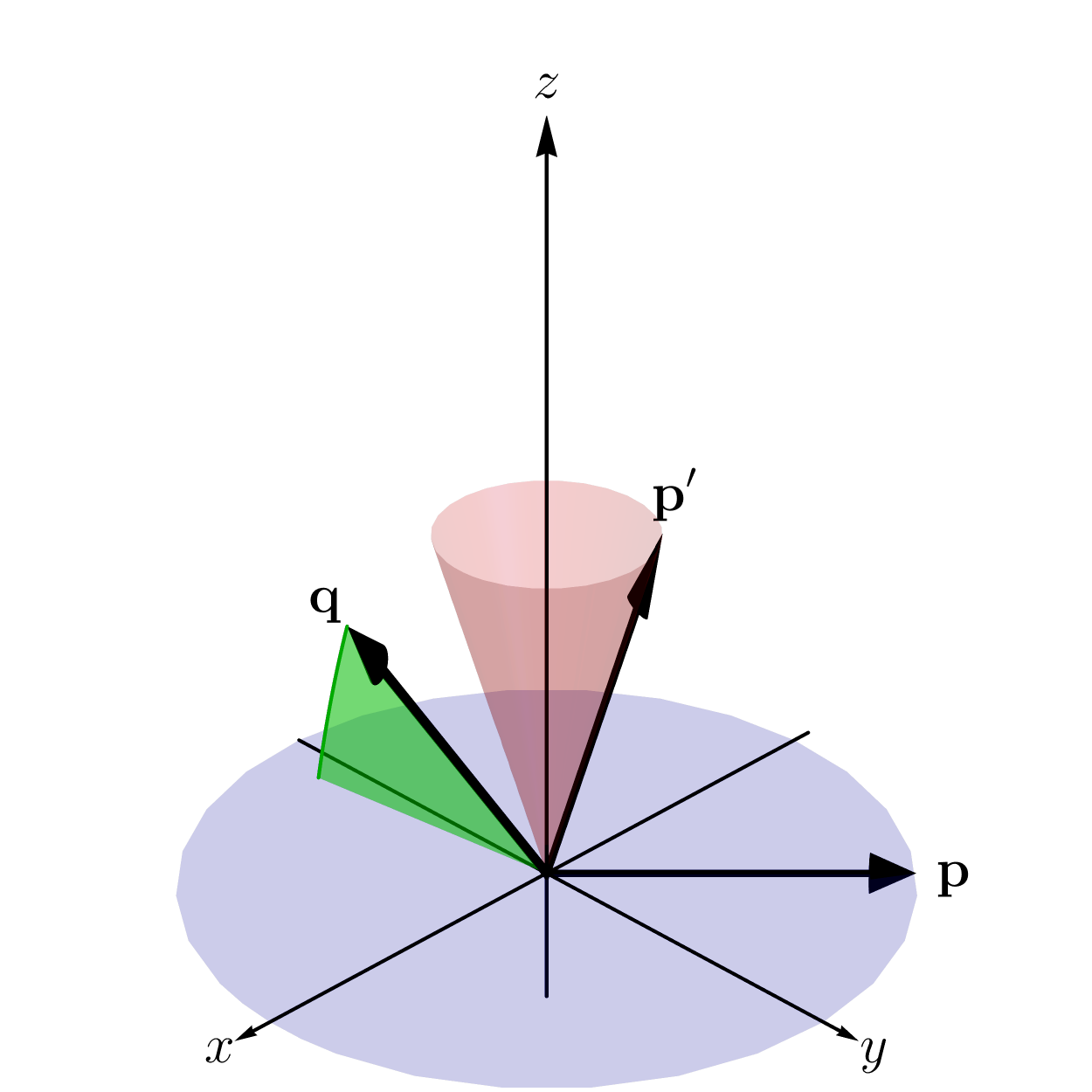}
    \label{fig:vectors-b}
  }
  \subfigure[$|\mathbf{p}|/m_f=3$, $|\mathbf{p}'|/m_f=5$, $\theta_p=\pi/2$, $\theta_{p'}=\pi/6$]
  {
    \includegraphics[scale=0.195]{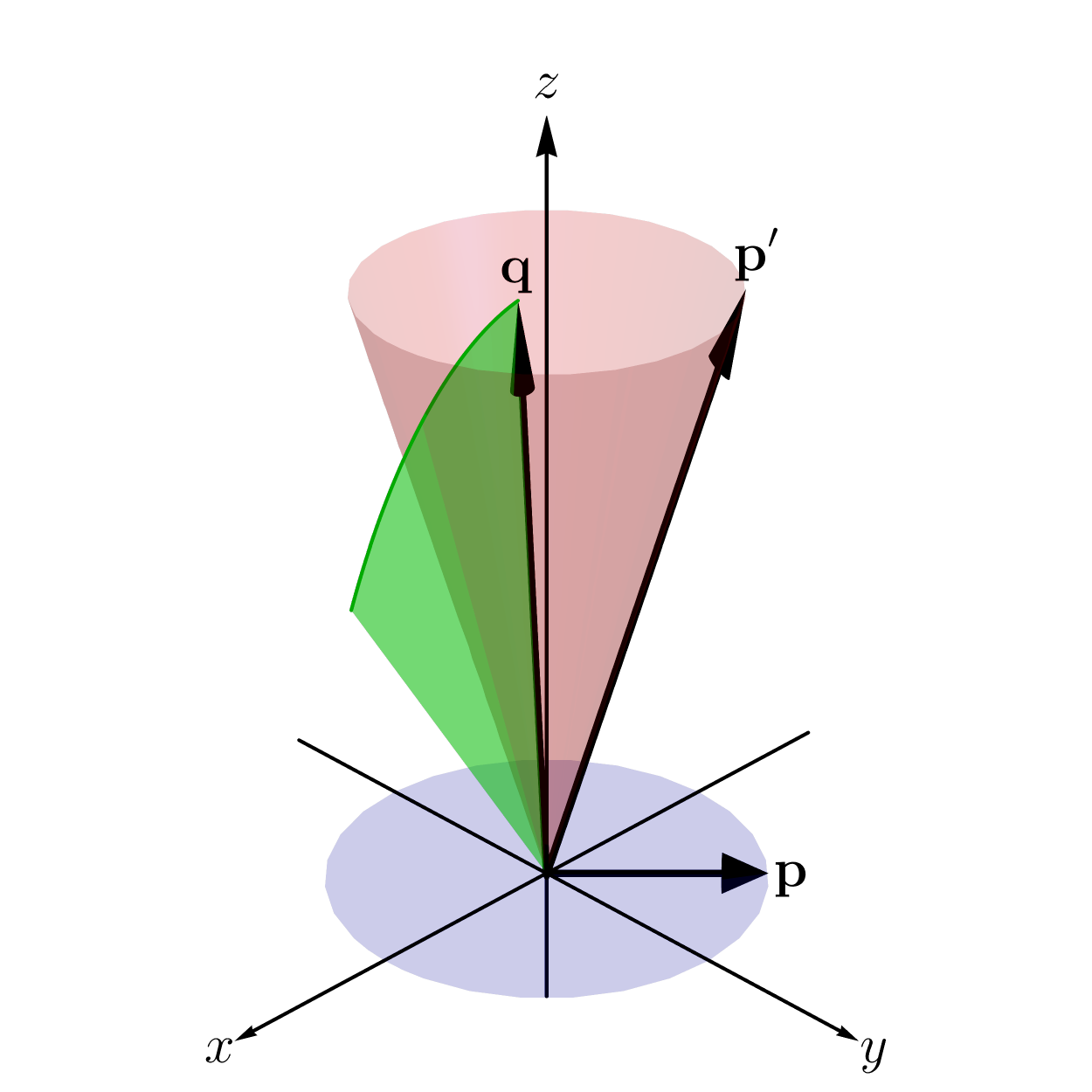}
    \label{fig:vectors-a}
  }\\
  \subfigure[$|\mathbf{p}|/m_f=1$, $|\mathbf{p}'|/m_f=10$, $\theta_p=\pi/2$, $\theta_{p'}=\pi/6$]
  {
    \includegraphics[scale=0.195]{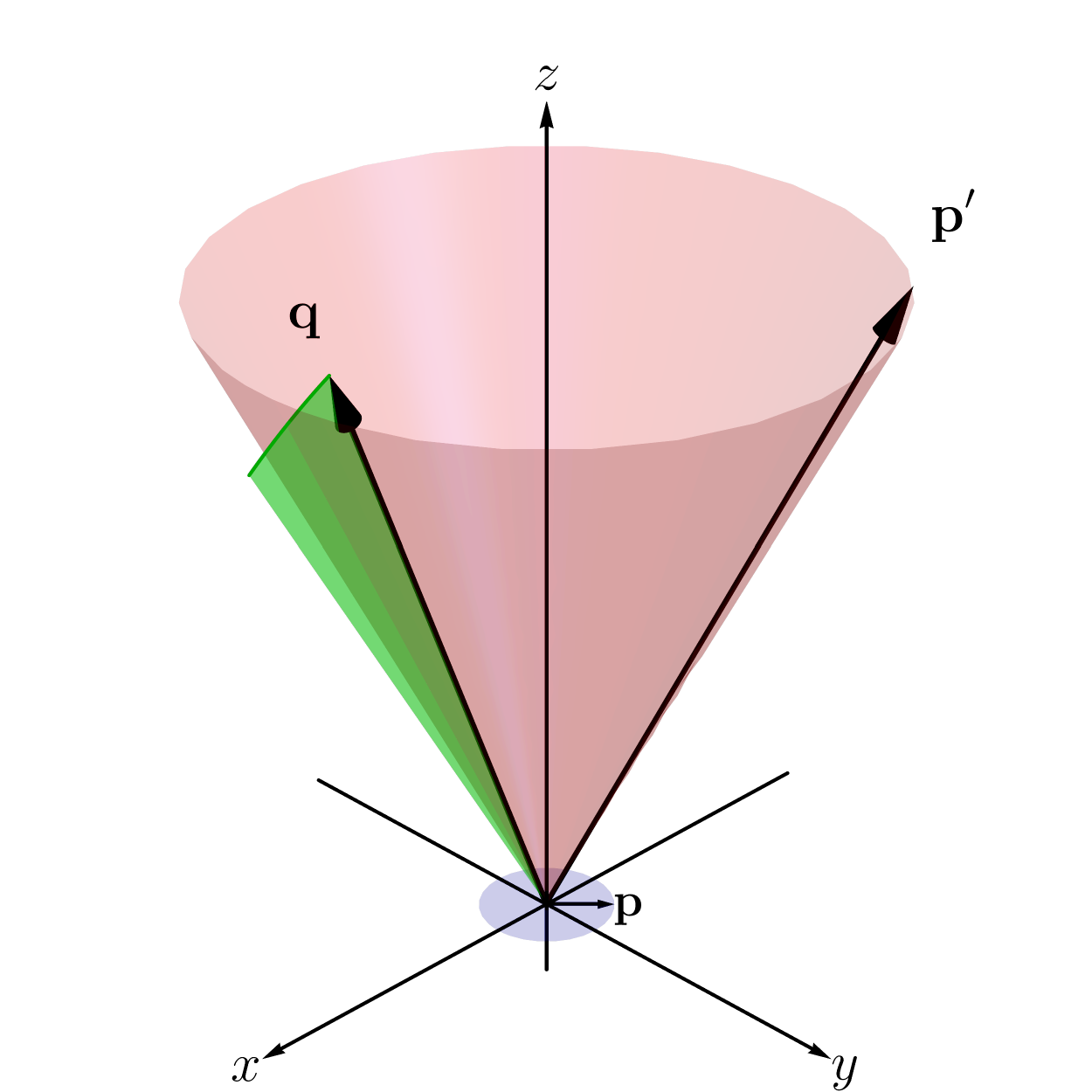}
    \label{fig:vectors-c}
  }
  \subfigure[$|\mathbf{p}|/m_f=3$, $|\mathbf{p}'|/m_f=5$, $\theta_p=\pi/3$, $\theta_{p'}=\pi/6$]
  {
    \includegraphics[scale=0.195]{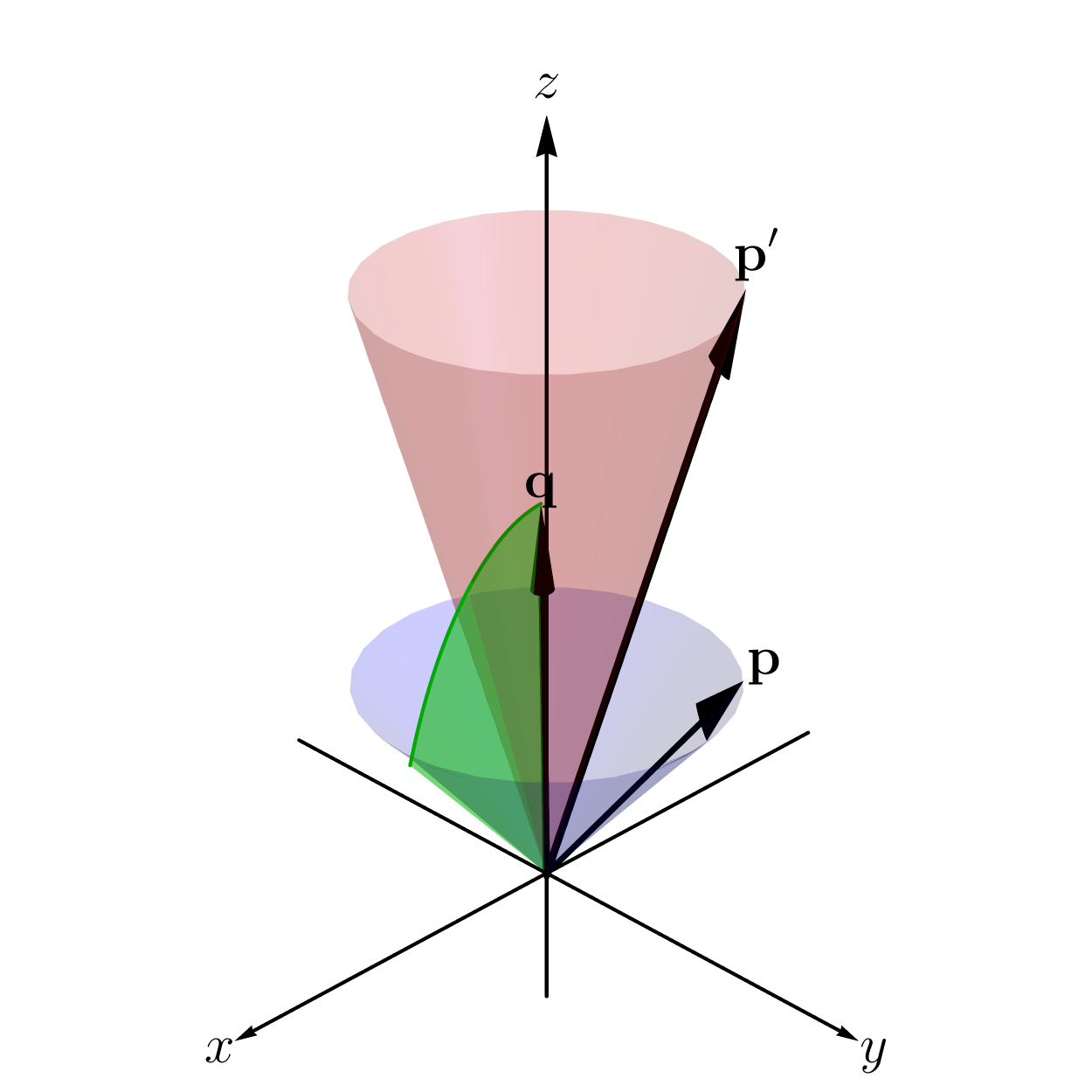}
    \label{fig:vectors-d}
  }
\caption{Kinematic support for $\boldsymbol{p}$, $\boldsymbol{p}'$, and $\boldsymbol{q}$ for different configurations. The blue and red surfaces represent the allowed configurations of $\boldsymbol{p}$ and $\boldsymbol{p}'$, respectively, generated by rotations around the $z$ axis of vectors with fixed magnitudes and polar angles $\theta_p$ and $\theta_{p'}$. The green sector represents the corresponding kinematic support for $\boldsymbol{q}$ determined by Eqs.~(\ref{KinematicMainQMagnitude}), and (\ref{KinematicMainTanThetaQ}). This support is invariant under independent azimuthal rotations of $\boldsymbol{p}$, $\boldsymbol{p}'$, and $\boldsymbol{q}$, since the corresponding transverse momentum magnitudes remain unchanged as discussed in Eq.~\eqref{IndependentAzimuthalVariations}. Here, $\phi_q=0$ is chosen as a reference.}
  \label{fig:Vectors}
\end{figure}

For comparable momentum magnitudes, the allowed angular window can vary
substantially across the $(\theta_p,\theta_{p'})$ plane. For example, in
panels (c1) and (c2), corresponding to $|\boldsymbol p|=5m_f$ and
$|\boldsymbol p'|=3m_f$, the configuration
$(\theta_p,\theta_{p'})=(\pi/2,\pi/6)$ gives
$53.4^\circ\lesssim\theta_q\lesssim68.2^\circ$, corresponding to a
relatively narrow interval of about $15^\circ$. In this configuration,
$\boldsymbol p$ lies entirely in the transverse plane, whereas
$\boldsymbol p'$ has both longitudinal and transverse components. The
longitudinal component $p_z'$ is therefore finite, while the transverse
Bessel support restricts $|\boldsymbol q_\perp|$ to a comparatively narrow
interval. As a result, the ratio $|\boldsymbol q_\perp|/|q_z|$, and
 $\theta_q$, has limited freedom. By contrast, configurations
for which $|q_z|$ becomes smaller while both transverse components remain
sizable can produce a substantially wider angular interval. The
corresponding three-dimensional geometry is illustrated in
Fig.~\ref{fig:vectors-b}, where the green sector occupies a restricted
range of directions around the transverse plane, consistent with the
narrow $\theta_q$ interval found in panels (c1) and (c2).

The angular window becomes particularly narrow when one of the transverse
momentum magnitudes approaches zero. For instance, as
$\theta_p\rightarrow0$ or $\theta_p\rightarrow\pi$, one has
$|\boldsymbol p_\perp|\rightarrow0$, and the transverse support condition
reduces to $|\boldsymbol q_\perp|\rightarrow|\boldsymbol p'_\perp|$.
The lower and upper bounds on $|\boldsymbol q_\perp|$ then coincide,
so that $\theta_q$ is fixed by the remaining transverse and longitudinal
kinematics. This explains the progressive narrowing of the allowed
$\theta_q$ interval near the boundaries of the nonwhite regions in
Fig.~\ref{fig:Tqmax}. 

The comparison between panels (b) and (c) further shows that the angular
freedom depends on the hierarchy between the momentum magnitudes, in
addition to their polar orientations. For the same angular configuration,
$(\theta_p,\theta_{p'})=(\pi/2,\pi/6)$, panel (b), with
$|\boldsymbol p|=3m_f$ and $|\boldsymbol p'|=5m_f$, gives approximately
$6.6^\circ\lesssim\theta_q\lesssim51.8^\circ$, corresponding to an
interval of about $45^\circ$. This is substantially broader than the
corresponding interval in panel (c), where
$|\boldsymbol p|=5m_f$ and $|\boldsymbol p'|=3m_f$. The difference is
associated with the simultaneous change in the longitudinal balance and
in the transverse momentum interval available to $\boldsymbol q$. The
geometric counterparts of these two configurations are shown in
Figs.~\ref{fig:vectors-a} and \ref{fig:vectors-b}: the green sector in
Fig.~\ref{fig:vectors-a} spans a substantially larger range of polar
directions, whereas in Fig.~\ref{fig:vectors-b} it is displaced toward
the transverse plane and is considerably narrower.

The limiting cases in panels (d) and (e) provide a useful interpretation of this behavior. For $|\boldsymbol p|\ll|\boldsymbol p'|$, the two transverse bounds become nearly equal. The allowed interval of $|\boldsymbol q_\perp|$, and  that of $\theta_q$, therefore becomes narrow, as reflected by the small separation between the two rows in panels (d1) and (d2). In this regime, the angular dependence is predominantly controlled by $\theta_{p'}$, as is shown in Fig.~\ref{fig:vectors-c}. Conversely, for $|\boldsymbol p|\gg|\boldsymbol p'|$, the corresponding restriction is controlled primarily by $\theta_p$, as shown in panels (e1) and (e2). In addition, the condition $q_z>0$ restricts the accessible configurations to sufficiently large $\theta_p$, since the longitudinal component associated with the dominant momentum $\boldsymbol p$ must be compensated by the smaller contribution from $\boldsymbol p'$.

The results shown in Fig.~\ref{fig:Tqmax} constrain only the polar geometry of the process. The corresponding transverse vectors may still have arbitrary azimuthal orientations. This is represented in Fig.~\ref{fig:Vectors}, where the blue and red surfaces collect the possible orientations of $\boldsymbol p$ and $\boldsymbol p'$ for fixed magnitudes and polar angles, while the green sector represents the corresponding allowed configurations of $\boldsymbol q$. Then, changing $\phi_p$ or $\phi_{p'}$ while keeping the momentum magnitudes and polar angles fixed does not modify the radial kinematic support. Thus, configurations with identical $|\boldsymbol p|$, $|\boldsymbol p'|$, $|\boldsymbol q|$, $\theta_p$, $\theta_{p'}$, and $\theta_q$ are indistinguishable at the level of the support condition, regardless of their azimuthal orientations.

\subsection{Matrix elements in rotating basis}\label{sec:factorfomrs}
As stated in Sec.~\ref{sec:treelevel}, since the physical photon states satisfy $A^0=0$, the vector structures entering Eq.~\eqref{Gammanu-general} can be organized in terms of the circular and longitudinal components $\Gamma_\pm^s$ and $\Gamma_z^s$, given by Eqs.~\eqref{eq:Gamma+}--\eqref{eq:Gammaz}. After performing the transverse integrations, the vertex for each photon-mode component $s$ can be expressed in terms of scalar form factors multiplying independent Dirac structures,
\bea
\Gamma_\nu^s&=&
\frac{F_+^s}{\sqrt{2}}(\gamma^1+\ii\gamma^2)
\left(\delta_{\nu1}-\ii\delta_{\nu2}\right)\nn\\
&+&\frac{F_-^s}{\sqrt{2}}(\gamma^1-\ii\gamma^2)
\left(\delta_{\nu1}+\ii\delta_{\nu2}\right)\nn\\
&+&
\left(F_z^s\gamma^3+F_{z\Sigma}^s\gamma^3\Sigma_3\right)\delta_{\nu3},
\label{Gamma-physical-decomposition}
\eea
for $\nu=1,2,3$.

Note that the spatial components of $\Gamma_\nu^s$ are naturally expressed in the circular and longitudinal polarization basis. It is therefore convenient to introduce the covariant basis vectors
\bea
\eta_\nu^{\pm}&\equiv&\delta_{\nu1}\pm\ii\delta_{\nu2},\nn\\
\eta_\nu^z&\equiv&\delta_{\nu3},
\eea
so that the physical spatial components of the effective vertex can be written compactly as
\bea
\Gamma_\nu^s&=&\frac{1}{\sqrt{2}}\sum_{\sigma=\pm}F_\sigma^s(\gamma^1+\sigma\ii\gamma^2)\eta_\nu^{-\sigma}\nn\\
&+&\left(F_z^s\gamma^3+F_{z\Sigma}^s\gamma^3\Sigma_3\right)\eta_\nu^z.
\label{Gamma-physical-decomposition}
\eea
The contraction with a physical photon polarization is then given by $\Gamma_\nu^s\epsilon_\beta^\nu$, with $\epsilon_\beta^0=0$.

 The selection rules obtained from the azimuthal integrations in Sec.~\ref{sec:master-integrals} are already incorporated into the definitions of the corresponding master integrals. Then, the Kronecker deltas do not appear explicitly in the form factors below. The non-vanishing coefficients multiplying the circular transverse structures are determined by the master integrals associated with the neighboring radial components of the fermion modes, namely
\bea
F_+^s&=&-\frac{\pi}{\sqrt{2}}\mathcal I_{l'l},
\label{Fp}
\eea
\bea
F_-^s&=&-\frac{\pi}{\sqrt{2}}\mathcal J_{l'l}.
\label{Fm}
\eea
Similarly, for the longitudinal structures we have
\bea
F_z^s&=&-\pi\left(\mathcal K_{l'l}+\mathcal K_{l'+1,l+1}\right),
\label{Fz}
\eea
\bea
F_{z\Sigma}^s&=&-\pi\left(\mathcal K_{l'l}-\mathcal K_{l'+1,l+1}\right).
\label{FzSigma}
\eea

The terms proportional to $\Sigma_3$ encode the dependence of the vertex on the spin projection along the rotation axis, with $\Sigma_3=\ii\gamma^1\gamma^2$. They arise from the spinor structure of the rotating fermion modes and from the decomposition of the covariant Dirac structures in the rotating background. Accordingly, the form factor $F_{z\Sigma}^s$ describes the spin-dependent contribution to the longitudinal vertex structure.

As discussed in Sec.~\ref{sec:master-integrals}, the master integrals $\mathcal I_{l'l}$, $\mathcal J_{l'l}$, and $\mathcal K_{l'l}$ are non-vanishing only in the region $\mathcal R_1$ where
\bea
\Big||\boldsymbol p_\perp'|-|\boldsymbol p_\perp|\Big|<|\boldsymbol q_\perp|<|\boldsymbol p_\perp'|+|\boldsymbol p_\perp|.
\label{eq:bessel_support_kinematics}
\eea

At tree level, the explicit rotational contribution contained in the temporal component of the covariant vertex does not contribute after projection onto the physical photon modes, while the rotational dependence of the effective vertex is encoded in the rotating fermion and photon modes and in the corresponding angular-momentum selection rules. Explicit $\Omega$-dependent structures may arise in higher-order vertex corrections through the rotating propagators.

{
\subsection{Tree-Level Kinematics and Consequences for Photon-Production Processes}
\label{sec:photonabsorption}
A particularly transparent limiting configuration is obtained when all three external states propagate in the transverse plane, $p_z=p'_z=q_z=0$. In this case, longitudinal momentum conservation is trivially satisfied, while the absence of transverse translational invariance persists: the transverse integration does not produce a two-dimensional momentum conservation delta function. Instead, the radial overlap requires the three transverse momentum magnitudes to satisfy
\begin{eqnarray}
    \big||\boldsymbol p'_\perp|-|\boldsymbol p_\perp|\big|<|\boldsymbol q_\perp|<|\boldsymbol p'_\perp|+|\boldsymbol p_\perp|.
\end{eqnarray}
The relative azimuthal orientations remain unconstrained by this radial support condition. Nevertheless, the absence of transverse vector-momentum conservation does not, by itself, imply that an otherwise forbidden radiative process becomes kinematically allowed. For an on-shell massive fermion absorbing a real photon, the angular-momentum selection rule cancels the rotational contribution to the conserved rotating-frame energy, so that energy conservation requires $|\boldsymbol q_\perp|=E'-E$. However, for $p_z=p'_z=0$, the fermion dispersion relations imply
\begin{eqnarray}
    |E'-E|<\big||\boldsymbol p'_\perp|-|\boldsymbol p_\perp|\big|,
\end{eqnarray}
for unequal transverse momentum magnitudes. This is incompatible with the lower bound imposed by the radial Bessel overlap. Hence, nontrivial photon absorption remains kinematically forbidden in this strictly transverse configuration, despite the absence of a transverse momentum-conservation delta function.

From a more general perspective, the kinematic restrictions associated with the tree-level fermion-photon vertex can be derived by combining the conservation laws encoded in the delta functions with the analytical properties of the radial overlap integrals. In the rotating-photon basis, the remaining conserved quantities appear explicitly in the delta functions
\bea
\delta\left[E'-E-|\mathbf q|+\Omega(j-(l'-l))\right]\delta(p_z-p'_z+q_z).
\label{eq:rotating_energy_longitudinal_deltas}
\eea

The longitudinal delta function in Eq.~\eqref{eq:rotating_energy_longitudinal_deltas} fixes the longitudinal photon momentum according to
\bea
q_z=p'_z-p_z,
\eea
so that the photon energy can be expressed in terms of its transverse momentum and the longitudinal momenta of the fermions as
\bea
|\mathbf q|^2=|\mathbf q_\perp|^2+(p'_z-p_z)^2,
\eea
from wich the lower bound in Eq.~\eqref{eq:bessel_support_kinematics} therefore implies
\bea
|\mathbf q|^2>\left(|\mathbf p'_\perp|-|\mathbf p_\perp|\right)^2+(p'_z-p_z)^2.
\label{eq:q_lower_bound_energy}
\eea

This inequality can be compared directly with the difference between the on-shell fermion energies. For the external fermion states, the dispersion relations read
\bea
E&=&\sqrt{m_f^2+|\mathbf p_\perp|^2+p_z^2},\nn\\
E'&=&\sqrt{m_f^2+|\mathbf p'_\perp|^2+p_z'^2}.
\label{eq:fermion_onshell_energies}
\eea
Their energy difference satisfies
\bea
(E'-E)^2&=&2m_f^2+|\mathbf p'_\perp|^2+|\mathbf p_\perp|^2+p_z'^2+p_z^2-2E'E\nn\\
&=&\left(|\mathbf p'_\perp|-|\mathbf p_\perp|\right)^2+(p'_z-p_z)^2\nn\\
&-&2\left(E'E-|\mathbf p'_\perp||\mathbf p_\perp|-p_z'p_z-m_f^2\right).
\eea
The last expression follows by completing the squares. Substituting it into Eq.~\eqref{eq:q_lower_bound_energy} gives
\bea
|\mathbf q|^2&>&(E'-E)^2\nn\\
&&+2\left(E'E-|\mathbf p'_\perp||\mathbf p_\perp|-p_z'p_z-m_f^2\right).
\label{inequality1}
\eea

The quantity in parentheses is non-negative. To establish this result, let us introduce the Euclidean vectors
\bea
\mathbf a&\equiv&(|\mathbf p'_\perp|,p'_z,m_f),\qquad |\mathbf a|=E',\nn\\
\mathbf b&\equiv&(|\mathbf p_\perp|,p_z,m_f),\qquad |\mathbf b|=E,
\eea
then, the Cauchy--Schwarz inequality gives
\bea
|\mathbf a|\,|\mathbf b|&\geq&\mathbf a\cdot\mathbf b,\nn\\
E'E&\geq&|\mathbf p'_\perp||\mathbf p_\perp|+p'_zp_z+m_f^2.
\eea

Therefore, it follows from Eq.~\eqref{inequality1} that
\bea
|\mathbf q|^2>(E'-E)^2,
\eea
and hence
\bea
|\mathbf q|>|E'-E|.
\label{eq:energy_photon_inequality_1}
\eea

On the other hand, the energy-conservation delta function in Eq.~\eqref{eq:rotating_energy_longitudinal_deltas} imposes
\bea
E'-E=|\mathbf q|-\Omega(j-(l'-l)).
\label{eq:rotating_energy_condition}
\eea
Combining this relation with Eq.~\eqref{eq:energy_photon_inequality_1} yields
\bea
\left||\mathbf q|-\Omega(j-(l'-l))\right|<|\mathbf q|.
\eea
For $\Omega>0$, the allowed angular-momentum channels must therefore satisfy
\bea
0<(j-(l'-l))<\frac{2|\mathbf q|}{\Omega}.
\label{eq:angular_energy_kinematic_constraint_1}
\eea
This condition is necessary for kinematic compatibility with the energy-conservation constraint. In particular, it excludes channels with $j-(l'-l)\leq0$.

A complementary restriction follows from the upper bound in Eq.~\eqref{eq:bessel_support_kinematics}. Combining that bound with the longitudinal momentum relation gives
\bea
|\mathbf q|^2&<&\left(|\mathbf p'_\perp|+|\mathbf p_\perp|\right)^2+(p'_z-p_z)^2\nn\\
&<&(E'+E)^2,
\eea
where the second inequality follows from the on-shell dispersion relations for massive fermions. Thus,
\bea
|\mathbf q|<E'+E.
\label{eq:energy_photon_inequality_2}
\eea
Using Eq.~\eqref{eq:rotating_energy_condition}, this upper bound implies
\bea
j-(l'-l)<\frac{2E}{\Omega}.
\label{eq:angular_energy_kinematic_constraint_2}
\eea
Equations~\eqref{eq:angular_energy_kinematic_constraint_1} and~\eqref{eq:angular_energy_kinematic_constraint_2} therefore provide complementary upper bounds on the angular-momentum difference $j-(l'-l)$, expressed in terms of the photon energy and the initial fermion energy, respectively. These conditions are necessary but not sufficient: the corresponding channels must also satisfy the angular selection rules and yield non-vanishing dynamical amplitudes.

The kinematic restrictions must consequently be combined with the analytical structure of the tree-level vertex. As shown in Sec.~\ref{sec:master-integrals}, the amplitude can be organized as a linear combination of the three master radial integrals $\mathcal{I}_{l'l}$, $\mathcal{J}_{l'l}$, and $\mathcal{K}_{l'l}$, evaluated in the region specified by Eq.~\eqref{eq:bessel_support_kinematics}. Their contributions are associated with the angular selection rules
\bea
\delta_{l'-l,1+j+\lambda_s}\mathcal{I}_{l'l};~~
\delta_{l'-l,j+\lambda_s-1}\mathcal{J}_{l'l};~~
\delta_{l'-l,j+\lambda_s}\mathcal{K}_{l'l}.
\label{eq:kronecker_master_integrals}
\eea

The values of $\lambda_s$, together with the kinematic restrictions in Eqs.~\eqref{eq:angular_energy_kinematic_constraint_1} and~\eqref{eq:angular_energy_kinematic_constraint_2}, determine which angular-momentum channels selected by the Kronecker deltas are kinematically compatible with energy and momentum conservation. Channels that violate either restriction cannot contribute to the corresponding on-shell process, irrespective of the detailed values of the radial integrals.

For instance, consider the longitudinal polarization, for which $\lambda_s=\lambda_z=0$. The three Kronecker deltas then select
\bea
\delta_{l'-l,1+j}&\to&j=l'-l-1\nn\\
\delta_{l'-l,j-1}&\to&j=l'-l+1\nn\\
\delta_{l'-l,j}&\to&j=l'-l.
\eea

Using the first kinematic constraint in Eq.~\eqref{eq:angular_energy_kinematic_constraint_1}, the three resulting values of $j$ give
\bea
0<-1<\frac{2|\mathbf q|}{\Omega}&\to&\text{kinematically forbidden}\nn\\
0<1<\frac{2|\mathbf q|}{\Omega}&\to&\text{kinematically allowed}\nn\\
0<0<\frac{2|\mathbf q|}{\Omega}&\to&\text{kinematically forbidden}.
\eea
In the third case, although $0<2|\mathbf q|/\Omega$ is automatically satisfied for $\Omega>0$, the chained inequality also requires $0<0$, which is false. Therefore, the third case is kinematically forbidden.

\begin{table}
\centering
\begin{tabular}{c|c|c|c|c}
\hline
$s$ & $\lambda_s$ & $\mathcal{I}_{l'l}$ & $\mathcal{J}_{l'l}$ & $\mathcal{K}_{l'l}$ \\
\hline

$+$
& $-1$
&
\begin{tabular}{c}
$0<0<|\mathbf q|/\Omega$ \\
$0<E/\Omega$ \\
\textbf{Forbidden}
\end{tabular}
&
\begin{tabular}{c}
$0<1<|\mathbf q|/\Omega$ \\
$1<E/\Omega$ \\
\textbf{Allowed}
\end{tabular}
&
\begin{tabular}{c}
$0<1/2<|\mathbf q|/\Omega$ \\
$1/2<E/\Omega$ \\
\textbf{Allowed}
\end{tabular}
\\
\hline

$-$
& $+1$
&
\begin{tabular}{c}
$0<-1<|\mathbf q|/\Omega$ \\
$-1<E/\Omega$ \\
\textbf{Forbidden}
\end{tabular}
&
\begin{tabular}{c}
$0<0<|\mathbf q|/\Omega$ \\
$0<E/\Omega$ \\
\textbf{Forbidden}
\end{tabular}
&
\begin{tabular}{c}
$0<-1/2<|\mathbf q|/\Omega$ \\
$-1/2<E/\Omega$ \\
\textbf{Forbidden}
\end{tabular}
\\
\hline

$z$
& $0$
&
\begin{tabular}{c}
$0<-1/2<|\mathbf q|/\Omega$ \\
$-1/2<E/\Omega$ \\
\textbf{Forbidden}
\end{tabular}
&
\begin{tabular}{c}
$0<1/2<|\mathbf q|/\Omega$ \\
$1/2<E/\Omega$ \\
\textbf{Allowed}
\end{tabular}
&
\begin{tabular}{c}
$0<0<|\mathbf q|/\Omega$ \\
$0<E/\Omega$ \\
\textbf{Forbidden}
\end{tabular}
\\
\hline

\end{tabular}
\caption{Kinematic selection of the master radial integrals for the different photon polarization components. The entries show the corresponding kinematic conditions and verdicts from Eqs.~\eqref{eq:angular_energy_kinematic_constraint_1} and \eqref{eq:angular_energy_kinematic_constraint_2}. Here, ``Allowed'' denotes kinematic compatibility only.}
\label{tab:kinematic_selection_master_integrals}
\end{table}

Table~\ref{tab:kinematic_selection_master_integrals} summarizes the kinematic selection of the three master radial integrals for each photon polarization component. The Kronecker deltas determine which angular momentum transitions are associated with each integral, while Eqs.~\eqref{eq:angular_energy_kinematic_constraint_1} and \eqref{eq:angular_energy_kinematic_constraint_2} determine whether those transitions are compatible with the energy and momentum constraints. The resulting pattern is highly restrictive: only the $\mathcal{J}_{l'l}$ and $\mathcal{K}_{l'l}$ contributions for $\lambda_s=-1$, and the $\mathcal{J}_{l'l}$ contribution for $\lambda_s=0$, remain kinematically allowed. The allowed channels are characterized by positive values of $j-(l'-l)$, so that the rotational contribution $\Omega(j-(l'-l))$ enters the energy balance with a positive sign. This term represents the energy that can be supplied by the rotating medium to the fermion-photon process. The table therefore makes explicit the connection between the angular momentum transfer encoded in the radial integrals and the possibility of extracting rotational energy from the medium. The ``Allowed'' entries refer only to these kinematic constraints; the corresponding amplitudes can still vanish due to the remaining dynamical factors. Nevertheless, the kinematic accessibility of individual radial-integral sectors does not necessarily imply a nonzero transition probability when all external particles are on shell in the rotating basis.

In the present case, the vanishing of the physical tree-level amplitude follows directly from its contraction with the physical photon polarization vectors,
\bea
\sum_{s=\pm,z}\Gamma^s_{\nu}\boldsymbol{\epsilon}^{\nu}_s
&=&F_+^+\left(\gamma^1+\ii\gamma^2\right)
+F_-^-\left(\gamma^1-\ii\gamma^2\right)
\nonumber\\
&+&\left(F_z^z\gamma^3+F_{z\Sigma}^z\gamma^3\Sigma_3\right).
\eea
This contraction selects only the diagonal components in the polarization labels. Moreover, the form factors are directly proportional to the master radial integrals, as given in Eqs.~\eqref{Fm}--\eqref{FzSigma}. Specifically, $F_+^+$ is proportional to $\mathcal{I}_{l'l}$, $F_-^-$ to $\mathcal{J}_{l'l}$, and both longitudinal form factors, $F_z^z$ and $F_{z\Sigma}^z$, involve linear combinations of $\mathcal{K}_{l'l}$ and $\mathcal{K}_{l'+1,l+1}$. According to the kinematic selection rules summarized in Table~\ref{tab:kinematic_selection_master_integrals}, all these diagonal contributions are forbidden for the on-shell processes considered here. Hence, the contracted tree-level amplitude vanishes identically.

This result shows that the opening of kinematic channels in the rotating basis does not necessarily lead to nonzero physical transition probabilities. The same cancellation upon contraction with the physical photon polarization vectors holds for the tree-level photon-production processes associated with pair annihilation and bremsstrahlung. These results apply to the on-shell processes considered here and do not imply that the vertex itself vanishes identically, since its analytic structure remains nontrivial for off-shell external states.

From a phenomenological perspective, it is important to distinguish the kinematic effects of describing the system in a rotating reference frame from the dynamical effects of rotation in an interacting medium. In the approximation adopted here, rotation enters through the rotating spatial coordinates, without specifying the microscopic mechanism sustaining the rotation or its dynamical coupling to the photon field. Under these assumptions, the tree-level probabilities vanish for the three processes
\bea
f_{\alpha}+\gamma(q)&\to&f_{\alpha'},\nn\\
f_{\alpha}+\bar{f}_{\alpha'}&\to&\gamma(q),\nn\\
f_{\alpha}&\to&f_{\alpha'}+\gamma(q),
\eea
when the photon is described by a rotating mode and all external states satisfy the corresponding on-shell conditions.

This result should not, however, be directly extrapolated to photon production in relativistic heavy-ion collisions. In such systems, the collective rotational motion of the quark-gluon plasma originates from the angular momentum deposited in non-central collisions and the subsequent evolution of the fluid velocity field. The dynamics is governed by the strong interactions, with shear viscosity influencing the evolution of velocity gradients and momentum transport~\cite{Shen:2021QGPFlow,Heinz:2013CollectiveFlow}. This physical setting therefore differs from the purely kinematic rotating-coordinate description adopted in the present work.} {Nevertheless, the rotational modification of the energy spectrum and the corresponding mode structure may have observable consequences, particularly through changes in particle distributions and transition rates in a thermal and dense medium.} {In particular, if rotation is understood as a property of the strongly interacting medium, it is the quark and gluon degrees of freedom that participate directly in the rotating background, whereas photons, being colorless electromagnetic probes, do not participate in the strong interactions responsible for the medium's collective motion. Furthermore, owing to their comparatively long mean free paths, direct photons produced within the fireball can generally escape without significant final-state interactions \cite{David:2020DirectPhotons,Geurts:2023ElectromagneticProbes}. It is thus physically motivated to distinguish the rotational dynamics of the strongly interacting constituents from the propagation of photons, whose angular distributions, polarization, and emission spectra may nevertheless retain signatures of the rotating source. Consequently, the rotation of the medium does not, by itself, require the photon to be described by a rotating mode. This distinction is essential} {when extending the present formalism to a physical medium, where rotation can affect the fermion degrees of freedom and their interactions while the emitted photon propagates as an asymptotic state in the laboratory frame.} 

{A natural extension of this work is to formulate the photon-production amplitude within this framework and determine how the rotational modifications of the fermion states and their interactions are reflected in observable photon-production rates and distributions.} In particular, the vertex structure and its contractions with the physical photon polarizations must be derived anew to establish whether the cancellation found in the present calculation persists when the external photon is described by a Minkowski plane wave. {Although a change of photon basis alone cannot open a forbidden channel, the inclusion of medium effects may modify the available transitions and their rates.} Resolving this issue is essential for assessing the possibility of non-vanishing photon-production amplitudes in a rotating, strongly interacting medium. This investigation is currently in progress and will be reported elsewhere.

\section{Summary and conclusions}\label{sec:conclusions}

We have studied the tree-level fermion-photon effective vertex for on-shell particles in a rigidly rotating background at zero temperature. The rotational effects are incorporated directly into the external fermion and photon modes, allowing the interaction to be evaluated without treating the rotation as a perturbation. The fermion modes are characterized by their orbital angular momentum, helicity, longitudinal momentum, and transverse momentum magnitude, while the physical photon modes are constructed from the circular polarization components in the transverse plane and the Cartesian component along the rotation axis.

The resulting matrix element exhibits the anisotropic structure induced by the rotating background. The integrations over time and the longitudinal coordinate produce the corresponding conservation laws for the rotating-frame energy and the longitudinal momentum. In contrast, the transverse integration does not yield a two-dimensional momentum-conservation delta function. Instead, the transverse dependence is encoded in radial integrals involving three Bessel functions. The azimuthal integrations lead to selection rules relating the orbital angular momentum of the fermions to the photon orbital label and its polarization-dependent spin projection. These relations reflect the angular-momentum structure of the rotating modes and replace the transverse momentum-conservation condition of the translationally invariant vacuum.

The radial integrals governing the different components of the vertex involve the same three transverse momentum magnitudes, $|\boldsymbol p_\perp|$, $|\boldsymbol p'_\perp|$, and $|\boldsymbol q_\perp|$. Their analytical evaluation shows that the nonvanishing contributions are restricted to the region in which these magnitudes can form an auxiliar triangle. This support condition constrains the magnitudes of the transverse projections of the momenta, but does not impose a vector relation among the corresponding transverse momenta. Then, the allowed configurations retain independent azimuthal orientations, while the longitudinal momentum remains fixed by the corresponding conservation law.

The vertex can be organized in terms of circular transverse and longitudinal Dirac structures. The transverse components are naturally associated with the combinations $\gamma^1\pm\ii\gamma^2$, whereas the longitudinal sector involves $\gamma^3$ and $\gamma^3\Sigma_3$, where $\Sigma_3=\ii\gamma^1\gamma^2$ describes the spin projection along the rotation axis. After the transverse integrations, these structures are parametrized by four scalar form factors for each photon-mode component. Including the three photon-mode components $s$ and the two independent physical photon states $\beta$, this corresponds to $4\times3=12$ form factors before the physical-state contraction and $4\times3\times2=24$ coefficients after projection onto the physical photon states. The explicit rotational contributions proportional to $\Omega$ arise in the temporal component of the covariant vertex through the coordinate-dependent gamma matrices. However, since the physical photon modes satisfy $A^0=0$, these terms do not contribute to the effective vertex after projection onto the physical photon states. The rotational dependence of the resulting tree-level form factors is therefore encoded in the rotating fermion and photon modes and in the associated angular-momentum selection rules.

We have also verified current conservation for the transition current constructed from on-shell rotating fermion solutions. The corresponding gauge variation of the tree-level matrix element reduces to a boundary contribution after using local current conservation. Under the boundary conditions and rigid-rotation prescription adopted here, this contribution vanishes, ensuring gauge invariance of the tree-level interaction. The resulting construction therefore provides a consistent framework for describing the fermion-photon vertex when translational invariance in the transverse plane is replaced by the symmetries of a rigidly rotating background.

{

Finally, we have examined the kinematic restrictions governing on-shell photon absorption and their implications for the corresponding tree-level amplitudes. Although the absence of transverse translational invariance relaxes the usual momentum-conservation constraints and permits kinematic channels that are forbidden in vacuum, the accessibility of individual radial-integral contributions does not guarantee a nonvanishing physical transition amplitude. In the processes considered here, the contraction of the vertex with the physical photon polarizations produces cancellations that render the corresponding on-shell tree-level amplitudes identically zero. The same cancellation mechanism can be established for the tree-level processes of bremsstrahlung and pair annihilation when the photon is described by a rotating mode. Thus, the kinematic restrictions derived from the radial Bessel integrals must be distinguished from the dynamical cancellations arising in the complete physical amplitude. These results establish that the opening of kinematically allowed channels in a rigidly rotating background does not, by itself, imply nonzero tree-level photon-production probabilities. 

A natural direction for future work is to extend the present analysis to phenomenologically relevant scenarios in which the rotational dynamics originate from the strongly interacting medium, as in the quark-gluon plasma produced in relativistic heavy-ion collisions. In this setting, the rotation would be incorporated into the fermion degrees of freedom and their interactions,} {including the effects of finite temperature and density on the relevant propagators} { while the emitted photon would be treated as a non-rotating asymptotic state propagating in the laboratory frame. This formulation would allow us to reassess the physical vertex and its polarization contractions, determine whether the cancellations found in the present calculation persist under these conditions, and establish how} {rotation, together with medium-induced modifications of the particle spectra and interactions,} { can influence observable photon-production rates. Such an extension is essential for connecting the microscopic structure of the fermion-photon interaction to phenomenological descriptions of electromagnetic emission from rotating QCD matter. This investigation is currently in progress and will be reported elsewhere.

}

\section*{Acknowledgments} 
JDC-Y acknowledges the kind support received from L.A. Hern\'andez and the Universidad Aut\'onoma Metropolitana, as well as from the Instituto de Ciencias Nucleares (ICN), Universidad Nacional Aut\'onoma de M\'exico, during a work visit when the final stages of this work were carried out. He also gratefully acknowledges the Universidad del Valle for authorizing the medium-term research leave within the framework of PPI No.~71433 of the Vicerrector\'ia de Investigaciones, which made this work visit to the ICN possible. E.M. acknowledges financial support from ANID Fondecyt Grant No 1230440. Support for this work has also been received in part by DGAPA-PAPIIT-UNAM grant number AG100826, and from Secretar\'ia de Ciencia, Humanidades, Tecnolog\'ia e Innovaci\'on (SECIHTI) M\'exico grant numbers CIORGANISMOS-2025-17, CBF-2025-G-1718, and CBF-2026-465.
\appendix

\section{Exact fermion modes in a rotating frame}
\label{app:RotatingFermionModes}

We derive in this appendix the exact fermion modes of the free Dirac equation in a uniformly rotating frame. The construction follows Ref.~\cite{CastanoYepes:2025}, retaining the complete set of orbital-angular-momentum channels. These modes will subsequently be used as the external fermion states and as the spectral basis of the free fermion propagator entering the vertex calculation.

For a constant angular velocity $\Omega$ directed along the $z$ axis, the free Dirac equation in the rotating frame is~\cite{Ayala:2021osy,CastanoYepes:2025}
\bea
\left[\ii\gamma^a\partial_a+\gamma^0\Omega\hat J_z-m_f\right]\psi(x)=0.
\label{AppFreeRotatingDiracEquation}
\eea
Multiplying by $\gamma^0$ gives the Hamiltonian form
\bea
\ii\partial_t\psi(x)=\hat H_\Omega\psi(x),
\label{AppHamiltonianEquation}
\eea
with
\bea
\hat H_\Omega=\hat H_0-\Omega\hat J_z,\qquad \hat H_0=-\ii\boldsymbol{\alpha}\cdot\boldsymbol{\nabla}+\beta m_f.
\label{AppRotatingHamiltonian}
\eea
The total angular momentum about the rotation axis is
\bea
\hat J_z=\hat L_z+\frac{\hat\Sigma_z}{2},\qquad \hat L_z=-\ii\partial_\phi,
\label{AppTotalAngularMomentum}
\eea
where
\bea
\hat\Sigma_z=\begin{pmatrix}\sigma_z&0\\0&\sigma_z\end{pmatrix}.
\label{AppSigmaZ}
\eea

The construction of the modes is most conveniently performed by first solving the spectral problem of the nonrotating Hamiltonian $\hat H_0$. Since
\bea
[\hat H_0,\hat J_z]=0,
\label{AppH0JzCommutator}
\eea
the resulting modes can be chosen as simultaneous eigenstates of $\hat H_0$ and $\hat J_z$. The effect of rotation is then incorporated exactly through the term $-\Omega\hat J_z$ in $\hat H_\Omega$.

We distinguish the positive- and negative-energy branches according to the spectrum of $\hat H_0$. For $E>0$, we define the positive- and negative-energy bispinors through
\bea
\hat H_0u_{\alpha}(x)=E u_{\alpha}(x),\qquad \hat H_0v_{\alpha}(x)=-E v_{\alpha}(x).
\label{AppEnergyBranches}
\eea
The corresponding dispersion relation is
\bea
E^2=|\boldsymbol{p}|^2+m_f^2,\qquad |\boldsymbol{p}|^2=\boldsymbol{p}_{\perp}^{\,2}+p_z^2.
\label{AppDispersionRelation}
\eea
where
\bea
\boldsymbol{p}_{\perp}^{\,2}\equiv p_x^2+p_y^2.
\label{AppTransverseMomentumDefinition}
\eea

The index $\alpha$ collectively denotes the quantum numbers required to specify a mode, including the orbital-angular-momentum channel $l$, the helicity $h$, and the continuous momentum variables.

We now construct the positive-energy solution. We decompose the four-component bispinor into two-component Pauli spinors,
\bea
u_{\alpha}(x)=\begin{pmatrix}\varphi_{\alpha}(x)\\\chi_{\alpha}(x)\end{pmatrix}.
\label{AppPositiveBispinorDecomposition}
\eea
Inserting this decomposition into the positive-energy eigenvalue
equation yields the following coupled equations:
\bea
m_f\varphi_{\alpha}+\left(\boldsymbol{\sigma}\cdot\hat{\boldsymbol{p}}\right)\chi_{\alpha}=E\varphi_{\alpha},
\label{AppPositiveUpperEquation}
\eea
and
\bea
\left(\boldsymbol{\sigma}\cdot\hat{\boldsymbol{p}}\right)\varphi_{\alpha}-m_f\chi_{\alpha}=E\chi_{\alpha}.
\label{AppPositiveLowerEquation}
\eea
The second equation determines the lower Pauli spinor in terms of the upper one,
\bea
\chi_{\alpha}=\frac{\left(\boldsymbol{\sigma}\cdot\hat{\boldsymbol{p}}\right)}{E+m_f}\varphi_{\alpha}.
\label{AppPositiveLowerSpinor}
\eea
Thus, only one independent two-component spinor is required to construct the positive-energy solution. The remaining structure follows from the eigenvalue equations for $\hat J_z$ and helicity.

We use cylindrical coordinates,
\bea
x=\rho\cos\phi,\qquad y=\rho\sin\phi.
\label{AppCylindricalCoordinates}
\eea
The two-component angular momentum operator acting on $\varphi_{\alpha}$ is
\bea
\hat J_z^{(2)}=-\ii\partial_\phi+\frac{\sigma_z}{2}.
\label{AppTwoComponentJz}
\eea
We choose the upper Pauli spinor to be an eigenstate of this operator,
\bea
\hat J_z^{(2)}\varphi_{\alpha}(\rho,\phi)=j\varphi_{\alpha}(\rho,\phi).
\label{AppPauliJzEigenvalue}
\eea
To determine its angular structure without making any prior assumption about the relation between the two components, we write
\bea
\varphi_{\alpha}(\rho,\phi)=\begin{pmatrix}f_m(\rho)e^{\ii m\phi}\\g_n(\rho)e^{\ii n\phi}\end{pmatrix}.
\label{AppGeneralPauliSpinor}
\eea
Single-valuedness under $\phi\rightarrow\phi+2\pi$ requires $m,n\in\mathbb{Z}$. Acting with $\hat J_z^{(2)}$ on the upper component gives
\bea
\hat J_z^{(2)}\left[f_m(\rho)e^{\ii m\phi}\chi_+\right]=\left(m+\frac12\right)f_m(\rho)e^{\ii m\phi}\chi_+.
\label{AppUpperAngularAction}
\eea
where $\chi_+$ is the eigenstate of $\sigma_z$ with eigenvalue $+1$. Therefore,
\bea
m+\frac12=j.
\label{AppUpperAngularEigenvalue}
\eea
For the lower component, the eigenvalue of $\sigma_z$ is $-1$, and hence
\bea
\hat J_z^{(2)}\left[g_n(\rho)e^{\ii n\phi}\chi_-\right]=\left(n-\frac12\right)g_n(\rho)e^{\ii n\phi}\chi_-.
\label{AppLowerAngularAction}
\eea
which requires
\bea
n-\frac12=j.
\label{AppLowerAngularEigenvalue}
\eea
Then,
\bea
n=m+1.
\label{AppAngularQuantumNumberRelation}
\eea
We identify the orbital-angular-momentum quantum number with the integer carried by the upper component,
\bea
l\equiv m,
\label{AppOrbitalAngularMomentum}
\eea
so that
\bea
j=l+\frac12.
\label{AppTotalAngularMomentumEigenvalue}
\eea
The upper Pauli spinor therefore has the definite angular structure
\bea
\varphi_{\alpha}(\rho,\phi)=\begin{pmatrix}f_l(\rho)e^{\ii l\phi}\\g_l(\rho)e^{\ii(l+1)\phi}\end{pmatrix}.
\label{AppAngularPauliSpinor}
\eea
The appearance of $l+1$ in the lower component is therefore a direct consequence of the requirement that both components have the same eigenvalue of the total angular momentum. No assumption about the radial functions has been made at this stage.

We next impose the helicity condition. The choice of helicity is natural because the free Hamiltonian $\hat H_0$ is rotationally invariant and helicity commutes with both $\hat H_0$ and $\hat J_z$. In discussions of rotating systems, vorticity is sometimes compared with a magnetic field because both introduce a preferred spatial direction. This analogy might suggest labeling the fermion states by the projection of the spin along the rotation axis, in analogy with the spin projection along the magnetic-field direction in a magnetized medium. However, the analogy is not direct. In a magnetic field, the field enters the fermion Hamiltonian through the minimal coupling to the vector potential and produces a distinguished spin operator associated with the magnetic-field direction, commonly represented by the corresponding spin projector. In the present problem, the rotation enters the free Hamiltonian through the coupling $-\Omega\hat J_z$, so the quantity selected dynamically by the rotating background is the total angular momentum $J_z$, rather than the spin projection alone. At the same time, because the free Hamiltonian remains rotationally invariant before the rotational shift is introduced, helicity provides a conserved quantum number that fixes the relative structure of the two Pauli components. It is therefore convenient to use $h$ to label the spinor modes, while the effect of the rotation is encoded independently through the eigenvalue of $J_z$.

The helicity operator of the free Dirac field is
\bea
\hat W_0=\frac{\hat{\boldsymbol{\Sigma}}\cdot\hat{\boldsymbol{p}}}{2|\boldsymbol{p}|},
\label{AppHelicityOperator}
\eea
and it commutes with the free Hamiltonian and the total angular momentum,
\bea
[\hat H_0,\hat W_0]=0,\qquad[\hat J_z,\hat W_0]=0.
\label{AppHelicityCommutators}
\eea
We may therefore choose the positive-energy modes to satisfy
\bea
\hat W_0u_{\alpha}=h u_{\alpha},\qquad h=\pm\frac12.
\label{AppPositiveHelicity}
\eea

Because $\hat{\boldsymbol{\Sigma}}$ is block diagonal, the upper Pauli spinor satisfies
\bea
\left(\boldsymbol{\sigma}\cdot\hat{\boldsymbol{p}}\right)\varphi_{\alpha}=2h|\boldsymbol{p}|\varphi_{\alpha}.
\label{AppPauliHelicityEquation}
\eea
The lower Pauli spinor then follows directly from Eq.~\eqref{AppPositiveLowerSpinor},
\bea
\chi_{\alpha}=\frac{2h|\boldsymbol{p}|}{E+m_f}\varphi_{\alpha}.
\label{AppPositiveLowerHelicityRelation}
\eea
Thus, once the upper Pauli spinor is determined, the complete positive-energy bispinor is fixed.

The transverse momentum operators in cylindrical coordinates are
\bea
\hat p_\pm=\hat p_x\pm\ii\hat p_y=-\ii e^{\pm\ii\phi}\left(\partial_\rho\pm\frac{\ii}{\rho}\partial_\phi\right).
\label{AppTransverseMomentumOperators}
\eea

Using Eq.~\eqref{AppAngularPauliSpinor} in the helicity equation gives
\bea
\left(p_z-2h|\boldsymbol{p}|\right)f_l(\rho)-\ii\left(\partial_\rho+\frac{l+1}{\rho}\right)g_l(\rho)=0.
\label{AppRadialEquationUpper}
\eea
and
\bea
-\ii\left(\partial_\rho-\frac{l}{\rho}\right)f_l(\rho)-\left(p_z+2h|\boldsymbol{p}|\right)g_l(\rho)=0.
\label{AppRadialEquationLower}
\eea
These two first-order equations contain the complete radial information. Eliminating $g_l(\rho)$ gives
\bea
\left[\partial_\rho^2+\frac{1}{\rho}\partial_\rho-\frac{l^2}{\rho^2}+|\boldsymbol{p}_\perp|^2\right]f_l(\rho)=0.
\label{AppBesselEquationUpper}
\eea
The solution regular at the origin is
\bea
f_l(\rho)=A_lJ_l\left(|\boldsymbol{p}_\perp|\rho\right).
\label{AppUpperRadialSolution}
\eea
Similarly, eliminating $f_l(\rho)$ gives
\bea
\left[\partial_\rho^2+\frac{1}{\rho}\partial_\rho-\frac{(l+1)^2}{\rho^2}+|\boldsymbol{p}_\perp|^2\right]g_l(\rho)=0.
\label{AppBesselEquationLower}
\eea
and therefore
\bea
g_l(\rho)=B_lJ_{l+1}\left(|\boldsymbol{p}_\perp|\rho\right).
\label{AppLowerRadialSolution}
\eea

The second-order equations determine the radial dependence but do not fix the relative coefficient between the two components. This information is contained in the original first-order helicity equation. Using the Bessel identity
\bea
\left(\partial_\rho+\frac{l+1}{\rho}\right)J_{l+1}\left(|\boldsymbol{p}_\perp|\rho\right)=|\boldsymbol{p}_\perp|J_l\left(|\boldsymbol{p}_\perp|\rho\right),
\label{AppBesselIdentity}
\eea
Eq.~\eqref{AppRadialEquationUpper} becomes
\bea
\left(p_z-2h|\boldsymbol{p}|\right)A_l-\ii|\boldsymbol{p}_\perp|B_l=0.
\label{AppRadialCoefficientEquation}
\eea
so that
\bea
\frac{B_l}{A_l}=-\ii\frac{p_z-2h|\boldsymbol{p}|}{|\boldsymbol{p}_\perp|}.
\label{AppRadialCoefficientRatio}
\eea

To express this ratio in a form adapted to the helicity quantum number, we introduce
\bea
N_\pm=\sqrt{1\pm\frac{p_z}{|\boldsymbol{p}|}}.
\label{AppNpmDefinition}
\eea
The origin of these factors is therefore purely kinematical. They follow from resolving the three-momentum into its longitudinal and transverse components and rewriting the ratio appearing in the helicity equation. Since
\bea
|\boldsymbol{p}_\perp|^2=|\boldsymbol{p}|^2-p_z^2,
\label{AppTransverseMomentumRelation}
\eea
we have
\bea
\frac{|\boldsymbol{p}_\perp|}{|\boldsymbol{p}|}=N_+N_-.
\label{AppNpmProduct}
\eea

For $2h=\pm1$, the coefficient relation can  be written as
\bea
A_l={\cal N}_lN_{2h},\qquad B_l=\ii(2h){\cal N}_lN_{-2h},
\label{AppRadialCoefficients}
\eea
where ${\cal N}_l$ is a common factor not fixed by the helicity equation. The two quantities $N_+$ and $N_-$ therefore determine the relative weight of the two components, while ${\cal N}_l$ fixes the overall normalization of the transverse mode.

The upper Pauli spinor is thus
\bea
\varphi_{\alpha}(\rho,\phi)=\frac{{\cal N}_l}{\sqrt{2}}\begin{pmatrix}N_{2h}J_l\left(|\boldsymbol{p}_\perp|\rho\right)e^{\ii l\phi}\\2\ii hN_{-2h}J_{l+1}\left(|\boldsymbol{p}_\perp|\rho\right)e^{\ii(l+1)\phi}\end{pmatrix}.
\label{AppNormalizedPauliSpinor}
\eea
The factor $1/\sqrt{2}$ is convenient because
\bea
N_+^2+N_-^2=2.
\label{AppNpmSum}
\eea
so that the momentum-dependent coefficients have a simple normalization. The remaining factor ${\cal N}_l$ must be fixed by the normalization convention for the complete cylindrical modes.

The positive-energy bispinor is now fully determined. We combine the
decomposition in Eq.~\eqref{AppPositiveBispinorDecomposition} with the
relations in Eqs.~\eqref{AppPositiveLowerSpinor} and
\eqref{AppNormalizedPauliSpinor} to obtain
\bea
u_{\alpha}(x)=e^{\ii p_zz}\begin{pmatrix}\varphi_{\alpha}(\rho,\phi)\\\dfrac{2h|\boldsymbol{p}|}{E+m_f}\varphi_{\alpha}(\rho,\phi)\end{pmatrix}.
\label{AppPositiveBispinor}
\eea
Using
\bea
\frac{|\boldsymbol{p}|}{E+m_f}=\sqrt{\frac{E-m_f}{E+m_f}},
\label{AppEnergyRatio}
\eea
the lower component may equivalently be written as
\bea
\chi_{\alpha}(\rho,\phi)=2h\sqrt{\frac{E-m_f}{E+m_f}}\varphi_{\alpha}(\rho,\phi).
\label{AppPositiveLowerComponentRatio}
\eea

The negative-energy branch is constructed from the negative eigenvalues of the free Hamiltonian. We write
\bea
v_{\alpha}(x)=\begin{pmatrix}\varphi_{\alpha}^{(v)}(x)\\\chi_{\alpha}^{(v)}(x)\end{pmatrix},
\label{AppNegativeBispinorDecomposition}
\eea
with
\bea
\hat H_0v_{\alpha}=-Ev_{\alpha},\qquad E>0.
\label{AppNegativeEnergyEquation}
\eea
These solutions are not interpreted as physical states with negative energy. They are the negative-energy solutions of the Dirac equation that, upon quantization, enter the field expansion through the antiparticle creation operators. The corresponding coupled equations are
\bea
m_f\varphi_{\alpha}^{(v)}+\left(\boldsymbol{\sigma}\cdot\hat{\boldsymbol{p}}\right)\chi_{\alpha}^{(v)}=-E\varphi_{\alpha}^{(v)},
\label{AppNegativeUpperEquation}
\eea
and
\bea
\left(\boldsymbol{\sigma}\cdot\hat{\boldsymbol{p}}\right)\varphi_{\alpha}^{(v)}-m_f\chi_{\alpha}^{(v)}=-E\chi_{\alpha}^{(v)}.
\label{AppNegativeLowerEquation}
\eea
The second equation gives
\bea
\chi_{\alpha}^{(v)}=-\frac{\left(\boldsymbol{\sigma}\cdot\hat{\boldsymbol{p}}\right)}{E-m_f}\varphi_{\alpha}^{(v)}.
\label{AppNegativeLowerSpinor}
\eea
The upper Pauli spinor can be chosen as an eigenstate of the helicity operator,
\bea
\left(\boldsymbol{\sigma}\cdot\hat{\boldsymbol{p}}\right)\varphi_{\alpha}^{(v)}=2hp\,\varphi_{\alpha}^{(v)},\qquad h=\pm\frac12.
\label{AppNegativePauliHelicityEquation}
\eea
and therefore
\bea
\chi_{\alpha}^{(v)}=-\frac{2hp}{E-m_f}\varphi_{\alpha}^{(v)}.
\label{AppNegativeLowerHelicityRelation}
\eea
Here $h$ labels the helicity eigenvalue of the negative-energy Dirac solution. Its identification with the physical helicity of the corresponding antiparticle requires the usual reinterpretation of the negative-energy modes in the quantized field expansion and should not be made at this stage.

The angular dependence of $\varphi_{\alpha}^{(v)}$ is fixed by the eigenvalue equation for the two-component total angular momentum operator. We therefore write
\bea
\hat J_z^{(2)}\varphi_{\alpha}^{(v)}(\rho,\phi)=j_v\varphi_{\alpha}^{(v)}(\rho,\phi).
\label{AppNegativePauliJzEigenvalue}
\eea
Following the same angular-momentum analysis used for the positive-energy branch, we write
\bea
\varphi_{\alpha}^{(v)}(\rho,\phi)=\begin{pmatrix}f_l^{(v)}(\rho)e^{\ii l\phi}\\g_l^{(v)}(\rho)e^{\ii(l+1)\phi}\end{pmatrix}.
\label{AppNegativeAngularPauliSpinor}
\eea
so that
\bea
j_v=l+\frac12.
\label{AppNegativeTotalAngularMomentum}
\eea
The radial functions are obtained by substituting this angular structure into the helicity equation. Eliminating $g_l^{(v)}(\rho)$ and $f_l^{(v)}(\rho)$, respectively, gives
\bea
\left[\partial_\rho^2+\frac{1}{\rho}\partial_\rho-\frac{l^2}{\rho^2}+|\boldsymbol{p}_\perp|^2\right]f_l^{(v)}(\rho)=0.
\label{AppNegativeBesselEquationUpper}
\eea
and
\bea
\left[\partial_\rho^2+\frac{1}{\rho}\partial_\rho-\frac{(l+1)^2}{\rho^2}+|\boldsymbol{p}_\perp|^2\right]g_l^{(v)}(\rho)=0.
\label{AppNegativeBesselEquationLower}
\eea
The solutions regular at the origin are therefore
\bea
f_l^{(v)}(\rho)=A_l^{(v)}J_l\left(|\boldsymbol{p}_\perp|\rho\right).
\label{AppNegativeUpperRadialSolution}
\eea
and
\bea
g_l^{(v)}(\rho)=B_l^{(v)}J_{l+1}\left(|\boldsymbol{p}_\perp|\rho\right).
\label{AppNegativeLowerRadialSolution}
\eea
The second-order equations determine the radial dependence but do not fix the relative coefficient between the two components. This coefficient is obtained from the original first-order helicity equation. Using the corresponding Bessel identity, one obtains
\bea
\left(p_z-2hp\right)A_l^{(v)}-\ii|\boldsymbol{p}_\perp|B_l^{(v)}=0.
\label{AppNegativeRadialCoefficientEquation}
\eea
and hence
\bea
\frac{B_l^{(v)}}{A_l^{(v)}}=-\ii\frac{p_z-2hp}{|\boldsymbol{p}_\perp|}.
\label{AppNegativeRadialCoefficientRatio}
\eea
We use the kinematical factors defined for the positive-energy branch:
\bea
N_\pm=\sqrt{1\pm\frac{p_z}{|\boldsymbol{p}|}},
\label{AppNegativeNpmDefinition}
\eea
the relative coefficients can be written as
\bea
A_l^{(v)}={\cal N}_l^{(v)}N_{2h},\qquad B_l^{(v)}=\ii(2h){\cal N}_l^{(v)}N_{-2h},
\label{AppNegativeRadialCoefficients}
\eea
where ${\cal N}_l^{(v)}$ is an overall normalization factor. The upper Pauli spinor therefore takes the form
\bea
\varphi_{\alpha}^{(v)}(\rho,\phi)=\frac{{\cal N}_l^{(v)}}{\sqrt{2}}\begin{pmatrix}N_{2h}J_l\left(|\boldsymbol{p}_\perp|\rho\right)e^{\ii l\phi}\\2\ii hN_{-2h}J_{l+1}\left(|\boldsymbol{p}_\perp|\rho\right)e^{\ii(l+1)\phi}\end{pmatrix}.
\label{AppNegativeNormalizedPauliSpinor}
\eea
The lower Pauli spinor is then obtained from Eq.~\eqref{AppNegativeLowerHelicityRelation}. Thus, the complete negative-energy bispinor is determined up to the overall normalization ${\cal N}_l^{(v)}$.

The normalization constants ${\cal N}_l$ and ${\cal N}_l^{(v)}$ are fixed by the normalization convention adopted for the complete four-component modes. The radial part is normalized using the orthogonality properties of the Bessel functions, while the longitudinal and angular dependences are normalized according to the continuous longitudinal momentum and the discrete angular-momentum quantum number. The explicit values of the normalization constants will therefore be fixed once the normalization convention for the fermion mode expansion is specified.

The rotating-frame time dependence is introduced only after the eigenfunctions of $\hat H_0$ have been constructed. The rotating Hamiltonian is
\bea
\hat H_\Omega=\hat H_0-\Omega\hat J_z.
\label{AppRotatingHamiltonianFinal}
\eea
For the positive-energy modes,
\bea
\hat H_0u_{\alpha}=E u_{\alpha},\qquad \hat J_zu_{\alpha}=j u_{\alpha},
\label{AppPositiveEigenvalues}
\eea
with
\bea
j=l+\frac12.
\label{AppPositiveAngularMomentum}
\eea
Then,
\bea
\hat H_\Omega u_{\alpha}=\left(E-\Omega j\right)u_{\alpha}.
\label{AppPositiveRotatingEigenvalue}
\eea
We define the corresponding rotating-frame energy by
\bea
\widetilde E_j=E-\Omega\left(l+\frac12\right).
\label{AppRotatingEnergy}
\eea
The positive-energy rotating mode is therefore
\bea
U_{\alpha}(x)=\frac{1}{2\pi}e^{-\ii\widetilde E_jt}u_{\alpha}(x).
\label{AppPositiveRotatingMode}
\eea
For the negative-energy modes constructed above,
\bea
\hat H_0v_{\alpha}=-Ev_{\alpha},\qquad \hat J_zv_{\alpha}=j_vv_{\alpha},
\label{AppNegativeEigenvalues}
\eea
with
\bea
j_v=l+\frac12.
\label{AppNegativeAngularMomentum}
\eea
It follows that
\bea
\hat H_\Omega v_{\alpha}=-\left(E+\Omega j_v\right)v_{\alpha}.
\label{AppNegativeRotatingEigenvalue}
\eea
The corresponding negative-energy rotating mode is therefore
\bea
V_{\alpha}(x)=\frac{1}{2\pi}e^{+\ii\left(E+\Omega j_v\right)t}v_{\alpha}(x).
\label{AppNegativeRotatingMode}
\eea
The adjoint modes are defined by
\bea
\overline{U}_{\alpha}(x)=U_{\alpha}^{\dagger}(x)\gamma^0,\qquad\overline{V}_{\alpha}(x)=V_{\alpha}^{\dagger}(x)\gamma^0.
\label{AppModeAdjoints}
\eea

The solutions constructed above form a complete basis of positive- and negative-energy solutions of the free Dirac equation. Upon quantization, the coefficients of the positive- and negative-energy modes become fermion annihilation and creation operators, respectively. The continuous quantum numbers $p_z$ and $|\boldsymbol{p}_\perp|$ are therefore integrated over, while the discrete angular-momentum and helicity quantum numbers are summed. Defining the measure
\bea
\int[d\alpha]\equiv\sum_{l=-\infty}^{\infty}\sum_{h=\pm1/2}\int_{-\infty}^{\infty}dp_z\int_0^\infty |\boldsymbol{p}_\perp|\,d|\boldsymbol{p}_\perp|,
\label{AppModeMeasure}
\eea
the field expansion can be written as
\bea
\psi(x)=\int[d\alpha]\left[\hat b_{\alpha}U_{\alpha}(x)+\hat d_{\alpha}^{\dagger}V_{\alpha}(x)\right].
\label{AppFermionFieldExpansionExplicit}
\eea
Here, $b_{\alpha}$ and $d_{\alpha}^{\dagger}$ are the fermion annihilation and creation operators associated with the positive- and negative-energy modes, respectively, with canonical fermion anticommutation relations given by
\bea
\left\{b_{\alpha},b_{\alpha' }^\dagger\right\}=\delta_{ll'}\delta_{hh'}\delta(p_z-p_z')\delta\left(|\boldsymbol{p}_\perp|-|\boldsymbol{p}_\perp'|\right),
\label{AppParticleAnticommutator}
\eea
and
\bea
\left\{d_{\alpha},d_{\alpha' }^\dagger\right\}=\delta_{ll'}\delta_{hh'}\delta(p_z-p_z')\delta\left(|\boldsymbol{p}_\perp|-|\boldsymbol{p}_\perp'|\right),
\label{AppAntiparticleAnticommutator}
\eea
with all other anticommutators vanishing,
\bea
\left\{b_{\alpha},b_{\alpha'}\right\}
&=&\left\{d_{\alpha},d_{\alpha'}\right\}
=\left\{b_{\alpha},d_{\alpha'}\right\}\nn\\
&=&\left\{b_{\alpha},d_{\alpha'}^\dagger\right\}=0.
\label{AppVanishingAnticommutators}
\eea

\onecolumngrid

\subsection{Normalization of the rotating fermion modes}
\label{app:FermionModeNormalization}

The normalization of the rotating fermion modes is fixed by the
momentum-space measure in Eq.~\eqref{AppModeMeasure} and the canonical
anticommutation relations in Eqs.~\eqref{AppParticleAnticommutator}--\eqref{AppVanishingAnticommutators}.
For the positive-energy modes, we require the orthonormality condition
\bea
\int d^3x\,U_{\alpha'}^\dagger(x)U_{\alpha}(x)=\delta_{ll'}\delta_{hh'}\delta(p_z-p_z')\frac{\delta\left(|\boldsymbol p_\perp|-|\boldsymbol p_\perp'|\right)}{|\boldsymbol p_\perp|}.
\label{AppModeNormalization}
\eea

Using Eq.~\eqref{FermionPositiveRotatingMode}, the spatial overlap can
be written as
\bea
\int d^3x\,U_{\alpha'}^\dagger(x)U_{\alpha}(x)=\frac{1}{(2\pi)^2}\int d^3x\,u_{\alpha'}^\dagger(x)u_{\alpha}(x).
\label{AppReducedModeNormalization}
\eea

The bispinors in Eq.~\eqref{FermionPositiveBispinor} give
\bea
u_{\alpha'}^\dagger(x)u_{\alpha}(x)=e^{\ii(p_z-p_z')z}\left[\varphi_{\alpha'}^\dagger(\rho,\phi)\varphi_{\alpha}(\rho,\phi)+\chi_{\alpha'}^\dagger(\rho,\phi)\chi_{\alpha}(\rho,\phi)\right].
\label{AppBispinorNormalizationOverlap}
\eea

Using Eq.~\eqref{FermionLowerPauliSpinor}, the lower-component
contribution becomes
\bea
\chi_{\alpha'}^\dagger\chi_{\alpha}=\frac{4h'h|\boldsymbol p'||\boldsymbol p|}{(E'+m_f)(E+m_f)}\varphi_{\alpha'}^\dagger\varphi_{\alpha}.
\label{AppLowerComponentNormalization}
\eea

For equal quantum numbers selected by the orthogonality condition,
$|\boldsymbol p'|=|\boldsymbol p|$ and $E'=E$, so that
\bea
u_{\alpha'}^\dagger(x)u_{\alpha}(x)=e^{\ii(p_z-p_z')z}\left(1+c_{\alpha}^2\right)\varphi_{\alpha'}^\dagger(\rho,\phi)\varphi_{\alpha}(\rho,\phi),
\label{AppBispinorNormalizationEqualMomentum}
\eea

where
\bea
c_{\alpha}=\frac{2h|\boldsymbol p|}{E+m_f}.
\label{AppNormalizationCoefficient}
\eea

The transverse Pauli-spinor overlap follows directly from
Eq.~\eqref{FermionUpperPauliSpinor},
\bea
\varphi_{\alpha'}^\dagger\varphi_{\alpha}=\frac{{\cal N}_{l'}{\cal N}_l}{2}e^{\ii(l-l')\phi}\left[N_{2h'}N_{2h}J_{l'}\left(|\boldsymbol p_\perp'|\rho\right)J_l\left(|\boldsymbol p_\perp|\rho\right)+4h'hN_{-2h'}N_{-2h}J_{l'+1}\left(|\boldsymbol p_\perp'|\rho\right)J_{l+1}\left(|\boldsymbol p_\perp|\rho\right)\right].
\label{AppPauliNormalizationOverlap}
\eea

The angular dependence gives
\bea
\int_0^{2\pi}d\phi\,e^{\ii(l-l')\phi}=2\pi\delta_{ll'}.
\label{AppAngularNormalization}
\eea

For $l'=l$, the two radial integrals are evaluated using the
orthogonality relation for Bessel functions,
\bea
\int_0^\infty\rho\,d\rho\,J_l\left(|\boldsymbol p_\perp'|\rho\right)J_l\left(|\boldsymbol p_\perp|\rho\right)=\frac{\delta\left(|\boldsymbol p_\perp'|-|\boldsymbol p_\perp|\right)}{|\boldsymbol p_\perp|},
\label{AppBesselNormalizationL}
\eea

and
\bea
\int_0^\infty\rho\,d\rho\,J_{l+1}\left(|\boldsymbol p_\perp'|\rho\right)J_{l+1}\left(|\boldsymbol p_\perp|\rho\right)=\frac{\delta\left(|\boldsymbol p_\perp'|-|\boldsymbol p_\perp|\right)}{|\boldsymbol p_\perp|}.
\label{AppBesselNormalizationLPlusOne}
\eea

The normalization factors defined in Eq.~\eqref{FermionNpmDefinition}
satisfy
\bea
N_{2h}^2+N_{-2h}^2=2.
\label{AppNpmNormalizationIdentity}
\eea

Then, the transverse overlap becomes
\bea
\int d^2x_\perp\,\varphi_{\alpha'}^\dagger\varphi_{\alpha}=2\pi{\cal N}_{l'}{\cal N}_l\delta_{ll'}\delta_{hh'}\frac{\delta\left(|\boldsymbol p_\perp'|-|\boldsymbol p_\perp|\right)}{|\boldsymbol p_\perp|}.
\label{AppTransverseNormalization}
\eea

The factor $\delta_{hh'}$ follows from the orthogonality of the helicity
components. In particular, for $h'=-h$, the two terms in
Eq.~\eqref{AppPauliNormalizationOverlap} have identical radial
structures and opposite coefficients, since $4h'h=-1$, and therefore
cancel.

The longitudinal integral is
\bea
\int_{-\infty}^{\infty}dz\,e^{\ii(p_z-p_z')z}=2\pi\delta(p_z-p_z').
\label{AppLongitudinalNormalization}
\eea

Combining the transverse and longitudinal contributions, one obtains
\bea
\int d^3x\,u_{\alpha'}^\dagger(x)u_{\alpha}(x)=4\pi^2{\cal N}_{l'}{\cal N}_l\left(1+c_{\alpha}^2\right)\delta_{ll'}\delta_{hh'}\delta(p_z-p_z')\frac{\delta\left(|\boldsymbol p_\perp'|-|\boldsymbol p_\perp|\right)}{|\boldsymbol p_\perp|}.
\label{AppCompleteReducedNormalization}
\eea

Substituting Eq.~\eqref{AppCompleteReducedNormalization} into
Eq.~\eqref{AppReducedModeNormalization}, the factors $4\pi^2$
cancel those generated by the prefactor in the normalized modes,
yielding
\bea
\int d^3x\,U_{\alpha'}^\dagger(x)U_{\alpha}(x)={\cal N}_{l'}{\cal N}_l\left(1+c_{\alpha}^2\right)\delta_{ll'}\delta_{hh'}\delta(p_z-p_z')\frac{\delta\left(|\boldsymbol p_\perp'|-|\boldsymbol p_\perp|\right)}{|\boldsymbol p_\perp|}.
\label{AppNormalizedModeOverlap}
\eea

The kinematic factor associated with the lower Pauli spinor can be
simplified using $h^2=1/4$ and the dispersion relation
$E^2=|\boldsymbol p|^2+m_f^2$. One finds
\bea
1+c_{\alpha}^2=1+\frac{|\boldsymbol p|^2}{(E+m_f)^2}=\frac{2E}{E+m_f}.
\label{AppNormalizationKinematicFactor}
\eea

The normalization condition in Eq.~\eqref{AppModeNormalization}
therefore requires
\bea
{\cal N}_{l'}{\cal N}_l\frac{2E}{E+m_f}=1.
\label{AppNormalizationConstraint}
\eea

Choosing the same normalization convention for all positive-energy
modes gives
\bea
{\cal N}_l=\sqrt{\frac{E+m_f}{2E}}.
\label{AppNormalizationFactor}
\eea

For the final state, the corresponding factor is
\bea
{\cal N}_{l'}=\sqrt{\frac{E'+m_f}{2E'}},
\label{AppNormalizationFactorFinal}
\eea

where
\bea
E'=\sqrt{|\boldsymbol p_\perp'|^2+p_z'^2+m_f^2}.
\label{AppNormalizationEnergyFinal}
\eea

The negative-energy modes are normalized analogously, with the same
momentum-space convention and the corresponding negative-energy
spinors defined in Eqs.~\eqref{AppNegativeRotatingMode}--\eqref{AppModeAdjoints}.

\twocolumngrid
\bibliography{BibVertexRotation}

@article{CastanoYepes2026FiniteSize,
  title         = {Finite-Size Effects on Symmetry Restoration in a Rotating Scalar System with Spatially Dependent Thermal Self-Energy},
  author        = {Casta{\~n}o-Yepes, Jorge David and Cer{\'o}n-Su{\'a}rez, Larry A.},
  journal       = {arXiv},
  year          = {2026},
  eprint        = {2609.34038},
  archivePrefix = {arXiv},
  primaryClass  = {hep-ph},
  version       = {v1}
}

@misc{Ayala:2026eja,
  author = {Ayala, Alejandro and Mu{\~n}oz, Enrique and Loewe, Marcelo and Rojas, Juan Cristobal and Scoccola, Norberto},
  title = {\mbox{QED} Vertex and Anomalous Magnetic Moment in the Presence of a Magnetic Field},
  year = {2026},
  eprint = {2607.10015},
  archivePrefix = {arXiv},
  primaryClass = {hep-ph}
}

@article{STAR:2017ckg,
    author = "Adamczyk, L. and others",
    collaboration = "STAR",
    title = "{Global $\Lambda$ hyperon polarization in nuclear collisions}",
    eprint = "1701.06657",
    archivePrefix = "arXiv",
    primaryClass = "nucl-ex",
    doi = "10.1038/nature23004",
    journal = "Nature",
    volume = "548",
    pages = "62--65",
    year = "2017"
}

@article{Liang:2004ph,
    author = "Liang, Zuo-Tang and Wang, Xin-Nian",
    title = "{Globally polarized quark-gluon plasma in non-central A+A collisions}",
    eprint = "nucl-th/0410079",
    archivePrefix = "arXiv",
    reportNumber = "LBNL-56383",
    doi = "10.1103/PhysRevLett.94.102301",
    journal = "Phys. Rev. Lett.",
    volume = "94",
    pages = "102301",
    year = "2005",
    note = "[Erratum: Phys.Rev.Lett. 96, 039901 (2006)]"
}

@article{Betz:2007kg,
    author = "Betz, Barbara and Gyulassy, Miklos and Torrieri, Giorgio",
    title = "{Polarization probes of vorticity in heavy ion collisions}",
    eprint = "0708.0035",
    archivePrefix = "arXiv",
    primaryClass = "nucl-th",
    doi = "10.1103/PhysRevC.76.044901",
    journal = "Phys. Rev. C",
    volume = "76",
    pages = "044901",
    year = "2007"
}

@article{Gao:2007bc,
    author = "Gao, Jian-Hua and Chen, Shou-Wan and Deng, Wei-tian and Liang, Zuo-Tang and Wang, Qun and Wang, Xin-Nian",
    title = "{Global quark polarization in non-central A+A collisions}",
    eprint = "0710.2943",
    archivePrefix = "arXiv",
    primaryClass = "nucl-th",
    reportNumber = "LBNL-63515",
    doi = "10.1103/PhysRevC.77.044902",
    journal = "Phys. Rev. C",
    volume = "77",
    pages = "044902",
    year = "2008"
}

@article{Becattini:2007sr,
    author = "Becattini, F. and Piccinini, F. and Rizzo, J.",
    title = "{Angular momentum conservation in heavy ion collisions at very high energy}",
    eprint = "0711.1253",
    archivePrefix = "arXiv",
    primaryClass = "nucl-th",
    doi = "10.1103/PhysRevC.77.024906",
    journal = "Phys. Rev. C",
    volume = "77",
    pages = "024906",
    year = "2008"
}

@article{Becattini:2013fla,
    author = "Becattini, F. and Chandra, V. and Del Zanna, L. and Grossi, E.",
    title = "{Relativistic distribution function for particles with spin at local thermodynamical equilibrium}",
    eprint = "1303.3431",
    archivePrefix = "arXiv",
    primaryClass = "nucl-th",
    doi = "10.1016/j.aop.2013.07.004",
    journal = "Annals Phys.",
    volume = "338",
    pages = "32--49",
    year = "2013"
}

@article{Becattini:2015ska,
    author = "Becattini, F. and Inghirami, G. and Rolando, V. and Beraudo, A. and Del Zanna, L. and De Pace, A. and Nardi, M. and Pagliara, G. and Chandra, V.",
    title = "{A study of vorticity formation in high energy nuclear collisions}",
    eprint = "1501.04468",
    archivePrefix = "arXiv",
    primaryClass = "nucl-th",
    doi = "10.1140/epjc/s10052-015-3624-1",
    journal = "Eur. Phys. J. C",
    volume = "75",
    number = "9",
    pages = "406",
    year = "2015",
    note = "[Erratum: Eur.Phys.J.C 78, 354 (2018)]"
}

@article{Becattini:2016gvu,
    author = "Becattini, F. and Karpenko, I. and Lisa, M. and Upsal, I. and Voloshin, S.",
    title = "{Global hyperon polarization at local thermodynamic equilibrium with vorticity, magnetic field and feed-down}",
    eprint = "1610.02506",
    archivePrefix = "arXiv",
    primaryClass = "nucl-th",
    doi = "10.1103/PhysRevC.95.054902",
    journal = "Phys. Rev. C",
    volume = "95",
    number = "5",
    pages = "054902",
    year = "2017"
}

@article{Florkowski:2018ahw,
    author = "Florkowski, Wojciech and Kumar, Avdhesh and Ryblewski, Radoslaw",
    title = "{Thermodynamic versus kinetic approach to polarization-vorticity coupling}",
    eprint = "1806.02616",
    archivePrefix = "arXiv",
    primaryClass = "hep-ph",
    doi = "10.1103/PhysRevC.98.044906",
    journal = "Phys. Rev. C",
    volume = "98",
    number = "4",
    pages = "044906",
    year = "2018"
}

@article{Pang:2016igs,
    author = "Pang, Long-Gang and Petersen, Hannah and Wang, Qun and Wang, Xin-Nian",
    title = "{Vortical Fluid and $\Lambda$ Spin Correlations in High-Energy Heavy-Ion Collisions}",
    eprint = "1605.04024",
    archivePrefix = "arXiv",
    primaryClass = "hep-ph",
    doi = "10.1103/PhysRevLett.117.192301",
    journal = "Phys. Rev. Lett.",
    volume = "117",
    number = "19",
    pages = "192301",
    year = "2016"
}

@article{Karpenko:2016jyx,
    author = "Karpenko, I. and Becattini, F.",
    title = "{Study of $\Lambda$ polarization in relativistic nuclear collisions at $\sqrt{s_\mathrm{NN}}=7.7$--200 GeV}",
    eprint = "1610.04717",
    archivePrefix = "arXiv",
    primaryClass = "nucl-th",
    doi = "10.1140/epjc/s10052-017-4765-1",
    journal = "Eur. Phys. J. C",
    volume = "77",
    number = "4",
    pages = "213",
    year = "2017"
}

@article{Sun:2017xhx,
    author = "Sun, Yifeng and Ko, Che Ming",
    title = "{$\Lambda$ hyperon polarization in relativistic heavy ion collisions from a chiral kinetic approach}",
    eprint = "1706.09467",
    archivePrefix = "arXiv",
    primaryClass = "nucl-th",
    doi = "10.1103/PhysRevC.96.024906",
    journal = "Phys. Rev. C",
    volume = "96",
    number = "2",
    pages = "024906",
    year = "2017"
}

@article{Li:2017slc,
    author = "Li, Hui and Pang, Long-Gang and Wang, Qun and Xia, Xiao-Liang",
    title = "{Global $\Lambda$ polarization in heavy-ion collisions from a transport model}",
    eprint = "1704.01507",
    archivePrefix = "arXiv",
    primaryClass = "nucl-th",
    doi = "10.1103/PhysRevC.96.054908",
    journal = "Phys. Rev. C",
    volume = "96",
    number = "5",
    pages = "054908",
    year = "2017"
}

@article{Xia:2018tes,
    author = "Xia, Xiao-Liang and Li, Hui and Tang, Ze-Bo and Wang, Qun",
    title = "{Probing vorticity structure in heavy-ion collisions by local $\Lambda$ polarization}",
    eprint = "1803.00867",
    archivePrefix = "arXiv",
    primaryClass = "nucl-th",
    doi = "10.1103/PhysRevC.98.024905",
    journal = "Phys. Rev. C",
    volume = "98",
    pages = "024905",
    year = "2018"
}

@article{Wei:2018zfb,
    author = "Wei, De-Xian and Deng, Wei-Tian and Huang, Xu-Guang",
    title = "{Thermal vorticity and spin polarization in heavy-ion collisions}",
    eprint = "1810.00151",
    archivePrefix = "arXiv",
    primaryClass = "nucl-th",
    doi = "10.1103/PhysRevC.99.014905",
    journal = "Phys. Rev. C",
    volume = "99",
    number = "1",
    pages = "014905",
    year = "2019"
}

@article{Fu:2020oxj,
    author = "Fu, Baochi and Xu, Kai and Huang, Xu-Guang and Song, Huichao",
    title = "{Hydrodynamic study of hyperon spin polarization in relativistic heavy ion collisions}",
    eprint = "2011.03740",
    archivePrefix = "arXiv",
    primaryClass = "nucl-th",
    doi = "10.1103/PhysRevC.103.024903",
    journal = "Phys. Rev. C",
    volume = "103",
    number = "2",
    pages = "024903",
    year = "2021"
}

@article{Liu:2021uhn,
    author = "Liu, Shuai Y. F. and Yin, Yi",
    title = "{Spin polarization induced by the hydrodynamic gradients}",
    eprint = "2103.09200",
    archivePrefix = "arXiv",
    primaryClass = "hep-ph",
    doi = "10.1007/JHEP07(2021)188",
    journal = "JHEP",
    volume = "2021",
    number = "07",
    pages = "188",
    year = "2021"
}

@article{Ayala:2021xrn,
    author = "Ayala, Alejandro and Dom{\'\i}nguez, Isabel and Maldonado, Ivonne and Tejeda-Yeomans, Mar{\'\i}a Elena",
    title = "{Rise and fall of $\Lambda$ and $\bar{\Lambda}$ global polarization in semi-central heavy-ion collisions at HADES, NICA and RHIC energies from the core-corona model}",
    eprint = "2106.14379",
    archivePrefix = "arXiv",
    primaryClass = "hep-ph",
    doi = "10.1103/PhysRevC.105.034907",
    journal = "Phys. Rev. C",
    volume = "105",
    number = "3",
    pages = "034907",
    year = "2022"
}

@article{Ayala:2020soy,
    author = "Ayala, Alejandro and others",
    title = "{Core meets corona: A two-component source to explain $\Lambda$ and $\bar{\Lambda}$ global polarization in semi-central heavy-ion collisions}",
    eprint = "2003.13757",
    archivePrefix = "arXiv",
    primaryClass = "hep-ph",
    doi = "10.1016/j.physletb.2020.135818",
    journal = "Phys. Lett. B",
    volume = "810",
    pages = "135818",
    year = "2020"
}

@article{Ayala:2023xyn,
    author = "Ayala, Alejandro and Dominguez, Isabel and Maldonado, Ivonne and Tejeda-Yeomans, Maria Elena",
    title = "{An Improved Core-Corona Model for $\Lambda$ and $\bar{\Lambda}$ Polarization in Relativistic Heavy-Ion Collisions}",
    doi = "10.3390/particles6010022",
    journal = "Particles",
    volume = "6",
    number = "1",
    pages = "405--415",
    year = "2023"
}

@article{Ayala:2026wro,
    author = "Ayala, Alejandro and Medina-Serna, Jos{\'e} Jorge and Dom{\'\i}nguez, Isabel and Tejeda-Yeomans, Mar{\'\i}a Elena",
    title = "{Excitation function for global \(\Lambda\) polarization in relativistic heavy-ion collisions with the core-corona model}",
    doi = "10.1103/fbt2-9gn1",
    journal = "Phys. Rev. D",
    volume = "114",
    number = "1",
    pages = "016034",
    year = "2026"
}

@article{Becattini:2020ngo,
    author = "Becattini, Francesco and Lisa, Michael A.",
    title = "{Polarization and Vorticity in the Quark--Gluon Plasma}",
    eprint = "2003.03640",
    archivePrefix = "arXiv",
    primaryClass = "nucl-ex",
    doi = "10.1146/annurev-nucl-021920-095245",
    journal = "Ann. Rev. Nucl. Part. Sci.",
    volume = "70",
    pages = "395--423",
    year = "2020"
}

@article{STAR:2018gyt,
    author = "Adam, Jaroslav and others",
    collaboration = "STAR",
    title = "{Global polarization of $\Lambda$ hyperons in Au+Au collisions at $\sqrt{s_{_{NN}}}$ = 200 GeV}",
    eprint = "1805.04400",
    archivePrefix = "arXiv",
    primaryClass = "nucl-ex",
    doi = "10.1103/PhysRevC.98.014910",
    journal = "Phys. Rev. C",
    volume = "98",
    pages = "014910",
    year = "2018"
}

@article{STAR:2021beb,
    author = "Abdallah, M. S. and others",
    collaboration = "STAR",
    title = "{Global $\Lambda$-hyperon polarization in Au+Au collisions at $\sqrt{s_{NN}}=3$ GeV}",
    eprint = "2108.00044",
    archivePrefix = "arXiv",
    primaryClass = "nucl-ex",
    doi = "10.1103/PhysRevC.104.L061901",
    journal = "Phys. Rev. C",
    volume = "104",
    number = "6",
    pages = "L061901",
    year = "2021"
}

@article{STAR:2023nvo,
    author = "Abdulhamid, M. I. and others",
    collaboration = "STAR",
    title = "{Global polarization of $\Lambda$ and $\bar{\Lambda}$ hyperons in Au+Au collisions at $\sqrt{s_{NN}}=19.6$ and 27 GeV}",
    eprint = "2305.08705",
    archivePrefix = "arXiv",
    primaryClass = "nucl-ex",
    doi = "10.1103/PhysRevC.108.014910",
    journal = "Phys. Rev. C",
    volume = "108",
    number = "1",
    pages = "014910",
    year = "2023"
}

@article{ALICE:2019onw,
    author = "Acharya, Shreyasi and others",
    collaboration = "ALICE",
    title = "{Global polarization of $\Lambda$ and $\bar{\Lambda}$ hyperons in Pb-Pb collisions at $\sqrt{s_{NN}}=2.76$ and 5.02 TeV}",
    eprint = "1909.01281",
    archivePrefix = "arXiv",
    primaryClass = "nucl-ex",
    reportNumber = "CERN-EP-2019-173",
    doi = "10.1103/PhysRevC.101.044611",
    journal = "Phys. Rev. C",
    volume = "101",
    number = "4",
    pages = "044611",
    year = "2020",
    note = "[Erratum: Phys.Rev.C 105, 029902 (2022)]"
}

@article{HADES:2022enx,
    author = "Abou Yassine, R. and others",
    collaboration = "HADES",
    title = "{Measurement of global polarization of $\Lambda$ hyperons in few-GeV heavy-ion collisions}",
    eprint = "2207.05160",
    archivePrefix = "arXiv",
    primaryClass = "nucl-ex",
    doi = "10.1016/j.physletb.2022.137506",
    journal = "Phys. Lett. B",
    volume = "835",
    pages = "137506",
    year = "2022"
}

@article{STAR:2020xbm,
    author = "Adam, J. and others",
    collaboration = "STAR",
    title = "{Global Polarization of $\Xi$ and $\Omega$ Hyperons in Au+Au Collisions at $\sqrt{s_{NN}}=200$ GeV}",
    eprint = "2012.13601",
    archivePrefix = "arXiv",
    primaryClass = "nucl-ex",
    doi = "10.1103/PhysRevLett.126.162301",
    journal = "Phys. Rev. Lett.",
    volume = "126",
    number = "16",
    pages = "162301",
    year = "2021",
    note = "[Erratum: Phys.Rev.Lett. 131, 089901 (2023)]"
}

@article{STAR:2019erd,
    author = "Adam, Jaroslav and others",
    collaboration = "STAR",
    title = "{Polarization of $\Lambda$ ($\bar{\Lambda}$) hyperons along the beam direction in Au+Au collisions at $\sqrt{s_{_{NN}}}=200$ GeV}",
    eprint = "1905.11917",
    archivePrefix = "arXiv",
    primaryClass = "nucl-ex",
    doi = "10.1103/PhysRevLett.123.132301",
    journal = "Phys. Rev. Lett.",
    volume = "123",
    number = "13",
    pages = "132301",
    year = "2019"
}

@article{STAR:2023eck,
    author = "Abdulhamid, Muhammad and others",
    collaboration = "STAR",
    title = "{Hyperon Polarization along the Beam Direction Relative to the Second and Third Harmonic Event Planes in Isobar Collisions at $\sqrt{s_{NN}}=200$ GeV}",
    eprint = "2303.09074",
    archivePrefix = "arXiv",
    primaryClass = "nucl-ex",
    doi = "10.1103/PhysRevLett.131.202301",
    journal = "Phys. Rev. Lett.",
    volume = "131",
    number = "20",
    pages = "202301",
    year = "2023"
}

@article{Sarkar:2022uyt,
    author = "Sarkar, Debojit",
    collaboration = "ALICE",
    title = "{Global and local polarization of $\Lambda$ ($\bar{\Lambda}$) hyperons in Pb-Pb collisions in ALICE at the LHC}",
    eprint = "2209.04798",
    archivePrefix = "arXiv",
    primaryClass = "hep-ex",
    doi = "10.1051/epjconf/202225906001",
    journal = "EPJ Web Conf.",
    volume = "259",
    pages = "06001",
    year = "2022"
}

@misc{ALICE:2026eje,
    author = "Ali Hassan Abdallah, Dana and others",
    collaboration = "ALICE",
    title = "{Hyperon ($\Lambda$) polarization along the beam axis in Pb-Pb collisions at $\sqrt{s_{\rm NN}}=5.36$ TeV}",
    eprint = "2606.18070",
    archivePrefix = "arXiv",
    primaryClass = "nucl-ex",
    reportNumber = "CERN-EP-2026-173",
    month = "6",
    year = "2026"
}

@article{STAR:2022fan,
    author = "Abdallah, M. S. and others",
    collaboration = "STAR",
    title = "{Pattern of global spin alignment of $\phi$ and $K^{*0}$ mesons in heavy-ion collisions}",
    eprint = "2204.02302",
    archivePrefix = "arXiv",
    primaryClass = "hep-ph",
    doi = "10.1038/s41586-022-05557-5",
    journal = "Nature",
    volume = "614",
    number = "7947",
    pages = "244--248",
    year = "2023"
}

@article{Ayala:2021osy,
    author = "Ayala, Alejandro and Hern{\'a}ndez, L. A. and Raya, K. and Zamora, R.",
    title = "{Fermion propagator in a rotating environment}",
    eprint = "2102.03476",
    archivePrefix = "arXiv",
    primaryClass = "hep-ph",
    doi = "10.1103/PhysRevD.103.076021",
    journal = "Phys. Rev. D",
    volume = "103",
    number = "7",
    pages = "076021",
    year = "2021",
    note = "[Erratum: Phys. Rev. D 104, 039901 (2021)]"
}

@article{Gaspar:2023nqk,
    author = "Gaspar, Irving I. and Hern{\'a}ndez, Luis A. and Zamora, Renato",
    title = "{Chiral symmetry restoration in a rotating medium}",
    eprint = "2305.00101",
    archivePrefix = "arXiv",
    primaryClass = "hep-ph",
    doi = "10.1103/PhysRevD.108.094020",
    journal = "Phys. Rev. D",
    volume = "108",
    number = "9",
    pages = "094020",
    year = "2023"
}

@article{Ayala:2026dmz,
  title = {Excitation function for global $\mathrm{\ensuremath{\Lambda}}$ polarization in relativistic heavy-ion collisions with the core-corona model},
  author = {Ayala, Alejandro and Medina-Serna, Jos\'e Jorge and Dom\'{\i}nguez, Isabel and Tejeda-Yeomans, Mar\'{\i}a Elena},
  journal = {Phys. Rev. D},
  volume = {114},
  issue = {1},
  pages = {016034},
  numpages = {16},
  year = {2026},
  month = {Jul},
  publisher = {American Physical Society},
  doi = {10.1103/fbt2-9gn1},
  url = {https://link.aps.org/doi/10.1103/fbt2-9gn1}
}

@article{CastanoYepes:2025,
  title = {Dilepton production in a rotating thermal medium: The rigid rotation approximation},
  author = {Casta\~no-Yepes, Jorge David and Mu\~noz, Enrique},
  journal = {Phys. Rev. D},
  volume = {113},
  issue = {11},
  pages = {116001},
  numpages = {26},
  year = {2026},
  month = {Jun},
  publisher = {American Physical Society},
  doi = {10.1103/fkfs-nlbm},
  url = {https://link.aps.org/doi/10.1103/fkfs-nlbm}
}

@article{Ambrus_PhysRevD.93.104014,
  title = {Rotating fermions inside a cylindrical boundary},
  author = {Ambru\ifmmode \mbox{\c{s}}\else \c{s}\fi{}, Victor E. and Winstanley, Elizabeth},
  journal = {Phys. Rev. D},
  volume = {93},
  issue = {10},
  pages = {104014},
  numpages = {43},
  year = {2016},
  month = {May},
  publisher = {American Physical Society},
  doi = {10.1103/PhysRevD.93.104014},
  url = {https://link.aps.org/doi/10.1103/PhysRevD.93.104014}
}

@article{GervoisNavelet1984,
  author  = {Gervois, Annie and Navelet, Henri},
  title   = {Some integrals involving three Bessel functions when their arguments satisfy the triangle inequalities},
  journal = {Journal of Mathematical Physics},
  volume  = {25},
  number  = {11},
  pages   = {3350--3356},
  year    = {1984},
  doi     = {https://doi.org/10.1063/1.526062}
}

@article{Geurts:2023ElectromagneticProbes,
  author        = {Geurts, Frank and Tripolt, Ralf-Arno},
  title         = {Electromagnetic probes: Theory and experiment},
  journal       = {Progress in Particle and Nuclear Physics},
  volume        = {128},
  pages         = {104004},
  year          = {2023},
  doi           = {10.1016/j.ppnp.2022.104004},
  eprint        = {2210.01622},
  archivePrefix = {arXiv},
  primaryClass  = {hep-ph}
}

@article{Heinz:2013CollectiveFlow,
  author  = {Heinz, Ulrich and Snellings, Raimond},
  title   = {Collective Flow and Viscosity in Relativistic Heavy-Ion Collisions},
  journal = {Annu. Rev. Nucl. Part. Sci.},
  volume  = {63},
  pages   = {123--151},
  year    = {2013},
  doi     = {10.1146/annurev-nucl-102212-170540}
}

@article{Shen:2021QGPFlow,
  author  = {Shen, Chun},
  title   = {Studying QGP with flow: A theory overview},
  journal = {Nucl. Phys. A},
  volume  = {1005},
  pages   = {121788},
  year    = {2021},
  doi     = {10.1016/j.nuclphysa.2020.121788}
}

@article{David:2020DirectPhotons,
  author        = {David, Gabor},
  title         = {Direct real photons in relativistic heavy ion collisions},
  journal       = {Reports on Progress in Physics},
  volume        = {83},
  number        = {4},
  pages         = {046301},
  year          = {2020},
  doi           = {10.1088/1361-6633/ab6f57},
  eprint        = {1907.08893},
  archivePrefix = {arXiv},
  primaryClass  = {nucl-ex}
}

@article{CastanoYepes:2026PhotonRotating,
  title         = {Photon Quantization and Propagator in Rotating Coordinates},
  author        = {Ayala, Alejandro and Bernal-Langarica, Santiago and Casta{\~n}o-Yepes, Jorge David and Medina-Serna, Jos{\'e} Jorge},
  journal       = {arXiv},
  year          = {2026},
  eprint        = {2610.02656},
  archivePrefix = {arXiv},
  primaryClass  = {hep-ph},
  version       = {v1}
}

\end{document}